\documentclass[pra,onecolumn,11pt,amsmath,amssymb,amsthm,superscriptaddress]{revtex4-2}

\usepackage{graphicx,amsmath,relsize,epstopdf,upgreek,color,mathtools,bm,times,tabularx,booktabs}
\usepackage[colorlinks=true,linkcolor=blue,citecolor=blue,urlcolor =blue]{hyperref}
\usepackage[hyphenbreaks]{breakurl}

\newcommand{\ket}[1]{|{#1}\rangle}
\newcommand{\bra}[1]{\langle{#1}|}

\newcommand{\opinner}[3]{\langle #1|#2|#3\rangle}

\newcommand{\rvec}[1]{\pmb{#1}}
\newcommand{\dyadic}[1]{{\bf#1}}
\newcommand{\tr}[1]{\mathrm{tr}\left\{#1\right\}}
\newcommand{\ptr}[2]{\mathrm{tr}_{#1}\left\{#2\right\}}

\newcommand{\E}[1]{\mathrm{e}^{\mbox{\footnotesize$#1$}}}

\newcommand{\MEAN}[2]{\left<{#1}\right>_{#2}}

\newcommand{\MSE}[1]{\mathrm{MSE}_{#1}}

\newcommand{\poly}{\mathrm{poly}}
\newcommand{\vacket}{\ket{\textsc{vac}}}
\newcommand{\vacbra}{\bra{\textsc{vac}}}
\newcommand{\appropto}{\mathrel{\vcenter{
			\offinterlineskip\halign{\hfil$##$\cr
				\propto\cr\noalign{\kern2pt}\sim\cr\noalign{\kern-2pt}}}}}

\begin{document}

\title{Virtual distillation with noise dilution}

\author{Yong Siah Teo}
\email{yong.siah.teo@gmail.com}
\affiliation{Department of Physics and Astronomy, 
	Seoul National University, 08826 Seoul, South Korea}

\author{Seongwook Shin}
\affiliation{Department of Physics and Astronomy, 
	Seoul National University, 08826 Seoul, South Korea}

\author{Hyukgun Kwon}
\affiliation{Department of Physics and Astronomy, 
	Seoul National University, 08826 Seoul, South Korea}

\author{Seok-Hyung Lee}
\affiliation{Department of Physics and Astronomy, 
	Seoul National University, 08826 Seoul, South Korea}

\author{Hyunseok Jeong}
\email{h.jeong37@gmail.com}
\affiliation{Department of Physics and Astronomy, 
	Seoul National University, 08826 Seoul, South Korea}

\begin{abstract}
	Virtual distillation is an error-mitigation technique that reduces quantum-computation errors without assuming the noise type. In scenarios where the user of a quantum circuit is required to additionally employ peripherals, such as delay lines, that introduce excess noise, we find that the error-mitigation performance can be improved if the peripheral, whenever possible, is split across the entire circuit; that is, when the noise channel is uniformly distributed in layers within the circuit. We show that under the multiqubit loss and Pauli noise channels respectively, for a given overall error rate, the average mitigation performance improves monotonically as the noisy peripheral is split~(diluted) into more layers, with each layer sandwiched between subcircuits that are sufficiently deep to behave as two-designs. For both channels, analytical and numerical evidence show that second-order distillation is generally sufficient for (near-)optimal mitigation. We propose an application of these findings in designing a quantum-computing cluster that houses realistic noisy intermediate-scale quantum circuits that may be shallow in depth, where measurement detectors are limited and delay lines are necessary to queue output qubits from multiple circuits.
\end{abstract}

\maketitle

\section{Introduction}

In principle, owing to the existence of universal quantum gate sets~\cite{Deutsch:1995universality,Barenco:1995elementary,Englert:2001universal,Bartlett:2002efficient,Sawicki:2022universality} for computation, full-fledged quantum computers~\cite{Chuang:2000fk,Ladd:2010aa,Campbell:2017aa,Lekitsch:2017aa} that obey the laws of quantum mechanics are promising devices that permit universal fault-tolerant quantum computation~\cite{Grover:1996fast,Shor:1997polynomial,Raussendorf:2001one-way,Kitaev:2003fault-tolerant,Raussendorf:2007topological,Sehrawat:2011test-state,Montanaro:2016quantum}, with the possibility of surpassing the performances of presently known classical algorithms. In practice, however, numerous challenges remain to be resolved before such a quantum vision can be realized. These include the qualities of qubit sources, gates and detectors~\cite{Knill:1998resilient,Franklin:2004challenges,Aharonov:2008fault-tolerant}, and the exponentially large number of components necessary to construct arbitrary quantum circuits~\cite{knill1995approximation}.

At present, one only has access to \emph{noisy intermediate-scale quantum}~(NISQ) devices~\cite{Preskill2018quantumcomputingin} that are capable of manipulating less than a thousand qubits using noisy unitary gates and measurements. These limitations motivated the development of several kinds of NISQ algorithms~\cite{Bromley:2020applications,Bharti:2022noisy,Finnila:1994quantum,Kadowaki:1998quantum,Aaronson:2011computational,Aaronson:2011linear-optical,Hamilton:2017gaussian,Trabesinger:2012quantum,Georgescu:2014quantum}. Most notably, variational quantum algorithms~(VQAs)~\cite{Biamonte:2021universal,Cerezo:2021variational,Cao:2019quantum,Endo:2021hybrid,McArdle:2020quantum,Teo:2022optimized} perform computations using both classical and NISQ devices in a hybrid fashion, which find applications in variational quantum eigensolvers designed for quantum-chemistry~\cite{Peruzzo:2014variational,Wecker:2015progress,McClean:2016theory} and combinatorial problems~\cite{Farhi:2014quantum,Zhou:2020quantum}, and quantum machine learning~\cite{Schuld:2015introduction,Schuld:2019quantum,Carleo:2019machine,date2020quantum,Perez-Salinas:2020aa,dutta2021singlequbit,Goto:2021universal,Shin:2022exponential}. 

Owing to noise accompanying any NISQ device, the output state of its circuit will deviate from the target state of interest to the computation procedure. Without resorting to quantum tomography~\cite{lnp:2004uq,Teo:2015introduction,Gross:2010quantum,Ahn:2019adaptive,Teo:2021modern} for certifying the circuit output quality, the techniques of \emph{error mitigation} may be considered as alternatives to directly reduce errors due to noise channels. The names of specific error-mitigation procedures have grown rapidly in recent years, which makes it implausible to cite them all. While interested readers may refer to~\cite{Bharti:2022noisy,Qin:2022overview} for a comprehensive survey, we shall highlight some representative techniques. Methods such as Richardson or zero-noise extrapolation and probabilistic error cancellation~\cite{Li:2017efficient,Temme:2017error,Giurgica-Tiron:2020digital} for estimating the circuit output in the near-noiseless regime requires knowledge about the noise model and precise control of the quantum-circuit parameters, the former of which could be replaced by gate-set tomography~\cite{Greenbaum:2015introduction,Chao:2019quantum,Zhang:2020error-mitigated,Kwon:2021hybrid}, which should be carried out prior to the quantum computation. Other methods, such as the subspace reduction~\cite{McClean:2017hybrid,McClean:2020decoding} method demand additional ancillary qubits appended to the main NISQ circuit, which can be very large especially when the noise model is unknown. Iterative power-series-based or perturbative methods hold promise to mitigate errors under noise-agnostic situations. Nevertheless, such methods may require a large number of measurements that may be reduced by employing manipulative tricks for certain target measurement observables~\cite{Suchsland:2021algorithmic}, and in other cases, may only work on invertible noise channels without bias~\cite{Wang:2021measurement}.

Virtual distillation~\cite{Koczor:2021exponential,Huggins:2021virtual,Yamamoto:2021error-mitigated} is yet another technique that can mitigate noise of small error rates in a model-agnostic manner. Moreover, mitigation happens on-the-fly either with an external correcting circuit~\cite{Huggins:2021virtual} or with efficient data post-processing using shadow tomography~\cite{Seif:2022shadow} that requires only randomized single-qubit unitary gates. Consequently, this technique can cope with noise drifts, which is an attractive feature in addition to its technical simplicity that permits accessible analysis.

While it is true that noise can influence a NISQ device anywhere in an uncontrollable way, in this work, we focus on scenarios where other than errors originating from ambient environment, certain external peripherals employed in addition to the main quantum-computation circuit could result in excess noise on the entire device. Examples of such peripherals could be delay lines and active switches. In these cases, one has the freedom of arranging peripherals within the circuit, thereby altering the effective noise-channel acting on the device. This work investigates both the multiqubit loss and Pauli noise channels, which respectively describe photon losses~\cite{Mogilevtsev:2010single-photon,Carpenter:2013nonlinear,Shomroni:2014all-optical,Bonneau:2015on-demand,Mendoza:2016active,Jones:2018PolarizationDL,Omkar:2020resource-efficient,Bartolucci:2021switch,Omkar:2022all-photonic,Kim:2022quantum} and polarization disturbances~\cite{Dragan:2005depolarization,Bayat:2006threshold,Karpinski:2008fiber,Amaral:2019characterization} in optical components. For small error rates, we analytically show that virtual distillation can exponentially mitigate errors due to the loss channel with increasing distillation order, whereas mitigation improvement stagnates beyond the second distillation order for the multiqubit Pauli channel. Furthermore, virtual distillation can, on average, better mitigate errors when the peripherals are split into more layers across the quantum circuit. The former is equivalent to diluting the noise channel into multiple layers within the circuit, where one such noise layer~(except for the last one) is sandwiched between two subcircuits. All analysis is carried out under the assumption that every sandwiching subcircuit is approximately a two-design~\cite{harrow_random_2009}. 

Finally, we give a practical application to this main result by studying the performance of virtual distillation on outputs of a quantum-computing cluster that contains several hardware-efficient circuits and limited detectors. In this situation, delay lines are necessary to queue the output qubits coming from concurrent circuits. According to the main result, under the same total delay time, it is evidently better to transmit qubits of uniformly-delayed unitary operations to the detectors instead of delaying the transmission of qubits after all unitary operations are performed quickly.

\section{Methods}

\subsection{Virtual distillation}

Suppose that an effective noise channel $\Phi_\mu$ acts on a pure state $\ket{\,\,\,}\bra{\,\,\,}$ of Hilbert-space dimension $d$, and is parametrized by $0\leq\mu\leq1$ that quantifies the ``strength'' of $\Phi$. Then, the resulting noisy state
\begin{equation}
	\rho'=\Phi_\mu[\rho]=\ket{\lambda_0(\mu)}\lambda_0(\mu)\bra{\lambda_0(\mu)}+\sum^{d-1}_{k=1}\ket{\lambda_k(\mu)}\lambda_k(\mu)\bra{\lambda_k(\mu)}
	\label{eq:noisy_eigen}
\end{equation}
possesses a spectral decomposition with eigenvalues ordered according to $\lambda_0(\mu)>\lambda_1(\mu)\geq\lambda_2(\mu)\geq\ldots\geq\lambda_{d-1}(\mu)$, where for sufficiently small $\mu$, the eigenstate $\ket{\lambda_0(\mu)}\bra{\lambda_0(\mu)}$ is in general close to the target pure state $\ket{\,\,\,}\bra{\,\,\,}$. On the other hand, $\rho'$ deviates significantly from $\ket{\,\,\,}\bra{\,\,\,}$ at maximal channel noise~($\mu=1$). Since $\Phi_\mu$ is completely-positive and trace-preserving, it may also be expressed as
\begin{equation}
	\rho'= \sum_m K_{m,\mu}\ket{\,\,\,}\bra{\,\,\,}K_{m,\mu}^\dag
\end{equation}
using a set of Kraus operators $\{K_{0,\mu},K_{1,\mu},K_{2,\mu},\ldots\}$ of the general property $K_{m,\mu=0}=\delta_{m,0}1$. Using a Hermitian operator basis $\{\Gamma_l\}^{d^2-1}_{l=0}$ such that $\Gamma_0=1/\sqrt{d}$, $\tr{\Gamma_{l>0}}=0$ and $\tr{\Gamma_l\Gamma_{l'}}=\delta_{l,l'}$, we have the expansion
\begin{equation}
	K_{m,\mu}=\frac{\gamma^{(m,\mu)}_0}{\sqrt{d}}1+\sum^{d^2-1}_{l=1}\gamma^{(m,\mu)}_l\Gamma_l
\end{equation}
in terms of \emph{complex} coefficients $\gamma^{(m,\mu)}_l$, where $\gamma^{(m,\mu=0)}_l=\sqrt{d}\,\delta_{l,0}\,\delta_{m,0}$. We may now group all nondominant terms together and give an equivalently exact representation for $\rho'$:
\begin{equation}
	\rho'=\ket{\,\,\,}[1-\epsilon(\mu)]\bra{\,\,\,}+\underbrace{\epsilon(\mu)\,\widetilde{\rho}_\mathrm{err}(\mu,\ket{\,\,\,}\bra{\,\,\,})}_{\displaystyle\mathclap{\text{nondominant}}}\,,
	\label{eq:easy_rep}
\end{equation}
with $\epsilon(\mu)=1-\left|\gamma^{(0,\mu)}_0\right|^2/d$, so that $\epsilon(\mu\rightarrow0)\rightarrow0$. The state $\widetilde{\rho}_\mathrm{err}(\mu,\ket{\,\,\,}\bra{\,\,\,})$ is the \emph{error component} owing to $\Phi_\mu=\Phi_\epsilon$ of an \emph{error rate} $\epsilon=\epsilon(\mu)$. Very generally, $\widetilde{\rho}_\mathrm{err}(\mu,\ket{\,\,\,}\bra{\,\,\,})$ can be a nonlinear operator function of $\mu$, and $[\widetilde{\rho}_\mathrm{err}(\mu,\ket{\,\,\,}\bra{\,\,\,}),\ket{\,\,\,}\bra{\,\,\,}]\neq0$.

A special case of \eqref{eq:easy_rep} arises in the regime $\mu,\epsilon\ll1$. In this case, it can be deduced~(see Appendix~\ref{app:small_err}) that
\begin{equation}
	\rho'=\ket{\,\,\,}(1-\epsilon)\bra{\,\,\,}+\epsilon\rho_\mathrm{err}(\ket{\,\,\,}\bra{\,\,\,})\,,
	\label{eq:small_err_rho_prime}
\end{equation}
where $\rho_\mathrm{err}$ is an $\epsilon$-independent error component, but could still depend on $\ket{\,\,\,}\bra{\,\,\,}$. Apart from this approximation, Eq.~\eqref{eq:small_err_rho_prime} is exact with \emph{any} $\epsilon$ for many commonly-known noise channels. 


There exists a simple error-mitigation procedure that utilizes a basic linear-algebraic principle. To illustrate such a procedure, let us first recall, which we have earlier assumed with a sleight of hand, that the \emph{dominant} eigenvalue $\lambda_0(\mu)$ in Eq.~\eqref{eq:noisy_eigen} is \emph{nondegenerate}, which is the typical case for noisy environments, as the situation of coincidentally having another dominant eigenstate of the same eigenvalue is unlikely for small~$\epsilon$. Then, because $\lambda_0(\mu)>\lambda_{k>0}(\mu)$, it is clear that
\begin{equation}
	\lim_{M\rightarrow\infty}\dfrac{\rho'^M}{\tr{\rho'^M}}=\ket{\lambda_0(\mu)}\bra{\lambda_0(\mu)}\,;
\end{equation} 
that is, raising $\rho'$ to a very large power $M$ (the \emph{distillation order}) and normalizing the answer amplifies the dominant eigenvalue $\lambda_0(\mu)$, thereby asymptotically leading to the singly-dominant eigenstate. This method may be traced back to von~Mises~\cite{vonMises:1929praktische}, and is a common numerical technique for finding the largest eigenvalue of a matrix. Hence, for a sufficiently small $\mu$ or $\epsilon$, this dominant pure state is close to the target state $\ket{\,\,\,}\bra{\,\,\,}$ under such a simple purification or \emph{distillation} scheme.

In a VQA setting, one is generally interested in measuring the expectation value $\MEAN{O}{}$ of a Hermitian observable $O$ with respect to some target $\ket{\,\,\,}\bra{\,\,\,}$. The use of such a distillation scheme would therefore entail the corresponding measurement of $\tr{\rho'^M\,O}/\tr{\rho'^M}$. Recipes to implement such a measurement from multiple copies of $\rho'$ using two-qubit entangling gates have been proposed~\cite{Koczor:2021exponential,Huggins:2021virtual,Lowe:2021unified,Cotler:2019quantum,Cai:2021resource-efficient,Huo:2022dual-state,Czarnik:2021qubit-efficient,Xiong:2022quantum}. More recently, a separate idea of using shadow tomography~\cite{Seif:2022shadow,Aaronson:2017shadow,Huang:2020predicting,Paini:2021estimating,Chen:2021robust} as an efficient post-processing protocol for estimating $\tr{\rho'^M\,O}/\tr{\rho'^M}$ with only randomized single-qubit unitary rotations in addition to the VQA circuit further enhances the feasibility of this scheme. The name \emph{virtual distillation} is fitting, since all practical implementations never directly generate the distilled state $\rho'^M/\tr{\rho'^M}$, but only estimate the corresponding observable measurements.

\subsection{Multiqubit loss channel}
\label{subsec:iid_loss}
Although virtual distillation is agnostic to any particular noise model, for the purpose of analysis and discussion, we shall investigate two particular classes of noise channels. The first class consists of the multiqubit loss channels. Let us start with the simplest case, that is the single-qubit loss channel defined by the completely-positive and trace-preserving (CPTP) map
\begin{equation}
	\Phi^\mathrm{loss}_\epsilon[\rho_\mathrm{qubit}]=(1-\epsilon)\rho_\mathrm{qubit}+\vacket\epsilon\vacbra
	\label{eq:1qubit_loss}
\end{equation}
for \emph{any} state $\rho_\mathrm{qubit}$. This map may be derived by solving for the response of the polarization degree of freedom with respect to the master equation~\cite{Phoenix:1990wave-packet}
\begin{equation}
	\dfrac{\partial\rho'_\mathrm{qubit}}{\partial t}=\gamma\sum^1_{j=0}\left(a_j\,\rho'_\mathrm{qubit}\, a_j^\dag-\frac{1}{2}\,a^\dag_j a_j\,\rho'_\mathrm{qubit}-\frac{1}{2}\,\rho'_\mathrm{qubit}\,a^\dag_j a_j\right)\,,
	\label{eq:loss_markov}
\end{equation}
where $a_j$ is the annihilation operator on the polarization ket $\ket{j}$, and $\epsilon=1-\E{-\gamma t}$ relates to the evolution time period $t$ and decay rate $\gamma$~(see Appendix~\ref{app:loss_channel}). The linearity of $\Phi^\mathrm{loss}_\epsilon$, or any CPTP map for that matter, implies the following operator actions:
\begin{align}
	\Phi^\mathrm{loss}_\epsilon[\ket{0}\bra{0}]=&\,\ket{0}(1-\epsilon)\bra{0}+\vacket\epsilon\vacbra\,,\nonumber\\
	\Phi^\mathrm{loss}_\epsilon[\ket{1}\bra{1}]=&\,\ket{1}(1-\epsilon)\bra{1}+\vacket\epsilon\vacbra\,,\nonumber\\
	\Phi^\mathrm{loss}_\epsilon[\ket{0}\bra{1}]=&\,\ket{0}(1-\epsilon)\bra{1}\,,\nonumber\\
	\Phi^\mathrm{loss}_\epsilon[\vacket\vacbra]=&\,\vacket\vacbra\,.
	\label{eq:pol_loss}
\end{align}

In the multiqubit product basis, an $n$-qubit noiseless state $\rho$ may be written as
\begin{align}
	\rho=&\,\sum^1_{l_1,l_2,\ldots,l_n=0}\,\sum^1_{l'_1,l'_2,\ldots,l'_n=0}\ket{l_1,l_2,\ldots,l_n}\,\rho_{l_1l_2\ldots l_n\,;\,l'_1l'_2\ldots l'_n}\,\bra{l'_1,l'_2,\ldots,l'_n}\,,
\end{align}
where each qubit basis ket is now extended to the three-dimensional space inasmuch as
\begin{equation}
	\ket{0}\,\widehat{=}\begin{pmatrix}
		1\\
		0\\
		0
	\end{pmatrix}\,,\,\,\ket{1}\,\widehat{=}\begin{pmatrix}
		0\\
		1\\
		0
	\end{pmatrix}\,,\,\,\vacket\,\widehat{=}\begin{pmatrix}
		0\\
		0\\
		1
	\end{pmatrix}\,.
\end{equation}
After subjecting each qubit to the loss channel in Eq.~\eqref{eq:1qubit_loss} under a \emph{common} small error rate $\epsilon$, we arrive at the noisy state
\begin{align}
	\rho'=&\,(1-\epsilon)^n\rho+\epsilon(1-\epsilon)^{n-1}\sum_{\rvec{l},\,\rvec{l}'}\rho_{\rvec{l};\,\rvec{l}'}\left(\begin{array}{rl}&\!\!\!\,\vacket\delta_{l_1,l_1'}\vacbra\otimes\ket{l_2}\bra{l'_2}\otimes\ldots\otimes\ket{l_n}\bra{l'_n}\\
	+&\!\!\!\,\ket{l_1}\bra{l'_1}\otimes\vacket\delta_{l_2,l_2'}\vacbra\otimes\ldots\otimes\ket{l_n}\bra{l'_n}\\
	+&\!\!\!\ldots\\
	+&\!\!\!\,\ket{l_1}\bra{l'_1}\otimes\ldots\otimes\ket{l_{n-1}}\bra{l'_{n-1}}\otimes\vacket\delta_{l_n,l_n'}\vacbra\\
	\end{array}\right)\nonumber\\
	&+\,\{\text{vacuum-related terms of higher $\epsilon$ orders}\}\nonumber\\
	\cong&\,(1-n\epsilon)\rho+\epsilon\left(\vacket\vacbra\otimes\ptr{1}{\rho}+\ldots+\ptr{n}{\rho}\otimes\vacket\vacbra\right)\,,
	\label{eq:nqubit_loss}
\end{align}
where we see that a photon-loss action on the $j$th qubit is equivalent to a partial trace ($\ptr{j}{\,\,\bm{\cdot}\,\,}$) applied to that qubit followed by a vacuum-state substitution. We emphasize here that all vacuum-related terms are mutually orthogonal with each other and the noiseless state $\rho$. Such mutual orthogonality shall prove advantageous in subsequent analysis for this channel.

\subsection{Multiqubit Pauli channel}
\label{subsec:iid_pauli}

The second class of multiqubit Pauli channels not only generates enormous interests in the fields of error correction and quantum computing~\cite{Fujiwara:2003quantum,Chiuri:2011experimental,Omkar:2013dissipative,Flammia:2020efficient,Terhal:2015quantum,Wallman:2016noise,Sanders:2015bounding,Kueng:2016comparing,Huang:2019performance,Siudziska:2020classical,Chen:2022quantum}, but is also relevant in the discussion of noise models generated from common quantum-optical peripherals, for instance, such as polarization disturbances in optical-fiber-based delay lines~\cite{Dragan:2005depolarization,Bayat:2006threshold,Karpinski:2008fiber}.

For a single-qubit, the Pauli channel is defined by the four Kraus operators $K_0=\sqrt{1-\epsilon}\,1$, $K_1=\sqrt{\epsilon_1}\,X$, $K_2=\sqrt{\epsilon_2}\,Y$ and $K_3=\sqrt{\epsilon_3}\,Z$, where $1\geq\epsilon=\epsilon_1+\epsilon_2+\epsilon_3$, resulting in the single-qubit CPTP map
\begin{equation}
	\Phi^{\mathrm{Pauli}}_\epsilon[\rho_\mathrm{qubit}]=(1-\epsilon)\,\rho_\mathrm{qubit}+\epsilon_1X\,\rho_\mathrm{qubit}\,X+\epsilon_2Y\,\rho_\mathrm{qubit}\,Y+\epsilon_3Z\,\rho_\mathrm{qubit}\,Z\,.
\end{equation}
The Pauli channel is isotropically~\emph{depolarizing} when $\epsilon_1=\epsilon_2=\epsilon_3=\epsilon/3$.

The multiqubit version naturally involves Kraus operators that are tensor-product. Given a pure target $n$-qubit state $\rho$, the corresponding noisy state $\rho'$ as a result of the multiqubit Pauli noise channel with identical qubit error rates, up to first order in $\epsilon_j$, is given by
\begin{equation}
	\rho'\cong(1-n\epsilon)\rho+\epsilon_1\sum^n_{j=1}X_j\rho X_j+\epsilon_2\sum^n_{j=1}Y_j\rho Y_j+\epsilon_3\sum^n_{j=1}Z_j\rho Z_j\,,
	\label{eq:nqubit_pauli}
\end{equation}
where $X_j$, for instance, refers to the $X$ operator for the $j$th qubit. For this channel, the corresponding error component $\rho_\mathrm{err}\propto\epsilon_1\sum^n_{j=1}X_j\rho X_j+\epsilon_2\sum^n_{j=1}Y_j\rho Y_j+\epsilon_3\sum^n_{j=1}Z_j\rho Z_j$ does not commute with $\rho$ in general.

\subsection{Noise dilution}
\label{subsec:noise_dilution}

Equations~\eqref{eq:nqubit_loss} and \eqref{eq:nqubit_pauli} are representatives of independent and identically distributed (i.i.d.)~noise models, namely that all qubit suffers from the same kind of noise of equal error rate $\epsilon$, where the resulting noisy $n$-qubit state $\rho'$ takes the form
\begin{equation}
	\rho'_1=\Phi^{\mathrm{i.i.d.}}_\epsilon[\rho]=(1-\epsilon)^n\rho+[1-(1-\epsilon)^n]\,\rho_\mathrm{err}
	\label{eq:Lerr_1}
\end{equation}
for any $0\leq\epsilon\leq1$ and some error component $\rho_\mathrm{err}$. 

That $\Phi^{\mathrm{i.i.d.}}_\epsilon$ is the i.i.d. noise channel due to a given quantum-circuit peripheral invites the concept of \emph{noise dilution}. Suppose that this peripheral is used in conjunction with a quantum circuit that is described by the unitary operator $U$, and can be split into several smaller peripherals. The corresponding noiseless output state $\rho=U\ket{\rvec{0}}\bra{\rvec{0}}U^\dag$, where $\ket{\rvec{0}}\bra{\rvec{0}}$ is some fixed initialized pure state. Given the possible decomposition $U=W_1W_2\ldots W_{L_\mathrm{err}}$ in terms of the unitary operators $W_1$, $W_2$, \ldots, $W_{L_\mathrm{err}}$, the user may choose to distribute the peripheral across $U$, so that the output state of the circuit is given by
\begin{align}
	\rho'_{L_\mathrm{err}}=&\,\Phi^\mathrm{i.i.d.}_{\epsilon/L_\mathrm{err}}\left[W_{L_\mathrm{err}}\ldots\Phi^\mathrm{i.i.d.}_{\epsilon/L_\mathrm{err}}\left[W_2\,\Phi^\mathrm{i.i.d.}_{\epsilon/L_\mathrm{err}}\left[W_1\ket{\rvec{0}}\bra{\rvec{0}}W_1^\dag\right] W_2^\dag\right]\ldots W_{L_\mathrm{err}}^\dag\right]\nonumber\\
	=&\,\left(1-\frac{\epsilon}{L_\mathrm{err}}\right)^{L_\mathrm{err}n}\rho+\left[1-\left(1-\frac{\epsilon}{L_\mathrm{err}}\right)^{L_\mathrm{err}n}\right]\,\rho^{(L_\mathrm{err})}_\mathrm{err}\,,
	\label{eq:Lerr}
\end{align}
where each of the $L_\mathrm{err}$ i.i.d. channels now acts with an error rate of $\epsilon/L_\mathrm{err}$. Example situations that are relevant for such a choice is the arrangement of delay lines that are necessary in many circumstances for the practical implementation of the NISQ device. In these situations, the user may choose to either use delay lines after the computation with $U$~\eqref{eq:Lerr_1} or distribute them uniformly across $U$~\eqref{eq:Lerr}. 

\begin{figure}[t]
	\centering
	\includegraphics[width=0.8\columnwidth]{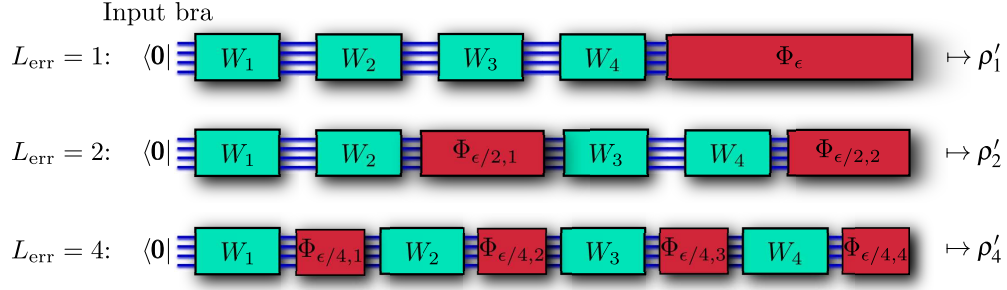}
	\caption{\label{fig:noise_dilution}Dilution of a peripheral i.i.d. noise channel $\Phi^{\mathrm{i.i.d.}}_\epsilon$ in a four-qubit~($d=2^4=16$) quantum circuit that accepts the initialized input state $\ket{\rvec{0}}\bra{\rvec{0}}$. Here, the number of noise layers, $L_\mathrm{err}$, are chosen to be 1~(no dilution), 2 and 4.}
\end{figure}

For small error rates $\epsilon$, 
\begin{align}
	\rho'_1\cong&\,(1-n\epsilon)\rho+n\epsilon\,\rho_\mathrm{err}\,,\nonumber\\
	\rho'_{L_\mathrm{err}}\cong&\,(1-n\epsilon)\rho+n\epsilon\,\rho^{(L_\mathrm{err})}_\mathrm{err}\,,
	\label{eq:dilution}
\end{align}
where we see that the noise level for both choices, measured as the weight attributed to $\rho$, is the same and only the error components differ~$\left[\rho^{(L_\mathrm{err}=1)}_\mathrm{err}=\rho_\mathrm{err}\right]$. Figure~\ref{fig:noise_dilution} illustrates \eqref{eq:dilution} for $L_\mathrm{err}=1,2,4$ with a four-qubit quantum circuit, where $W_1$, $W_2$, $W_3$ and $W_4$ are unitary operators of subcircuits that make up $U$. We shall compare the performance of virtual distillation on noisy output states for various values of $L_\mathrm{err}$.

There is a technical exception to \eqref{eq:Lerr_1}, \eqref{eq:Lerr} and \eqref{eq:dilution}, namely with the i.i.d. loss channels as described in Sec.~\ref{subsec:iid_loss}. For such a channel, it turns out that
\begin{equation}
	(\text{I.i.d. loss, trace unpreserved})\quad\rho'_{L_\mathrm{err}}\cong\left(1-\frac{\epsilon}{L_\mathrm{err}}\right)^{L_\mathrm{err}n}\rho+\frac{n\epsilon}{L_\mathrm{err}}\left(1-\frac{\epsilon}{L_\mathrm{err}}\right)^{L_\mathrm{err}n-1}\rho^{(L_\mathrm{err})}_\mathrm{err}
	\label{eq:lossy_trace}
\end{equation}
for a small loss error rate $\epsilon$~[see also either \eqref{eq:small_err_loss_state} or \eqref{eq:full_loss_product}], with a trace that is unpreserved even up to first order in $\epsilon$ whenever $L_\mathrm{err}>1$, unlike the second equality in~\eqref{eq:Lerr}. Upon a trace renormalization,
\begin{equation}
	(\text{I.i.d. loss, trace preserved})\quad\rho'_{L_\mathrm{err}}\cong\left(1-\frac{n\epsilon}{L_\mathrm{err}}\right)\rho+\frac{n\epsilon}{L_\mathrm{err}}\,\rho^{(L_\mathrm{err})}_\mathrm{err}\,.
	\label{eq:lossy_trace_renorm}
\end{equation}
The reason for the trace-lossy form in~\eqref{eq:lossy_trace} is that while the trace of $\rho'$ in \eqref{eq:Lerr} is preserved throughout the noise dilution procedure under typical noise channels [even up to first-order approximation in $\epsilon$ as in \eqref{eq:dilution}], this is not the case for the i.i.d. loss channel. If after the measurement phase, data corresponding to detector clicks with missing qubits are to be discarded anyway, then the effective action of the subcircuit $W_l$ on a lossy state at every dilution layer amounts to losing information about the error component, unless $L_\mathrm{err}=1$. In this sense, the vacuum state is invisible to circuit operations in practice.

Physically, the noise dilution strategy outlined here is equivalent to a redistribution of noisy peripherals such that certain aspects of the peripherals are conserved. As a working example which shall be the main theme of this article, consider a realistic physical situation where the lossy peripheral is a delay line of a certain decay rate $\gamma$ for which the error rate $\epsilon=1-\E{-\gamma \tau}$ after some delay time period $\tau$. For a small~$\tau$, we find that $\epsilon\cong\gamma \tau$, so that the $L_\mathrm{err}$-layered noise dilution scheme outlined here is equivalent to splitting the delay line into $L_\mathrm{err}$ equal delay times $\tau/L_\mathrm{err}$ and distributing them evenly throughout the quantum circuit, whilst preserving the total delay time $\tau$.

\subsection{Figure of merit and circuit averaging}

To compare the mitigative power of virtual distillation in noise-diluted scenarios of various $L_\mathrm{err}$, we take the figure of merit to be the Hilbert--Schmidt distance or \emph{mean squared-error}~(MSE) between a target pure state $\rho=\ket{\,\,\,}\bra{\,\,\,}=U\ket{\rvec{0}}\bra{\rvec{0}}U^\dag$ and the corresponding noisy state $\rho'$ subjected to some noise-channel map $\Phi^{\mathrm{i.i.d.}}_\epsilon$ of error rate $\epsilon$. This is defined as
\begin{equation}
	\MSE{}=\MEAN{\tr{(\rho-\rho')^2}}{}\,,
	\label{eq:mse_def}
\end{equation}
where the average $\MEAN{\,\,\bm{\cdot}\,\,}{}$ is taken over all possible independent circuit unitary operators. For instance, in Fig.~\ref{fig:noise_dilution}, based on that particular decomposition of circuit unitary $U$, the average is taken over all possible $W_1$, $W_2$, $W_3$ and $W_4$. Such a circuit-averaged figure of merit quantifies the average accuracy over all possible randomly-chosen circuit parameters that define $U$.

To obtain analytical formulas, we shall assume that all unitary operators $W_j$ are two-designs~\cite{Dankert:2009exact}, that is their first and second-moment averages, such as $\MEAN{W_j\,\bm{\cdot}\,W_j^\dag}{}$ and $\MEAN{W_j^{\otimes2}\,\bm{\cdot}\,W_j^{\dag\,\otimes2}}{}$, are those of the \emph{Haar measure} over the unitary group~\cite{Collins:2006integration,Puchala_Z._Symbolic_2017}. Useful identities that apply for any $d$-dimensional two-design unitary $V$, $d$-dimensional observable $O_1$, and $d^2$-dimensional observables~$O_2$, $A$, $B$ and $C$ include~\cite{Holmes:2022connecting,Teo:2022optimized}
\begin{align}
	\MEAN{VO_1 V^\dag}{}=&\,\dfrac{\tr{O_1}}{d}\,,\label{eq:useful_2design_1}\\
	\MEAN{V^{\otimes2}O_2 V^{\dag\,\otimes2}}{}=&\,\left[\dfrac{\tr{O_2}}{d^2-1}-\dfrac{\tr{O_2\tau}}{d(d^2-1)}\right]1+\left[\dfrac{\tr{O_2\tau}}{d^2-1}-\dfrac{\tr{O_2}}{d(d^2-1)}\right]\tau\,,\label{eq:useful_2design_2}\\
	\MEAN{VAV^\dag BVCV^\dag}{}=&\,\dfrac{1}{d^2-1}\left(\tr{A}\tr{C}B+\tr{B}\tr{AC}1\right)\nonumber\\
	&\,-\dfrac{1}{d(d^2-1)}\left(\tr{A}\tr{B}\tr{C}1+\tr{AC}B\right)\,,\label{eq:useful_2design_3}
\end{align}
where $\tau$ is the $d\times d$ bipartite swap operator. A popular two-design distribution that we shall adopt for more general simulation runs is the Haar measure itself. According to~\cite{Mezzadri:2007qr}, one may generate random unitary operators ($U_\mathrm{Haar}$) of dimension $d$ that are distributed according to this measure from the following numerical recipe:
\begin{center}
	\begin{minipage}[c][7cm][c]{0.9\columnwidth}
		\noindent
		\rule{\columnwidth}{1ex}
		\begin{enumerate}
			\item Generate a random $d\times d$ matrix $\dyadic{A}$ with entries i.i.d. standard Gaussian distribution.
			\item Compute the matrices $\dyadic{Q}$ and $\dyadic{R}$ from the QR decomposition $\dyadic{A}=\dyadic{Q}\dyadic{R}$.
			\item Define $\dyadic{R}_\text{diag}=\mathrm{diag}\{\dyadic{R}\}$.
			\item Define $\dyadic{L}=\dyadic{R}_\text{diag} \oslash |\dyadic{R}_\text{diag}|$ ($\oslash$ refers to the Hadamard division).
			\item Define $U_\mathrm{Haar}\,\widehat{=}\,\dyadic{Q}\dyadic{L}$.
		\end{enumerate}		
		\rule{\columnwidth}{1ex}
	\end{minipage}
\end{center}  

\section{Results}
\subsection{Two-design networks}

\subsubsection{I.i.d. loss channel}

Because the vacuum state $\vacket\vacbra$ resides in the Hilbert-space sector that is orthogonal to that of $\rho$, for small error rate $\epsilon$, it is possible to obtain the complete expression for the MSE between the target pure state $\rho=\ket{\,\,\,}\bra{\,\,\,}$ and the mitigated state $\rho'$ using virtual distillation, which reads
\begin{equation}
	\mathrm{MSE}^{\mathrm{i.i.d.\,loss}}_{M,L_\mathrm{err}}=\left(\frac{\epsilon}{{L_\mathrm{err}}}\right)^{2M}\left[n\MEAN{\tr{\ptr{1}{\ket{\,\,\,}\bra{\,\,\,}}^{2M}}}{}+\MEAN{\left(\sum^n_{j=1}\tr{\ptr{j}{\ket{\,\,\,}\bra{\,\,\,}}^M}\right)^2}{}\right]\,,
	\label{eq:mse_loss}
\end{equation}
where the averages $\MEAN{\tr{\ptr{1}{\ket{\,\,\,}\bra{\,\,\,}}^{2M}}}{}$ and $\MEAN{\left(\sum^n_{j=1}\tr{\ptr{j}{\ket{\,\,\,}\bra{\,\,\,}}^M}\right)^2}{}$ would depend specifically on the kind of two-design circuit \emph{ansatz} employed in the application. For $M=1$, it can be shown that
\begin{equation}
	\mathrm{MSE}^{\mathrm{i.i.d.\,loss}}_{M,L_\mathrm{err}}=\left(\frac{\epsilon}{{L_\mathrm{err}}}\right)^{2}\left[\dfrac{n(d+4)}{2(d+1)}+n^2\right]\,,
	\label{eq:mse_loss_M2}
\end{equation}
which universally holds for all two-design circuits.

We may now extract some key behaviors concerning virtual distillation with peripheral noise dilution. The first observation is that the MSE scales exponentially with $M$ in the error rate $\epsilon$ for a fixed $L_\mathrm{err}$. If the magnitude of $\epsilon$ is about $0.01$, say, then a second-order virtual distillation results in an MSE that is about $10^{-8}$ in orders of magnitude. We therefore see that for such error rates, $M=2$ is typically sufficient for practical purposes. 

\begin{figure}[t]
	\centering
	\includegraphics[width=0.7\columnwidth]{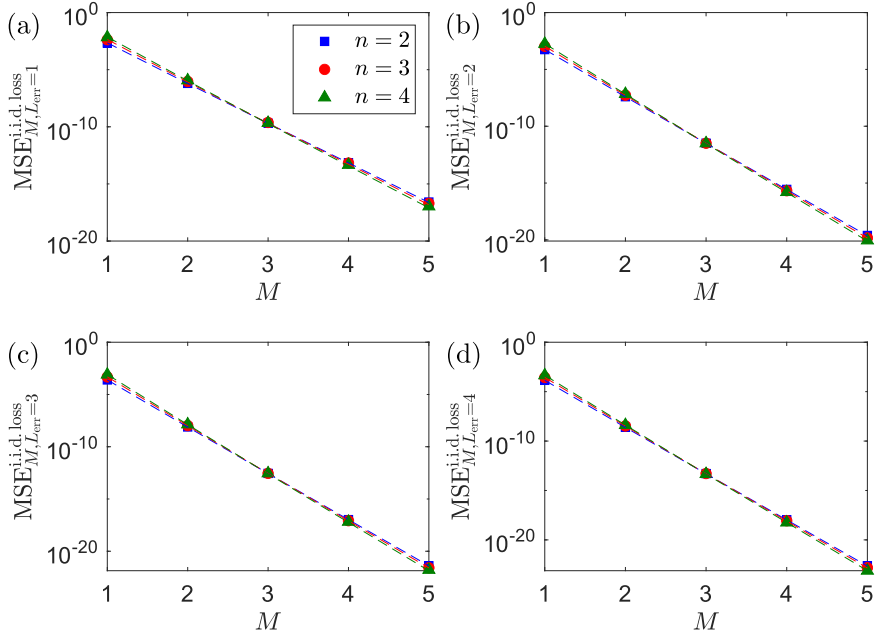}
	\caption{\label{fig:loss_varM_fixederr}Comparison between Monte~Carlo simulations~(markers) and theoretical predictions~(dashed curves) of $\mathrm{MSE}^{\mathrm{i.i.d.\,loss}}_{M,L_\mathrm{err}}$ for $\epsilon=0.02$, (a)~$L_\mathrm{err}=1$, (b)~2, (c)~3, (d)~4, and various numbers of qubits $n$ and virtual-distillation order $M$. For each value of $L_\mathrm{err}$, one hundred sets of Haar unitary operators $\{W_l\}^{L_\mathrm{err}}_{l=1}$ are used to average every plot marker. All graphs are plotted with a vertical logarithmic scale.}
\end{figure} 

\begin{figure}[t]
	\centering
	\includegraphics[width=0.7\columnwidth]{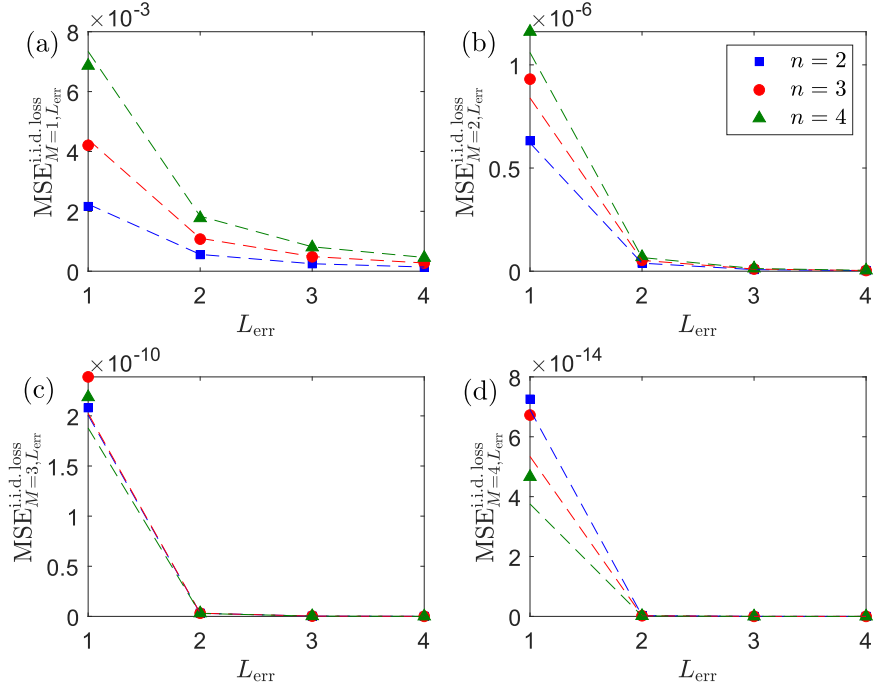}
	\caption{\label{fig:loss_varLerr_fixederr}Comparison between Monte~Carlo simulations~(markers) and theoretical predictions~(dashed curves) of $\mathrm{MSE}^{\mathrm{i.i.d.\,loss}}_{M,L_\mathrm{err}}$ for $\epsilon=0.02$, (a)~$M=1$, (b)~2, (c)~3, (d)~4, and various numbers of qubits $n$ and diluted layers $L_\mathrm{err}$. For each value of $M$, 100 sets of Haar unitary operators $\{W_l\}^{L_\mathrm{err}}_{l=1}$ are used to average every plot marker.}
\end{figure}

\begin{figure}[h!]
	\centering
	\includegraphics[width=0.7\columnwidth]{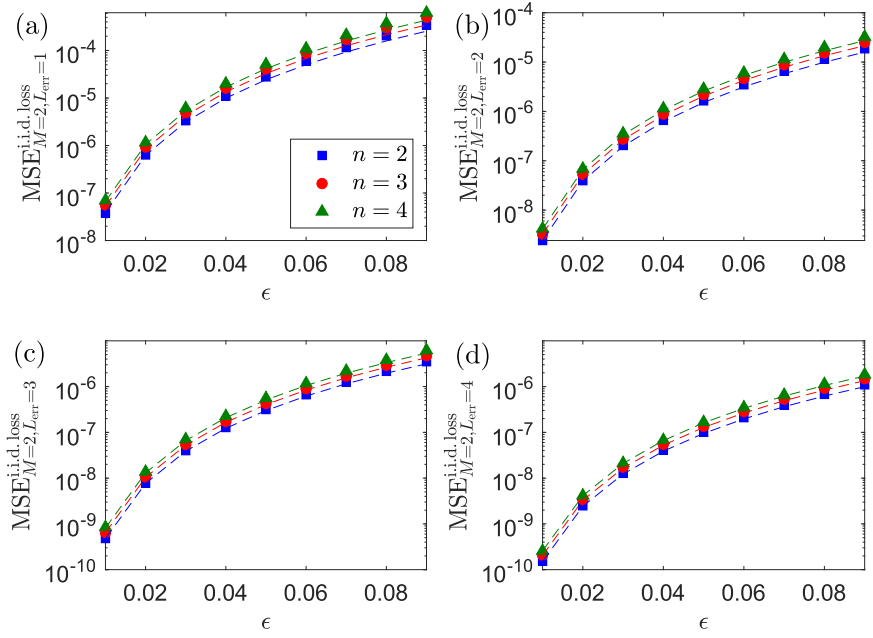}
	\caption{\label{fig:loss_fixedM_varerr}Comparison between Monte~Carlo simulations~(markers) and theoretical predictions~(dashed curves) of $\mathrm{MSE}^{\mathrm{i.i.d.\,loss}}_{M,L_\mathrm{err}}$ for a virtual-distillation order of $M=2$, (a)~$L_\mathrm{err}=1$, (b)~2, (c)~3, (d)~4, and various numbers of qubits~$n$ and error rate~$\epsilon$. For each value of $L_\mathrm{err}$, one hundred sets of Haar unitary operators $\{W_l\}^{L_\mathrm{err}}_{l=1}$ are used to average every plot marker. All graphs are plotted with a vertical logarithmic scale.}
\end{figure}

The next important finding is that for fixed $\epsilon$ and order $M$, the MSE decreases with increasing $L_\mathrm{err}$ according to a power law. This verifies the intuition that buffering the peripheral noise channel using the existing circuit components indeed result in higher mitigative power at least with virtual distillation. Figures~\ref{fig:loss_varM_fixederr}, \ref{fig:loss_varLerr_fixederr} and \ref{fig:loss_fixedM_varerr} graphically showcase the precision of Eq.~\eqref{eq:mse_loss} relative to Monte~Carlo simulations. In all simulations, the circuit unitary operators are distributed according to the Haar measure.

Moreover, for the i.i.d. loss channel, both Eq.~\eqref{eq:mse_loss} and Fig.~\ref{fig:loss_varM_fixederr} tell us that virtual distillation results in mitigated states that are asymptotically unbiased. That is, in the limit of large $M$, $\mathrm{MSE}^{\mathrm{i.i.d.\,loss}}_{M\rightarrow\infty,L_\mathrm{err}}\rightarrow0$. One can understand this property by inspecting the structure of $\rho'^M/\tr{\rho'^M}$ in \eqref{eq:rhopM_loss}, which possesses only the target-only term and the $\epsilon^M$-dependent term, with no other cross terms of lower $\epsilon$ orders. These cross terms are responsible for a nonzero bias in $\lim_{M\rightarrow\infty}\rho'^{M}/\tr{\rho'^{M}}$, which are missing for this channel, a consequence of the orthogonality between $\vacket$ and the circuit Hilbert-space sector.

\subsubsection{I.i.d. Pauli channel}

The i.i.d.~Pauli channel takes on a very different structure than the i.i.d.~loss channel. Most notably, its action described by \eqref{eq:nqubit_pauli} entails an error component that typically does not commute with the target state $\rho$. Furthermore, the nonorthogonality of this error component makes it \emph{persistent} regardless of the virtual-distillation order~[see Eq.~\eqref{eq:MSE_pauli_M1} in Appendix~\ref{app:Pauli_channel}]. That is, the error component of the distilled noisy state $\rho'^M/\tr{\rho'^M}$ carries a coefficient that is always of the leading order $O(\epsilon_1,\epsilon_2,\epsilon_3)$.

More specifically, the respective MSE expressions of distillation orders $M=1$ and $M\geq2$ for small $\epsilon$ and arbitrary $L_\mathrm{err}$ are 
\begin{align}
	\mathrm{MSE}^{\mathrm{i.i.d.\,Pauli}}_{M=1,L_\mathrm{err}}=&\,(n\epsilon)^2\left[\dfrac{d^3}{(d+1)(d^2-1)}-\dfrac{d}{L_\mathrm{err}(d+1)(d^2-1)}\right]+\dfrac{n\,d}{L_\mathrm{err}(d+1)}\sum^3_{l=1}\epsilon^2_l\,,\label{eq:mse_pauli}\\
	\mathrm{MSE}^{\mathrm{i.i.d.\,Pauli}}_{M\geq2,L_\mathrm{err}}=&\,2\sum^3_{l,l'=1}\frac{\epsilon_l\epsilon_{l'}}{L_\mathrm{err}^2}\sum^n_{j,j'=1}\left[\MEAN{\tr{\rho\,T^{(l)}_{j}T^{(l')}_{j'}}}{}-\MEAN{\tr{\rho\,T^{(l)}_{j}}\tr{\rho\,T^{(l')}_{j'}}}{}\right]\,,\label{eq:mse_pauli2}\\
	T^{(l)}_{j}=&\,P^{(l)}_j\rho P^{(l)}_j+W_{L_\mathrm{err}}P^{(l)}_j\rho_{L_\mathrm{err}-1}P^{(l)}_jW_{L_\mathrm{err}}^\dag+\ldots\nonumber\\
	&\,+W_{L_\mathrm{err}}W_{L_\mathrm{err}-1}\ldots W_2P^{(l)}_j\rho_1P^{(l)}_jW_2^\dag\ldots W_{L_\mathrm{err}-1}^\dag W_{L_\mathrm{err}}^\dag\,,\label{eq:mse_pauli3}
\end{align}
where $P^{(l)}_j$ is a single-qubit Pauli operator. A significant departure from the i.i.d. loss channel is that the MSE for $M\geq2$ is independent of $M$. This may appear counter-intuitive at first, but a closer inspection establishes consistency with \eqref{eq:MSE_pauli_M1}, that is perfect error mitigation is not possible in the presence of a persistent error component for any $M$, since it is always accompanied by error coefficients of unit leading order regardless of the value of $M$. Without loss of generality, we shall subsequently take the i.i.d. Pauli channel to be depolarizing, or $\epsilon_1=\epsilon_2=\epsilon_3=\epsilon/3$.

\begin{figure}[t]
	\centering
	\includegraphics[width=0.7\columnwidth]{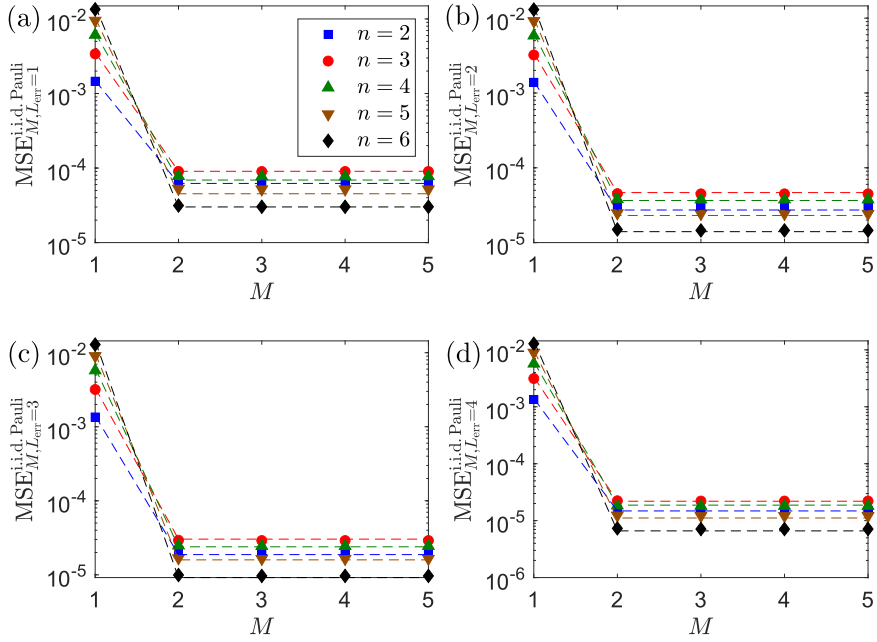}
	\caption{\label{fig:pauli_varM_fixederr}Comparison between Monte~Carlo simulations~(markers) and theoretical predictions~(dashed curves) of $\mathrm{MSE}^{\mathrm{i.i.d.\,Pauli}}_{M,L_\mathrm{err}}$ for $\epsilon=0.02$, (a)~$L_\mathrm{err}=1$, (b)~2, (c)~3, (d)~4, and various numbers of qubits $n$ and virtual-distillation order $M$. The channel is depolarizing with $\epsilon_1=\epsilon_2=\epsilon_3=\epsilon/3$. For each value of $L_\mathrm{err}$, 1000 sets of Haar unitary operators $\{W_l\}^{L_\mathrm{err}}_{l=1}$ are used to average every plot marker. All graphs are plotted with a vertical logarithmic scale.}
\end{figure}

\begin{figure}[h!]
	\centering
	\includegraphics[width=0.7\columnwidth]{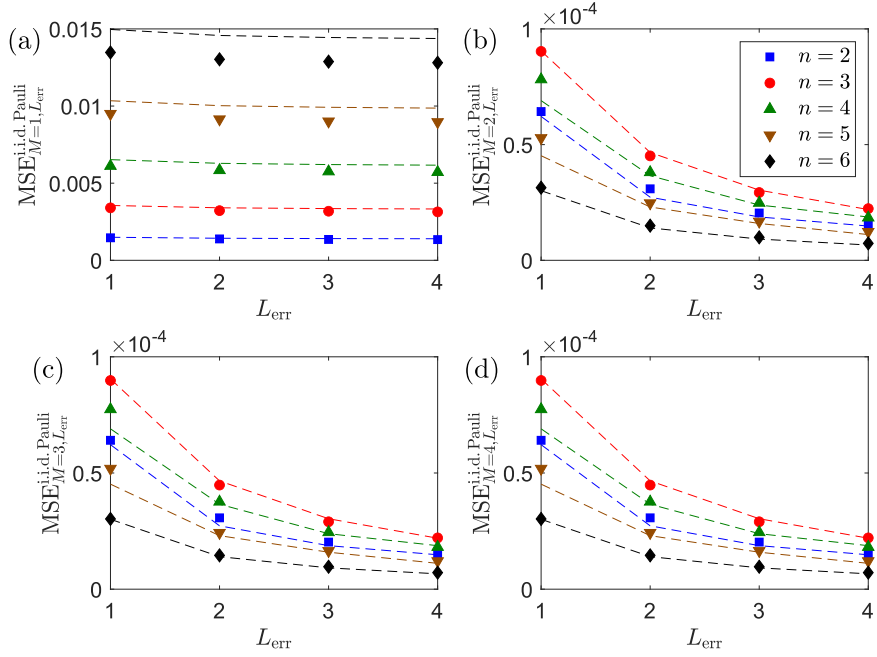}
	\caption{\label{fig:pauli_varLerr_fixederr}Comparison between Monte~Carlo simulations~(markers) and theoretical predictions~(dashed curves) of  $\mathrm{MSE}^{\mathrm{i.i.d.\,Pauli}}_{M,L_\mathrm{err}}$ for $\epsilon=0.02$, (a)~$M=1$, (b)~2, (c)~3, (d)~4, and various qubit numbers $n$ and $L_\mathrm{err}$. The noise channel is depolarizing. For each $M$, 1000 sets of Haar unitary $\{W_l\}^{L_\mathrm{err}}_{l=1}$ are used to average every plot marker.}
\end{figure}

\begin{figure}[h!]
	\centering
	\includegraphics[width=0.7\columnwidth]{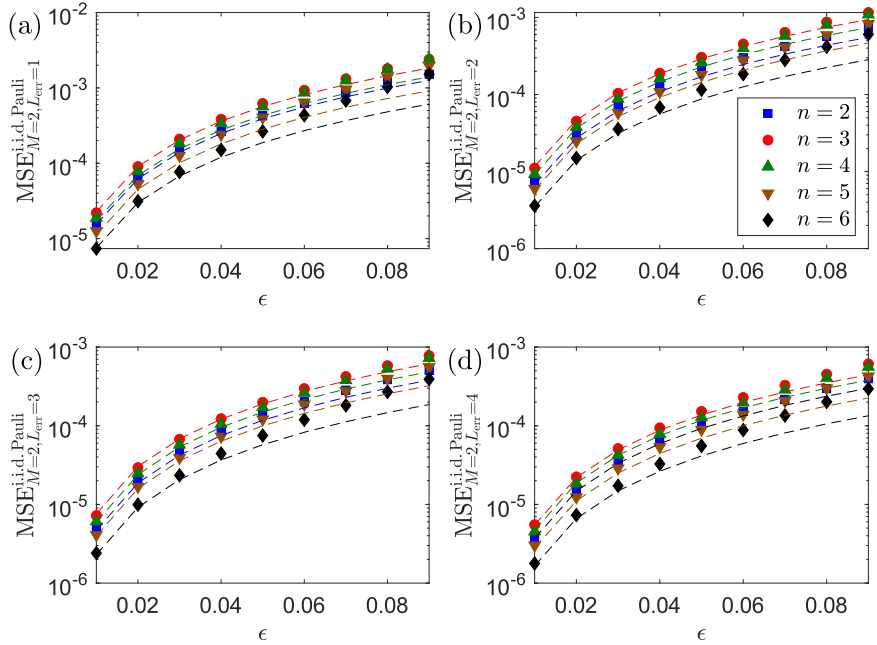}
	\caption{\label{fig:pauli_fixedM_varerr}Comparison between Monte~Carlo simulations~(markers) and theoretical predictions~(dashed curves) of $\mathrm{MSE}^{\mathrm{i.i.d.\,Pauli}}_{M,L_\mathrm{err}}$ for a distillation order of $M=2$, (a)~$L_\mathrm{err}=1$, (b)~2, (c)~3, (d)~4, and various $n$ and $\epsilon$. The noise channel is depolarizing. For each $L_\mathrm{err}$, 1000 sets of Haar unitary $\{W_l\}^{L_\mathrm{err}}_{l=1}$ are used to average every plot marker.}
\end{figure}

Another observation we can make is that for fixed $\epsilon\ll1/n$ and very large circuits ($n,d\gg2$), $\mathrm{MSE}^{\mathrm{i.i.d.\,Pauli}}_{M=1,L_\mathrm{err}}\rightarrow n\sum^3_{l=1}\epsilon_l^2/L_\mathrm{err}$, that is the unmitigated distance increases with $n$ and asymptotically approaches a finite limit. On the other hand, Figure~\ref{fig:pauli_varM_fixederr} shows that the mitigated MSE for $M\geq2$ drops gradually with increasing $n$~(excluding $n=2$). As with all simulations up to this section, all circuit unitary operators are assumed to possess a Haar distribution. Figure~\ref{fig:pauli_varLerr_fixederr} confirms the mitigation improvement as $L_\mathrm{err}$ increases. Figure~\ref{fig:pauli_fixedM_varerr} supply the MSE graphs with respect to different error rates whilst fixing $M$ and $L_\mathrm{err}$.


\subsection{Eigenvalue distribution of the error component}

To understand matters properly, let us, for the moment, consider the following special case~\cite{Koczor:2021exponential} of a noise-channel map
\begin{equation}
	\rho'=\ket{\lambda_0}(1-\epsilon_0)\bra{\lambda_0}+\epsilon_0\sum^{d-1}_{k=1}\ket{\lambda_k}p_k\bra{\lambda_k}
\end{equation}
that brings the ideal target pure state $\ket{\,\,\,}\bra{\,\,\,}\equiv\ket{\lambda_0}\bra{\lambda_0}$ to a noisy mixed state $\rho'$, where the error component
\begin{equation}
	\rho_\mathrm{err}=\sum^{d-1}_{k=1}\ket{\lambda_k}p_k\bra{\lambda_k}
\end{equation}
resides in the orthogonal subspace of the target~$\left(\sum^{d-1}_{k=1}p_k=1\,\,\text{and}\,\,\sum^{d-1}_{k=0}\ket{\lambda_k}\bra{\lambda_k}=1\right)$. Under a \emph{fixed} error rate $\epsilon_0$ and virtual-distillation order $M$, it is interesting to search for the optimal distribution of $\rvec{p}=(p_1\,\,p_2\,\,\ldots\,\,p_{d-1})$ that minimizes the MSE defined in \eqref{eq:mse_def} (without circuit averaging for this situation). This finding would at least serve as a guide towards the optimal noise-coping strategy under this special case. 

After some simple manipulation, upon denoting $\mathcal{N}_{M,\epsilon_0}=(1-\epsilon_0)^M+\epsilon^M_0\sum^{d-1}_{k=1}p_k^M$, we find that for \emph{any} $0\leq\epsilon_0\leq1$,
\begin{equation}
	\MSE{M}=\tr{\left(\dfrac{\rho'^M}{\tr{\rho'^M}}-\ket{\,\,\,}\bra{\,\,\,}\right)^2}=\left[1-\frac{(1-\epsilon_0)^M}{\mathcal{N}_{M,\epsilon_0}}\right]^2+\dfrac{\epsilon_0^{2M}}{\mathcal{N}_{M,\epsilon_0}^2}\sum^{d-1}_{k=1}p_k^{2M}\,.
\end{equation}
Since $\{p_k\}^{d-1}_{k=1}$ is a discrete set of normalized probabilities, it may be parametrized as $p_k=a^2_k/\sum^{d-1}_{k'=1}a_{k'}^2$, so that the variation
\begin{equation}
	\updelta p_k=\dfrac{2}{\sum^{d-1}_{k''=1}a^2_{k''}}\left(a_k\updelta a_k-p_k\sum^{d-1}_{k'=1}a_{k'}\updelta a_{k'}\right)
\end{equation}
implies that
\begin{align}
	\dfrac{\updelta\MSE{M}}{\updelta a_k}\propto&\,a_k\left[A_{M,k}(\rvec{p})-\sum^{d-1}_{k'=1}A_{M,k'}(\rvec{p})\,p_{k'}\right]\,,\nonumber\\
	A_{M,k}(\rvec{p})=&\,\left[(1-\epsilon_0)^M-\dfrac{(1-\epsilon_0)^{2M}}{\mathcal{N}_{M,\epsilon0}}-\dfrac{\epsilon_0^{2M}}{\mathcal{N}_{M,\epsilon0}}\,\sum^{d-1}_{k'=1}p^{2M}_{k'}\right]p_k^{M-1}+\epsilon_0^{M}\,p_k^{2M-1}\,,
	\label{eq:ext_eqn}
\end{align}
after some straightforward calculations. Upon setting the gradient to zero and solving the extremal equation $p_k\sum^{d-1}_{k'=1}A_{M,k'}(\rvec{p})\,p_{k'}=p_k A_{M,k}(\rvec{p})$, it is easy to verify that $p_k=1/(d-1)$ is the solution.~\footnote{The interested reader may even try taking the gradient in \eqref{eq:ext_eqn} to set up a steepest-descent method and obtain the uniform probability distribution after many iterations.} So, the optimal strategy should give a rank-$(d-1)$ $\rho_\mathrm{err}$ that possesses a uniform eigenspectrum.

\begin{figure}[t]
	\centering
	\includegraphics[width=0.8\columnwidth]{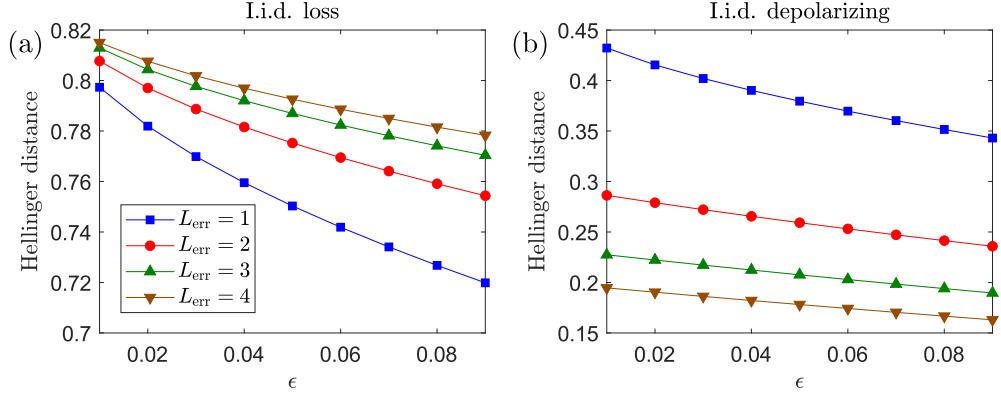}
	\caption{\label{fig:hellinger}The Hellinger distance between the error-component eigenspectrum and uniform distribution for respectively the i.i.d. (a)~loss and (b)~depolarizing channels. All graphs are averaged over the respective numbers of Haar circuit-unitary operators as stated in all previous simulation figures.}
\end{figure} 

The purpose of this subsection is to emphasize that, while the special situation in which the user can freely change noise-coping strategies, thereby varying the $p_k$s, and simultaneously fix the error rate is an especially easy one to analyze, there are practical scenarios where the error rate will also be inevitably varied as a consequence of such a strategy change. An example is noise dilution on the i.i.d.~loss channel, as cautioned at the end of Sec.~\ref{subsec:noise_dilution}. Increasing the number of diluted noise layers $L_\mathrm{err}$ not only changes the error component, but also alters the effective error rate as a result of the non-trace-preserving character of circuit operations on lossy states~[see Eq.~\eqref{eq:lossy_trace_renorm}]. Therefore, all previous arguments leading to \eqref{eq:ext_eqn} no longer fly with this channel, and so a better noise-coping strategy does not necessarily lead to error components of eigenspectra closer to a uniform distribution.

To systematically compare the difference between the eigenspectrum~$\rvec{\lambda}$ of a $d$-dimensional error component $\rho_\mathrm{err}$ with the uniform distribution $\rvec{\lambda}_\mathrm{unif}=(1\,\,1\,\,\ldots\,\,1)^\top/d$, we take the \emph{Hellinger distance},
\begin{equation}
	H(\rvec{\lambda},\rvec{\lambda}_\mathrm{unif})=\dfrac{1}{\sqrt{2}}\sqrt{\sum^{d-1}_{k=0}\left(\sqrt{\lambda_k}-1/\sqrt{d}\right)^2}
	\label{eq:hellinger}
\end{equation}
as the figure of merit, where for i.i.d.~lossy scenarios, $d=3^n$. If the noiseless $n$-qubit target $\rho=\ket{\,\,\,}\bra{\,\,\,}$ is a product state~($\ket{\,\,\,}=\ket{\psi_1}\ket{\psi_2}\ldots\ket{\psi_n}$), then the \emph{exact} Hellinger distance is given by
\begin{equation}
	H^{\mathrm{i.i.d.\,loss}}_x=\left\{1+\dfrac{1}{2\cdot 3^{n}}-\dfrac{(1-x)^{n/2}}{3^{n/2}\sqrt{1-(1-x)^n}}\left[\left(1+\sqrt{\dfrac{x}{1-x}}\right)^n-1\right]\right\}^{1/2}\,,\,\,x=\frac{\epsilon}{L_\mathrm{err}}\,.
	\label{eq:hellinger_iid_loss}
\end{equation}
One can refer to Appendix~\ref{app:loss_hellinger} for the derivation, and verify, indeed, that $H^{\mathrm{i.i.d.\,loss}}_{\epsilon/L_\mathrm{err}}$ increases with increasing $L_\mathrm{err}$, which is in direct contrast with the special case discussed previously.

Figure~\ref{fig:hellinger}(a) shows very similar behaviors in the Hellinger distances for general random states. On the other hand, the i.i.d. depolarizing channel~[see Fig.~\ref{fig:hellinger}(b)] yields distances that agree in trend with that for the special case. An important distinction from the i.i.d. loss channel is that the overall error rate in~\eqref{eq:dilution} remains as $n\epsilon$ for any $L_\mathrm{err}$, as opposed to that in~\eqref{eq:lossy_trace_renorm}.

\subsection{Quantum clusters of hardware-efficient networks}
\label{eq:quantum_cluster}

\begin{figure}[t]
	\centering
	\includegraphics[width=0.8\columnwidth]{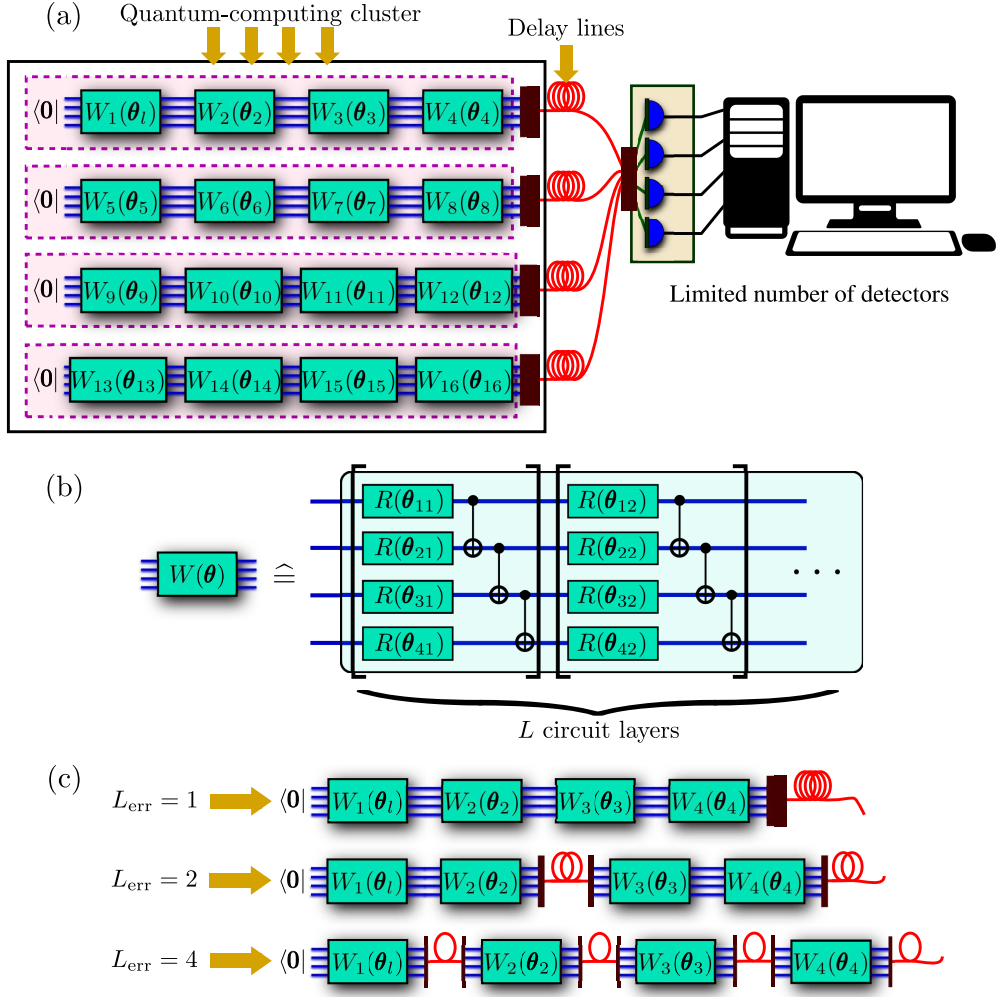}
	\caption{\label{fig:quantum_cluster}(a)~A quantum-computing cluster consisting of four separate NISQ circuits, where each circuit accepts a maximum of four qubits, and may be employed independently by a user at any given instance. Delay lines are used to queue the output qubits for the final measurement with a (limited) set of four qubit detectors. (b)~Each circuit unitary module $W_l(\rvec{\theta}_l)$ is made up of $L$ layers of single-qubit rotations and CNOT gates. (c)~Noise dilution is done splitting the delay lines and distribute them equally amongst the circuit modules. For small error rates, dividing the delay length by $L_\mathrm{err}$ is equivalent to dividing the unsplit error rate by $L_\mathrm{err}$ in each diluted noise layer.}
\end{figure} 

Finally, we give a practical application to virtual distillation with noise dilution. Suppose one is interested in designing a quantum-computer cluster~(see Fig.~\ref{fig:quantum_cluster}) that houses several independent quantum circuits that are accessible by the public domain. At any given instance, a user may login to the cluster and use one such $n$-qubit circuit. Additionally, we assume a limited number of detectors, so that the output qubits need to be queued with delay lines for the final measurement. In a typical NISQ cluster, each unitary operator $W_l$ describes a circuit comprising, say, $L$ layers of passive components, which are single-qubit and controlled-NOT (CNOT) gates. From~\cite{harrow_random_2009}, randomized circuits of this kind are approximately two-designs if \mbox{$L=O(\poly(n))$}.

We shall illustrate the results of noise dilution under such a practical situation by investigating two separate scenarios. The first scenario is where photon loss is a dominant source of noise in the delay lines, which could be the case when the cluster is integrated into a photonic chip. For more concrete simulations, we take the decay rate $\gamma_\mathrm{loss}=0.2$ dB\,cm$^{-1}$~\cite{Arrazola:2021quantum} or equivalently $\gamma_\mathrm{loss}=6$ dB\,ns$^{-1}$, where $\epsilon_\mathrm{loss}=1-10^{-\gamma_\mathrm{loss}\tau/10}$ is related to the delay time $\tau$~(in ns). The second scenario is where polarization drifts occur more frequently in regular optical-fiber-based delay lines~\cite{Dragan:2005depolarization,Bayat:2006threshold,Karpinski:2008fiber}, such that the depolarizing channel serves as an appropriate noise model. As an example, we quote from~\cite{Weihs:1998violation} the depolarizing rate of about $\gamma_\mathrm{depol}=1.3\times10^{-4}$~s$^{-1}$, which is equivalent to a depolarization error rate of about $0.01$ for a 400-meter optical fiber. In other words, $\epsilon_\mathrm{depol}=1-\E{-\gamma_\mathrm{depol}\tau}$.

\begin{figure}[t]
	\centering
	\includegraphics[width=0.7\columnwidth]{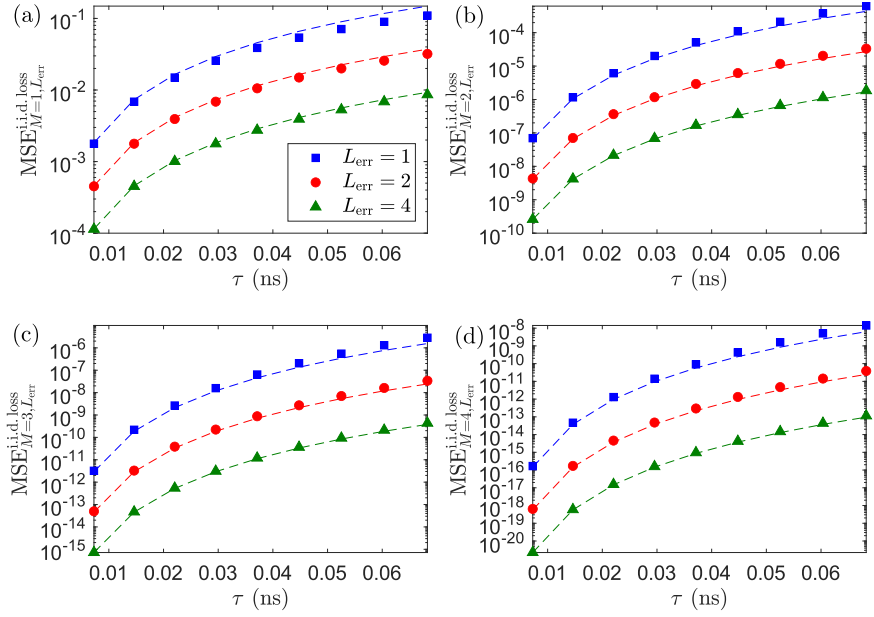}
	\caption{\label{fig:photon_loss}Virtual-distillation MSE curves for four-qubit hardware-efficient circuits~($L=2$) with respect to the total delay time~$\tau$ under the i.i.d. loss channel. Performances for various $M$ and $L_\mathrm{err}$ are illustrated.}
\end{figure}

\begin{figure}[t]
	\centering
	\includegraphics[width=0.7\columnwidth]{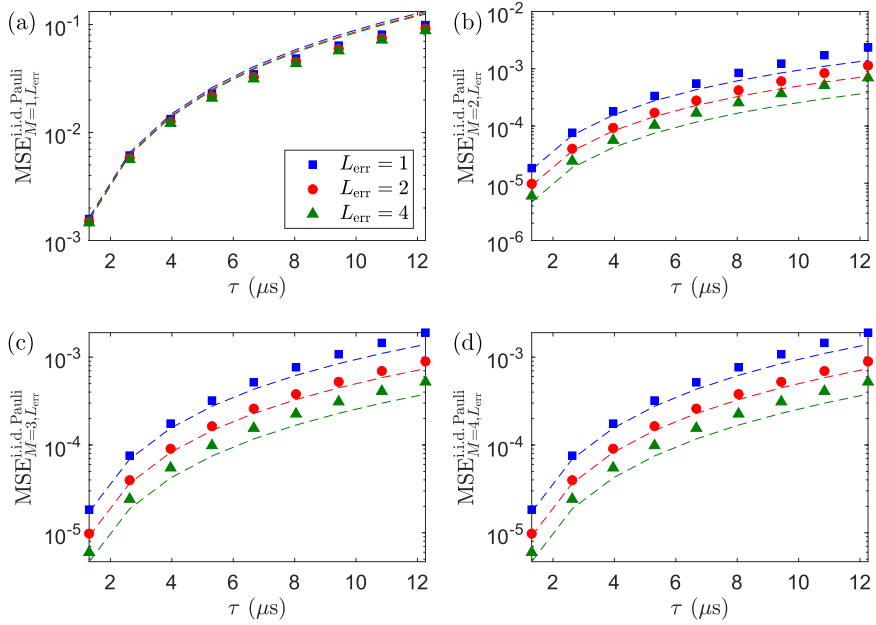}
	\caption{\label{fig:depolarization}Virtual-distillation MSE curves for four-qubit hardware-efficient circuits~($L=2$) with respect to the total delay time~$\tau$ under the i.i.d. depolarizing channel. Performances for various $M$ and $L_\mathrm{err}$ are illustrated.}
\end{figure}

To present the simulation findings, we study virtual distillation performances on four-qubit hardware-efficient circuits, where the number of single-qubit-CNOT layers $L=2$, such that a total of eight circuit layers is considered for each entire computation circuit. Hence, $L_\mathrm{err}=2$, for instance, implies that one of the diluted noise layers is sandwiched between two unitary subcircuits, each with $2L=4$ circuit layers~[see Fig.~\ref{fig:quantum_cluster}(c)]. Figures~\ref{fig:photon_loss} and \ref{fig:depolarization} plot the virtual-distillation performances on both i.i.d. channels based on the aforementioned specifications.

An interesting conclusion out of these numerical findings is that even though each circuit unitary operator $W_l$ is shallow ($L=2$), noise dilution can still effectively enhance error mitigation as $L_\mathrm{err}$ increases. Therefore, whenever situation permits, the lesson here is that a uniform distribution of peripherals can, in practice, improve the performance of virtual distillation, at least for the i.i.d. loss and Pauli channels. For the case where the peripherals are delay lines, within the same total delay time, our results suggest that transmitting output qubits from uniformly-delayed circuit operations to the detectors is a better error-mitigation strategy than delaying qubits from instantaneous circuit operations.

\section{Discussion}

Virtual distillation is a technically easy-to-use technique that mitigates errors in a noise-model-agnostic manner. Recent developments have further enhanced the feasibility of this error-mitigation technique. We apply this distillation procedure to practical situations in which general quantum-computing circuits are used in conjunction with additional peripherals that could introduce excess noise. 

Under the general two-design assumption about the quantum circuits, and multiqubit loss and Pauli channels as models for the excess noise, we analytically show~(supported with numerical simulations) that for the same total error rate, distributing the peripherals homogeneously across the entire quantum circuit improves the quality of the error-mitigated output state as opposed to collecting all peripherals at one place---the more noise layers the peripheral excess-noise channel is diluted into, the better the mitigative power. Furthermore, it turns out that virtual distillation exponentially reduces errors due to losses as the order increases, but gives the same magnitude of error reduction under the Pauli channel for all distillation orders greater than one \emph{up to the leading order of the error rate}. These two different behaviors stem from the types of error component generated by each noise model, where the one from losses is orthogonal to the multiqubit Hilbert-space sector and that from the Pauli channel generally does not commute with the target state. The latter error component is therefore persistent against virtual distillation.

Such findings come in handy when designing a quantum-computing cluster consisting of several noisy intermediate-scale quantum-circuit networks, where having a huge number of measurement detectors accommodating all these networks is economically difficult. In this case, delay lines are generally used to queue output qubits for acquiring the final measurement dataset. Our results suggest that splitting the delay lines homogeneously across the circuits can typically improve the error-mitigative power with virtual distillation, even when circuit depths are not deep.

\begin{acknowledgments}
	We thank Yosep Kim for fruitful discussions. This work is supported by the National Research Foundation of Korea (NRF) grants funded by the Korea government~(Grant Nos.~NRF-2020R1A2C1008609, NRF-2020K2A9A1A06102946, NRF-2019R1A6A1A10073437 and NRF-2022M3E4A1076099) \emph{via} the Institute of Applied Physics at Seoul National University, and by the Institute of Information \& Communications Technology Planning \& Evaluation (IITP) grant funded by the Korea government (MSIT) (IITP-2022-2020-0-01606).
\end{acknowledgments}

\appendix
\section{Small-error regime}
\label{app:small_err}

The error component $\rho_\mathrm{err}$, being a quantum state itself, may be parametrized with the auxiliary complex operator $A_\mathrm{err}$ inasmuch as $\widetilde{\rho}_\mathrm{err}(\mu,\ket{\,\,\,}\bra{\,\,\,})=A_\mu^\dag A_\mu/\tr{A_\mu^\dag A_\mu}$. When $\mu\ll1$, we have the following Taylor expansion about $\mu=0^+$
\begin{equation}
	A_\mu=\Delta_0+\mu\Delta_1\,,
\end{equation}
so that 
\begin{equation}
	\widetilde{\rho}_\mathrm{err}(\mu,\ket{\,\,\,}\bra{\,\,\,})\cong\dfrac{\Delta_0^\dag \Delta_0}{\tr{\Delta_0^\dag \Delta_0}}+\mu\left(\dfrac{\Delta_1^\dag\Delta_0+\Delta_0^\dag\Delta_1}{\tr{\Delta_0^\dag\Delta_0}}-\dfrac{\Delta_0^\dag\Delta_0\tr{\Delta_1^\dag\Delta_0+\Delta_0^\dag\Delta_1}}{\tr{\Delta_0^\dag\Delta_0}^2}\right)\,.
\end{equation}
Similarly, by taking $\gamma_0^{(0,\mu)}$ to be real without loss of generality, $\gamma_0^{(0,\mu)}\cong\sqrt{d}+\mu\delta_1$, where we note that $\delta_1=\partial \gamma_0^{(0,\mu)}/\partial\mu|_{\mu=0^+}<0$. This gives us
\begin{equation}
	\epsilon(\mu)\cong-\frac{2\mu\delta_1}{\sqrt{d}}\equiv\epsilon\,.
\end{equation}
Upon defining the constant $\rho_\mathrm{err}\equiv\Delta_0^\dag \Delta_0/\tr{\Delta_0^\dag \Delta_0}$, one obtains Eq.~\eqref{eq:small_err_rho_prime}.

\section{Virtual distillation under i.i.d. loss channel}
\label{app:loss_channel}

As a brief review demonstration, the solution to \eqref{eq:loss_markov} is given by
\begin{equation}
	\rho'=\E{t\mathcal{L}}[\rho]\,,\quad\mathcal{L}[\rho]=\allowbreak\gamma\sum^1_{j=0}\left(a_j\rho a_j^\dag-\frac{1}{2}a^\dag_j a_j\rho-\frac{1}{2}\rho a^\dag_j a_j\right)\,.
\end{equation}
Now, taking the number ket $\ket{n_0=1,n_1=0}\equiv\ket{1,0}\leftrightarrow\ket{0}$ for the $0$th (horizontal) polarization ket as the initial ket for $\rho$ at $t=0$, it is easy to see that
\begin{align}
	\mathcal{L}[\ket{0}\bra{0}]\leftrightarrow&\,\gamma\Bigg(a_0\ket{1,0}\bra{1,0} a_0^\dag-a^\dag_0 a_0\ket{1,0}\frac{1}{2}\bra{1,0}-\ket{1,0}\frac{1}{2}\bra{1,0} a^\dag_0 a_0\Bigg)\nonumber\\
	=&\,-\gamma\left(\ket{0}\bra{0}-\vacket\vacbra\right)\,,\nonumber\\
	\mathcal{L}[\vacket\vacbra]=&\,0\,.
\end{align}
The corresponding solution for the noisy state $\rho'$ is therefore given by
\begin{align}
	\rho'=&\,\ket{0}\bra{0}+\sum^\infty_{l=1}\frac{t^l}{l!}\mathcal{L}^l[\ket{0}\bra{0}]\nonumber\\
	=&\,\ket{0}\bra{0}+\sum^\infty_{l=1}\frac{(-\gamma t)^l}{l!}(\ket{0}\bra{0}-\vacket\vacbra)\nonumber\\
	=&\,\vacket\vacbra+\E{-\gamma t}(\ket{0}\bra{0}-\vacket\vacbra)\nonumber\\
	=&\,\ket{0}(1-\epsilon)\bra{0}+\vacket\epsilon\vacbra\,.
\end{align}
The reader may feel free to go through the same exercise and obtain all other actions stated in \eqref{eq:pol_loss}.

When $L_\mathrm{err}=1$ and $\rho=\ket{\,\,\,}\bra{\,\,\,}$ of dimension $d=2^n$, \eqref{eq:nqubit_loss} leads to
\begin{align}
	\rho'^M\cong(1-Mn\epsilon)\rho+\epsilon^M\left(\vacket\vacbra\otimes\ptr{1}{\rho}^M+\ldots+\ptr{n}{\rho}^M\otimes\vacket\vacbra\right)\,,
\end{align}
so that
\begin{align}
	\dfrac{\rho'^M}{\tr{\rho'^M}}\cong&\,\rho+\epsilon^M\Bigg(\vacket\vacbra\otimes\ptr{1}{\rho}^M+\ldots+\ptr{n}{\rho}^M\otimes\vacket\vacbra\nonumber\\
	&\,-\rho\sum^n_{j=1}\tr{\ptr{j}{\rho}^M}\Bigg)\,.
	\label{eq:rhopM_loss_Lerr1}
\end{align}
The MSE for $L_\mathrm{err}=1$ therefore reads
\begin{equation}
	\mathrm{MSE}^{\mathrm{i.i.d.\,loss}}_{M,L_\mathrm{err}=1}=\epsilon^{2M}\left[n\MEAN{\tr{\ptr{1}{\rho}^{2M}}}{}+\MEAN{\left(\sum^n_{j=1}\tr{\ptr{j}{\rho}^M}\right)^2}{}\right]\,.
\end{equation}

A simplified expression is available for $M=1$, namely beginning with
\begin{equation}
	\mathrm{MSE}^{\mathrm{i.i.d.\,loss}}_{M=1,L_\mathrm{err}=1}=\epsilon^{2}\left(n\MEAN{\tr{\ptr{1}{\rho}^2}}{}+n^2\right)\,,
\end{equation}
we realize that
\begin{align}
	\tr{\ptr{1}{\rho}^2}=&\,\sum^1_{m_1,m_1'=0}\sum^{2^{n-1}-1}_{m_2=0}\tr{\rho\,F_{m_1,m_1'}\,\rho\,G_{m_1,m_1',m_2}}\,,
\end{align}
with
\begin{align}
	F_{m_1,m_1'}=&\,\ket{m_1}\bra{m_1'}\otimes1_{n-1}\,,\nonumber\\
	G_{m_1,m_1',m_2}=&\,\ket{m_1'}\ket{m_2}\bra{m_1}\bra{m_2}\,.
\end{align}
Therefore, in order to average
\begin{equation}
	\tr{\rho\,F_{m_1,m_1'}\,\rho\,G_{m_1,m_1',m_2}}=\tr{U\ket{0}\bra{0}U^\dag\,F_{m_1,m_1'}\,U\ket{0}\bra{0}U^\dag\,G_{m_1,m_1',m_2}}\,,
\end{equation}
we make use of Eq.~\eqref{eq:useful_2design_3} for any two-design unitary $U$ and find that
\begin{equation}
	\MEAN{\tr{\rho\,F_{m_1,m_1'}\,\rho\,G_{m_1,m_1',m_2}}}{}=\dfrac{1}{d(d+1)}\left(\frac{d}{2}\delta_{m_1,m_1'}+1\right)\,,
\end{equation}
so that
\begin{equation}
	\MEAN{\tr{\ptr{1}{\rho}^2}}{}=\dfrac{d+4}{2(d+1)}
\end{equation}
and
\begin{equation}
	\mathrm{MSE}^{\mathrm{i.i.d.\,loss}}_{M=1,L_\mathrm{err}=1}=\epsilon^{2}\left[\dfrac{n(d+4)}{2(d+1)}+n^2\right]
\end{equation}
for $M=1$.

For any $L_\mathrm{err}\geq1$ case is governed by the action
\begin{equation}
	\rho'_{L_\mathrm{err}}=\Phi^\mathrm{multiqubit\, loss}_{\epsilon/L_\mathrm{err}}\left[W_{L_\mathrm{err}}\ldots\Phi^\mathrm{multiqubit\, loss}_{\epsilon/L_\mathrm{err}}\left[W_2\,\Phi^\mathrm{multiqubit\, loss}_{\epsilon/L_\mathrm{err}}\left[W_1\ket{\rvec{0}}\bra{\rvec{0}}W_1^\dag\right] W_2^\dag\right]\ldots W_{L_\mathrm{err}}^\dag\right]\,.
\end{equation}
By reminding ourselves that the unitary operators $W_l$ act on the Hilbert-space sector that is orthogonal to $\vacket\vacbra$, it is straightforward to see that
\begin{align}
	\rho'_{L_\mathrm{err}}\cong&\,\left(1-\frac{\epsilon}{{L_\mathrm{err}}}\right)^{{L_\mathrm{err}}n}\rho+\frac{\epsilon}{{L_\mathrm{err}}}\left(1-\frac{\epsilon}{{L_\mathrm{err}}}\right)^{L_\mathrm{err}n-1}\nonumber\\
	&\qquad\qquad\qquad\qquad\qquad\qquad\times\left(\vacket\vacbra\otimes\ptr{1}{\rho}+\ldots+\ptr{n}{\rho}\otimes\vacket\vacbra\right)\nonumber\\
	\cong&\,(1-n\epsilon)\rho+\frac{\epsilon}{{L_\mathrm{err}}}\left(\vacket\vacbra\otimes\ptr{1}{\rho}+\ldots+\ptr{n}{\rho}\otimes\vacket\vacbra\right)\,.
	\label{eq:small_err_loss_state}
\end{align} 
Then, as all vacuum-related terms and the noiseless state $\rho$ are mutually orthogonal to each other, raising $\rho'$ to the $M$th power amounts simply to
\begin{equation}
	\rho'^M_{L_\mathrm{err}}\cong(1-Mn\epsilon)\rho+\left(\frac{\epsilon}{{L_\mathrm{err}}}\right)^M\left(\vacket\vacbra\otimes\ptr{1}{\rho}^M+\ldots+\ptr{n}{\rho}^M\otimes\vacket\vacbra\right)\,,
\end{equation}
along with its resulting normalized state
\begin{align}
	\dfrac{\rho'^M_{L_\mathrm{err}}}{\tr{\rho'^M_{L_\mathrm{err}}}}\cong&\,\rho+\left(\frac{\epsilon}{{L_\mathrm{err}}}\right)^M\Bigg(\vacket\vacbra\otimes\ptr{1}{\rho}^M+\ldots+\ptr{n}{\rho}^M\otimes\vacket\vacbra\nonumber\\
	&\,-\rho\sum^n_{j=1}\tr{\ptr{j}{\rho}^M}\Bigg)\,.
	\label{eq:rhopM_loss}
\end{align}
Hence, one obtains the more general MSE expression for small $\epsilon$,
\begin{equation}
	\mathrm{MSE}^{\mathrm{i.i.d.\,loss}}_{M,L_\mathrm{err}}=\left(\frac{\epsilon}{{L_\mathrm{err}}}\right)^{2M}\left[n\MEAN{\tr{\ptr{1}{\rho}^{2M}}}{}+\MEAN{\left(\sum^n_{j=1}\tr{\ptr{j}{\rho}^M}\right)^2}{}\right]\,,
	\label{eq:MSE_loss}
\end{equation}
or
\begin{equation}
	\mathrm{MSE}^{\mathrm{i.i.d.\,loss}}_{M=1,L_\mathrm{err}}=\left(\frac{\epsilon}{{L_\mathrm{err}}}\right)^{2}\left[\dfrac{n(d+4)}{2(d+1)}+n^2\right]
	\label{eq:MSE_loss_M1}
\end{equation}
for $M=1$. This gives us the ratio $\mathrm{MSE}^{\mathrm{i.i.d.\,loss}}_{M,L_\mathrm{err}}/\mathrm{MSE}^{\mathrm{i.i.d.\,loss}}_{M,L_\mathrm{err}=1}=O(L^{-2M}_\mathrm{err})$ for any $M$. For completeness, we tabulate values of $\MEAN{\tr{\ptr{1}{\rho}^{2M}}}{}$ and $\MEAN{\left(\sum^n_{j=1}\tr{\ptr{j}{\rho}^M}\right)^2}{}$ in Tabs.~\ref{tab:val1} and \ref{tab:val2}, which are used to generate Figs.~\ref{fig:loss_varM_fixederr} through~\ref{fig:loss_fixedM_varerr}. To this end, all random unitary operators are assumed to follow the Haar distribution for $M>1$, which is arguably the most common two-design distribution.

\begin{table}[h!] 
	\caption{Values of $\MEAN{\tr{\ptr{1}{\rho}^{2M}}}{}$ for various $n$ and $M$. For $M>1$, Monte~Carlo simulations with 10000 samples are conducted for each $n,M$ pair. Random circuit unitary operators are assumed to follow the Haar distribution.\label{tab:val1}}
	\newcolumntype{C}{>{\centering\arraybackslash}X}
	\begin{tabularx}{\textwidth}{CCCCCC}
		\toprule
		$n$\textbackslash $M$	& 1	& 2 & 3 & 4 & 5\\
		\midrule
		2		& 4/5			& 0.6283 & 0.5191 & 0.4490 & 0.3885\\
		3		& 2/3			& 0.3934 & 0.2605 & 0.1808 & 0.1311\\
		4		& 10/17			& 0.2648 & 0.1356 & 0.0754 & 0.0431\\
		5		& 6/11			& 0.1940 & 0.0794 & 0.0352 & 0.0167\\
		6		& 34/65			& 0.1604 & 0.0544 & 0.0201 & 0.0075\\
		\bottomrule
	\end{tabularx}
\end{table}

\begin{table}[h!] 
	\caption{Values of $\MEAN{\left(\sum^n_{j=1}\tr{\ptr{j}{\rho}^M}\right)^2}{}$ for various $n$ and $M$. For $M>1$, Monte~Carlo simulations with 10000 samples are conducted for each $n,M$ pair. Random circuit unitary operators are assumed to follow the Haar distribution.\label{tab:val2}}
	\newcolumntype{C}{>{\centering\arraybackslash}X}
	\begin{tabularx}{\textwidth}{CCCCCC}
		\toprule
		$n$\textbackslash $M$	& 1	& 2 & 3 & 4 & 5\\
		\midrule
		2		& 4 & 2.6278 & 2.0963 & 1.7994 & 1.5547\\
		3		& 9 & 4.0678 & 2.3808 & 1.5438 & 1.0734\\
		4		& 16 & 5.5635 & 2.3935 & 1.1653 & 0.6196\\
		5		& 25 & 7.4374 & 2.5432 & 0.9705 & 0.3969\\
		6		& 36 & 9.8520 & 2.9229 & 0.9272 & 0.3104\\
		\bottomrule
	\end{tabularx}
\end{table}

\section{Virtual distillation under i.i.d. Pauli channel}
\label{app:Pauli_channel}

The general MSE expression for arbitrary $\epsilon=\sum^3_{l=1}\epsilon_l$ and $L_\mathrm{err}$ may be obtained from the map action in \eqref{eq:nqubit_pauli} by sketching a flowchart of how $\rho'$ evolves as noise dilution proceeds for $L_\mathrm{err}$ layers, each with a diluted error rate of $\epsilon/L_\mathrm{err}$.

\begin{figure}[h!]
	\centering
	\includegraphics[width=1\columnwidth]{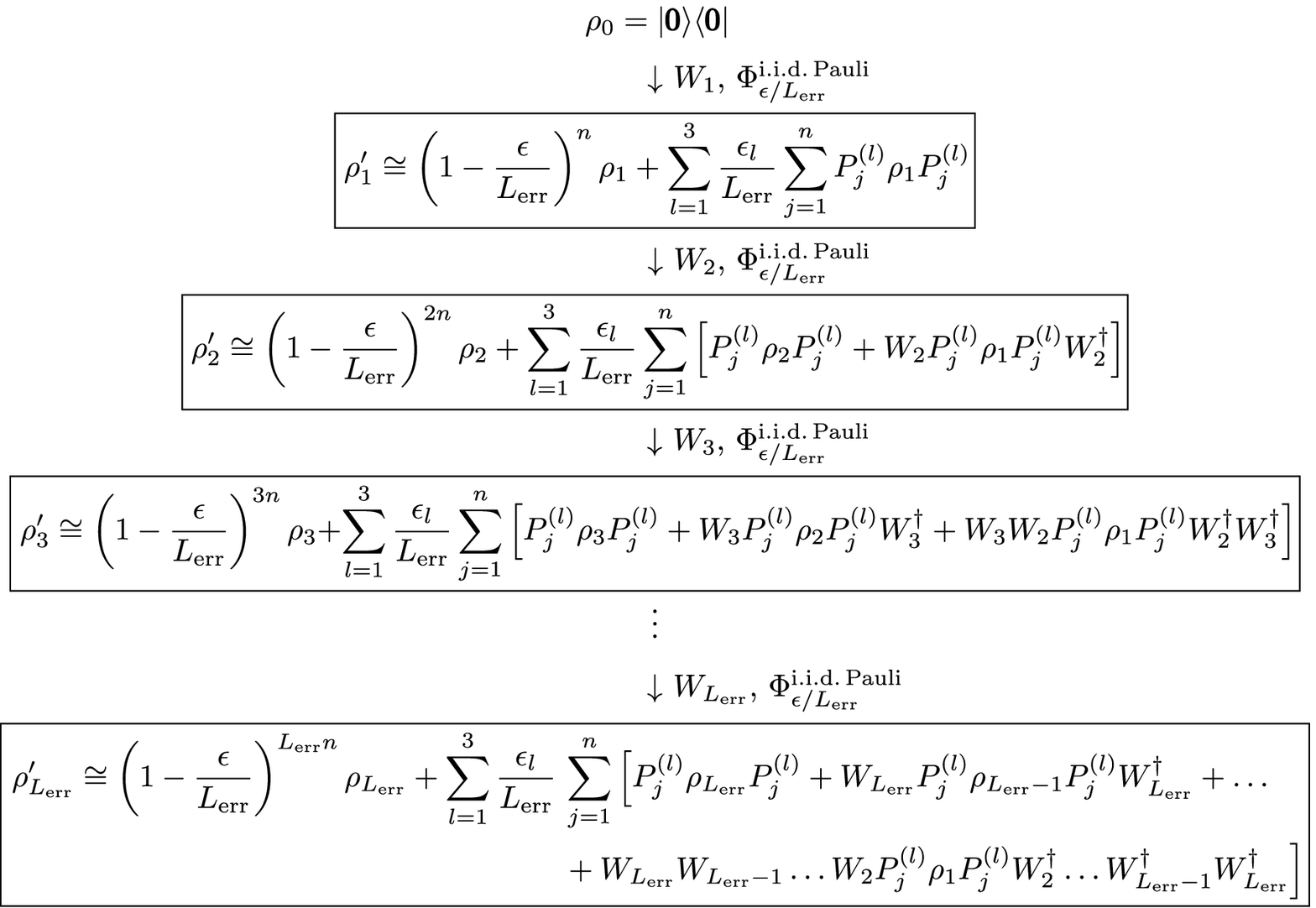}
	\caption{\label{fig:pauli_flowchart}An evolution flowchart of $\rho'$ in an $L_\mathrm{err}$-layered dilution setting. Here $\rho_l\equiv W_l\ldots W_2W_1\ket{\rvec{0}}\bra{\rvec{0}}W_1^\dag W_2^\dag\ldots W_l^\dag$ and $P_{j}^{(l)}$ is one of the single-qubit Pauli operators.}
\end{figure} 

We therefore find the following general noisy state $\rho'\equiv\rho'_{L_\mathrm{err}}$, with respect to the target $\rho\equiv\rho_{L_\mathrm{err}}$, expanded up to first order in all the error rates $\epsilon_1$, $\epsilon_2$ and $\epsilon_3$:
\begin{align}
	\rho'\cong&\,\left(1-n\epsilon\right)\rho+\sum^3_{l=1}\frac{\epsilon_l}{L_\mathrm{err}}\sum^n_{j=1}T^{(l)}_{j}\,,\nonumber\\
	T^{(l)}_{j}=&\,P^{(l)}_j\rho P^{(l)}_j+W_{L_\mathrm{err}}P^{(l)}_j\rho_{L_\mathrm{err}-1}P^{(l)}_jW_{L_\mathrm{err}}^\dag+\ldots\nonumber\\
	&\,+W_{L_\mathrm{err}}W_{L_\mathrm{err}-1}\ldots W_2P^{(l)}_j\rho_1P^{(l)}_jW_2^\dag\ldots W_{L_\mathrm{err}-1}^\dag W_{L_\mathrm{err}}^\dag\,.
\end{align}
After raising $\rho'_{L_\mathrm{err}}$ to the $M$th power, upon a further trace normalization, we get
\begin{equation}
	\dfrac{\rho'^M}{\tr{\rho'^M}}\cong\rho+\sum^3_{l=1}\frac{\epsilon_l}{L_\mathrm{err}}\sum^n_{j=1}\left[\rho\,T^{(l)}_j+(M-2)\,\rho \,T^{(l)}_j\,\rho+T^{(l)}_j\,\rho-M\,\rho\,\tr{\rho T^{(l)}_j}\right]\,.
\end{equation}
Hence, unlike the loss channel, where the error term goes as $\epsilon^M$, the Pauli channel gives rise to a persistent error term that \emph{does not} go away by simply increasing $M$ to infinity. This implies the MSE expressions
\begin{align}
	\mathrm{MSE}^{\mathrm{i.i.d.\,Pauli}}_{M=1,L_\mathrm{err}}=&\,(n\epsilon)^2-2n\epsilon\sum^3_{l=1}\frac{\epsilon_l}{L_\mathrm{err}}\sum^n_{j=1}\MEAN{\tr{\rho\,T^{(l)}_{j}}}{}+\sum^3_{l,l'=1}\frac{\epsilon_l\epsilon_{l'}}{L_\mathrm{err}^2}\sum^n_{j,j'=1}\MEAN{\tr{T^{(l)}_{j}T^{(l')}_{j'}}}{}\,,\nonumber\\
	\mathrm{MSE}^{\mathrm{i.i.d.\,Pauli}}_{M\geq2,L_\mathrm{err}}=&\,2\sum^3_{l,l'=1}\frac{\epsilon_l\epsilon_{l'}}{L_\mathrm{err}^2}\sum^n_{j,j'=1}\left[\MEAN{\tr{\rho\,T^{(l)}_{j}T^{(l')}_{j'}}}{}-\MEAN{\tr{\rho\,T^{(l)}_{j}}\tr{\rho\,T^{(l')}_{j'}}}{}\right]\,.
\end{align}
In other words, there is a difference in MSE in raising the virtual distillation order from $M=1$ to $M=2$. However, up to first order in error rates, carrying out virtual distillation with orders beyond $M=2$ offers no further reduction in MSE. This is a manifestation of the noncommutativity between $T^{(l)}_j$ and $\rho$.

The remaining task is the evaluation of circuit averages. First, recall that $T^{(l)}_j$ is a sum of $L_\mathrm{err}$ pure states. The average
\begin{equation}
	\MEAN{\tr{\rho\,T^{(l)}_{j}}}{}=\MEAN{\tr{\rho\,P^{(l)}_{j}\,\rho\,P^{(l)}_{j}}}{}+\ldots
\end{equation}
consists of the term
\begin{equation}
	\MEAN{\tr{\rho\,P^{(l)}_{j}\,\rho\,P^{(l)}_{j}}}{}=\MEAN{\tr{W_{L_\mathrm{err}}\rho_{L_\mathrm{err}-1}W_{L_\mathrm{err}}^\dag\,P^{(l)}_{j}\,W_{L_\mathrm{err}}\rho_{L_\mathrm{err}-1}W_{L_\mathrm{err}}^\dag\,P^{(l)}_{j}}}{}=\dfrac{1}{d+1}\,,
\end{equation}
where we have put \eqref{eq:useful_2design_3} to good use, and all other $L_\mathrm{err}-1$ terms contribute precisely the same result,
so that
\begin{equation}
	\MEAN{\tr{\rho\,T^{(l)}_{j}}}{}=\dfrac{L_\mathrm{err}}{d+1}\,.
	\label{eq:avgrhoT}
\end{equation}
The next average, $\MEAN{\tr{T^{(l)}_{j}T^{(l')}_{j'}}}{}$, would depend on the indices $j$, $j'$, $l$ and $l'$. Generally speaking, this term is consists of two types of averages, the self-terms such as $\MEAN{\tr{P^{(l)}_{j}\,\rho\,P^{(l)}_{j}P^{(l')}_{j'}\,\rho\,P^{(l')}_{j'}}}{}$, and cross-terms like $\MEAN{\tr{P^{(l)}_{j}\,\rho\,P^{(l)}_{j}W_{L_\mathrm{err}}P^{(l')}_{j'}\,\rho_{L_\mathrm{err}-1}\,P^{(l')}_{j'}W_{L_\mathrm{err}}^\dag}}{}$. Since every term of the same type gives the same result, we simply calculate one of each. Starting with the latter, for any indices, the combined use of \eqref{eq:useful_2design_2} and \eqref{eq:useful_2design_3} leads to
\begin{align}
	&\,\MEAN{\tr{P^{(l)}_{j}\,\rho\,P^{(l)}_{j}W_{L_\mathrm{err}}P^{(l')}_{j'}\,\rho_{L_\mathrm{err}-1}\,P^{(l')}_{j'}W_{L_\mathrm{err}}^\dag}}{}\nonumber\\
	=&\,\MEAN{\tr{P^{(l)}_{j}\,W_{L_\mathrm{err}}\rho_{L_\mathrm{err}-1}W_{L_\mathrm{err}}^\dag\,P^{(l)}_{j}\,W_{L_\mathrm{err}}P^{(l')}_{j'}\,\rho_{L_\mathrm{err}-1}\,P^{(l')}_{j'}W_{L_\mathrm{err}}^\dag}}{}\nonumber\\
	=&\,\dfrac{d}{d^2-1}-\dfrac{1}{d^2-1}\MEAN{\tr{\rho_{L_\mathrm{err}-1}\,P^{(l')}_{j'}\,\rho_{L_\mathrm{err}-1}\,P^{(l')}_{j'}}}{}\nonumber\\
	=&\,\dfrac{d^2+d-1}{(d+1)(d^2-1)}\,.
\end{align}
For the former,
\begin{equation}
	\MEAN{\tr{P^{(l)}_{j}\,\rho\,P^{(l)}_{j}P^{(l')}_{j'}\,\rho\,P^{(l')}_{j'}}}{}=\begin{cases}
		\quad1\!\!\!\!\!\!&,\,\,j=j'\text{ and }l=l'\,;\nonumber\\
		\dfrac{1}{d+1}\!\!\!\!\!\!&,\,\,\text{otherwise}\,.
	\end{cases}
\end{equation}
So,
\begin{equation}
	\MEAN{\tr{T^{(l)}_{j}T^{(l')}_{j'}}}{}=L_\mathrm{err}\,\dfrac{d\,\delta_{j,j'}\delta_{l,l'}+1}{d+1}+L_\mathrm{err}(L_\mathrm{err}-1)\dfrac{d^2+d-1}{(d+1)(d^2-1)}\,.
	\label{eq:avgTT}
\end{equation}
Equations~\eqref{eq:avgrhoT} and \eqref{eq:avgTT}, therefore, supply the exact answer
\begin{equation}
	\mathrm{MSE}^{\mathrm{i.i.d.\,Pauli}}_{M=1,L_\mathrm{err}}=(n\epsilon)^2\left[\dfrac{d^3}{(d+1)(d^2-1)}-\dfrac{d}{L_\mathrm{err}(d+1)(d^2-1)}\right]+\dfrac{n\,d}{L_\mathrm{err}(d+1)}\sum^3_{l=1}\epsilon^2_l
	\label{eq:MSE_pauli_M1}
\end{equation}
for any two-design unitary operators $W_l$.

When $M\geq2$, only certain averages are calculable solely from the two-design properties. For instance, the term $\MEAN{\tr{\rho\,T^{(l)}_{j}T^{(l')}_{j'}}}{}$ may again be found by considering different index conditions. If $j=j'$ and $l=l'$, then the same steps as before produce
\begin{equation}
	\MEAN{\tr{\rho\,T^{(l)\,2}_{j}}}{}=\dfrac{L_\mathrm{err}}{d+1}+\dfrac{L_\mathrm{err}(L_\mathrm{err}-1)}{(d+1)^2}\,.
\end{equation}
Otherwise, if $l\neq l'$, there exists an interesting analytical observation for $\MEAN{\tr{\rho\,T^{(l)}_{j}T^{(l')}_{j}}}{}$, that is when $j=j'$, one finds that each of the $L_\mathrm{err}$ self-terms, say
\begin{align}
	\MEAN{\tr{\rho\,P^{(l)}_{j}\,\rho\,P^{(l)}_{j}P^{(l')}_{j}\,\rho\,P^{(l')}_{j}}}{}=&\,\MEAN{\opinner{\,\,\,}{P^{(l)}_{j}}{\,\,\,}\opinner{\,\,\,}{P^{(l')}_{j}}{\,\,\,}\opinner{\,\,\,}{P^{(l)}_{j}P^{(l')}_{j}}{\,\,\,}}{}\nonumber\\
	=&\,-\MEAN{\opinner{\,\,\,}{P^{(l)}_{j}}{\,\,\,}\opinner{\,\,\,}{P^{(l')}_{j}}{\,\,\,}\opinner{\,\,\,}{P^{(l')}_{j}P^{(l)}_{j}}{\,\,\,}}{}\,,
\end{align}
is purely imaginary owing to the anticommutativity of the single-qubit Pauli operators. Apart from this, both $\MEAN{\tr{\rho\,T^{(l)}_{j}T^{(l')}_{j'}}}{}$ and $\MEAN{\tr{\rho\,T^{(l)}_{j}}\tr{\rho\,T^{(l')}_{j'}}}{}$ appearing in $\mathrm{MSE}^{\mathrm{i.i.d.\,Pauli}}_{M\geq2,L_\mathrm{err}}$ involve third-moments that depend on the two-design distribution.

By assuming the Haar distribution for all circuit unitary operators, we present some lists of values for these two averages in order to compare the small-$\epsilon$ MSE analytical formulas with the simulation result for $M\geq2$ in Tabs.~\ref{tab:avgrhoTT} and \ref{tab:avgrhoTrhoT}. These are used to produce Figs.~\ref{fig:pauli_varM_fixederr} through~\ref{fig:pauli_fixedM_varerr}.

\begin{table}[h!] 
	\caption{Values of $\MEAN{\tr{\rho\,T^{(l)}_{j}T^{(l')}_{j'}}}{}$ for various $n$ and $L_\mathrm{err}$. Monte~Carlo simulations with 10000 samples are conducted for each $n,L_\mathrm{err}$ pair. Random circuit unitary operators are assumed to follow the Haar distribution.\label{tab:avgrhoTT}}
	\newcolumntype{C}{>{\centering\arraybackslash}X}
	
	\flushleft{$l\neq l'$, $j=j'$}
	\begin{tabularx}{\textwidth}{CCCCC}
		\toprule
		$n$\textbackslash $L_\mathrm{err}$	& 1	& 2 & 3 & 4\\
		\midrule
		2 & 0.0000 & 0.0791 & 0.2422 & 0.4802\\
		3 & 0.0000 & 0.0246 & 0.0739 & 0.1486\\
		4 & 0.0000 & 0.0071 & 0.0204 & 0.0415\\
		5 & 0.0000 & 0.0019 & 0.0054 & 0.0110\\
		6 & 0.0000 & 0.0005 & 0.0014 & 0.0028\\
		\bottomrule
	\end{tabularx}\\[2ex]
	\flushleft{Any $l$ and $l'$, $j\neq j'$}
	\begin{tabularx}{\textwidth}{CCCCC}
		\toprule
		$n$\textbackslash $L_\mathrm{err}$	& 1	& 2 & 3 & 4\\
		\midrule
		2 & 0.0696 & 0.2090 & 0.4363 & 0.7526\\
		3 & 0.0226 & 0.0703 & 0.1417 & 0.2336\\
		4 & 0.0061 & 0.0201 & 0.0397 & 0.0679\\
		5 & 0.0017 & 0.0053 & 0.0109 & 0.0175\\
		6 & 0.0005 & 0.0014 & 0.0027 & 0.0044\\
		\bottomrule
	\end{tabularx}
\end{table}

\begin{table}[h!] 
	\caption{Values of $\MEAN{\tr{\rho\,T^{(l)}_{j}}\tr{\rho\,T^{(l')}_{j'}}}{}$ for various $n$ and $L_\mathrm{err}$. Monte~Carlo simulations with 10000 samples are conducted for each $n,L_\mathrm{err}$ pair. Random circuit unitary operators are assumed to follow the Haar distribution.\label{tab:avgrhoTrhoT}}
	\newcolumntype{C}{>{\centering\arraybackslash}X}
	\flushleft{$l=l'$, $j=j'$}
	\begin{tabularx}{\textwidth}{CCCCC}
		\toprule
		$n$\textbackslash $L_\mathrm{err}$	& 1	& 2 & 3 & 4\\
		\midrule
		2 & 0.0867 & 0.2560 & 0.5036 & 0.8145\\
		3 & 0.0298 & 0.0862 & 0.1649 & 0.2715\\
		4 & 0.0092 & 0.0268 & 0.0489 & 0.0809\\
		5 & 0.0026 & 0.0070 & 0.0134 & 0.0217\\
		6 & 0.0007 & 0.0018 & 0.0034 & 0.0056\\
		\bottomrule
	\end{tabularx}\\[2ex]
	\flushleft{$l\neq l'$, $j=j'$}
	\begin{tabularx}{\textwidth}{CCCCC}
		\toprule
		$n$\textbackslash $L_\mathrm{err}$	& 1	& 2 & 3 & 4\\
		\midrule
		2 & 0.0295 & 0.1371 & 0.3296 & 0.5993\\
		3 & 0.0101 & 0.0445 & 0.1027 & 0.1868\\
		4 & 0.0030 & 0.0131 & 0.0301 & 0.0541\\
		5 & 0.0009 & 0.0036 & 0.0080 & 0.0144\\
		6 & 0.0002 & 0.0009 & 0.0021 & 0.0037\\
		\bottomrule
	\end{tabularx}\\[2ex]
	\flushleft{Any $l$ and $l'$, $j\neq j'$}
	\begin{tabularx}{\textwidth}{CCCCC}
	\toprule
	$n$\textbackslash $L_\mathrm{err}$	& 1	& 2 & 3 & 4\\
	\midrule
	2 & 0.0480 & 0.1744 & 0.3848 & 0.6821\\
	3 & 0.0142 & 0.0518 & 0.1161 & 0.2021\\
	4 & 0.0038 & 0.0144 & 0.0316 & 0.0559\\
	5 & 0.0010 & 0.0038 & 0.0084 & 0.0150\\
	6 & 0.0002 & 0.0010 & 0.0021 & 0.0038\\
	\bottomrule
	\end{tabularx}
\end{table}

\section{Hellinger distance for i.i.d. losses on a product state}
\label{app:loss_hellinger}

If $\ket{\,\,\,}=\ket{\psi_1}\ket{\psi_2}\ldots\ket{\psi_n}$ is an $n$-qubit product ket, then remembering once again that $\vacket\vacbra$ is invisible to \emph{all} circuit unitary operators, the exact lossy state
\begin{align}
	\rho'_{L_\mathrm{err}}=&\,\ket{\,\,\,}\left(1-\frac{\epsilon}{L_\mathrm{err}}\right)^{L_\mathrm{err}n}\bra{\,\,\,}+\frac{\epsilon}{L_\mathrm{err}}\left(1-\frac{\epsilon}{L_\mathrm{err}}\right)^{L_\mathrm{err}n-1}\underbrace{\left(\Psi_1+\ldots+\Psi_n\right)}_{\displaystyle\mathclap{\text{$\binom{n}{1}$ terms}}}\nonumber\\
	&\,+\left(\frac{\epsilon}{L_\mathrm{err}}\right)^2\left(1-\frac{\epsilon}{L_\mathrm{err}}\right)^{L_\mathrm{err}n-2}\underbrace{\left(\Psi_{1,2}+\Psi_{1,3}+\ldots+\Psi_{n-1,n-2}+\Psi_{n-1,n}\right)}_{\displaystyle\mathclap{\text{$\binom{n}{2}$ terms}}}+\ldots\nonumber\\
	&\,\left(\frac{\epsilon}{L_\mathrm{err}}\right)^{n-1}\left(1-\frac{\epsilon}{L_\mathrm{err}}\right)^{L_\mathrm{err}n-n+1}\underbrace{\left(\Psi_{1,2,\ldots,n-1}+\Psi_{1,2,\ldots,n-2,n}+\ldots+\Psi_{2,3,\ldots,n}\right)}_{\displaystyle\mathclap{\text{$\binom{n}{n-1}$ terms}}}\nonumber\\
	&\,+\vacket\left(\frac{\epsilon}{L_\mathrm{err}}\right)^{n}\left(1-\frac{\epsilon}{L_\mathrm{err}}\right)^{L_\mathrm{err}n-n}\vacbra\nonumber\\
	=&\,\ket{\,\,\,}\left(1-\frac{\epsilon}{L_\mathrm{err}}\right)^{L_\mathrm{err}n}\bra{\,\,\,}+\left(1-\frac{\epsilon}{L_\mathrm{err}}\right)^{(L_\mathrm{err}-1)n}\left[1-\left(1-\frac{\epsilon}{L_\mathrm{err}}\right)^{n}\right]\rho_\mathrm{err}
	\label{eq:full_loss_product}
\end{align}
is a linear combination of \emph{orthonormal} vacuum-substituted projectors $\Psi_{\,\cdot\,}$, where $\Psi_{1,2}=\vacket\vacbra\otimes\vacket\vacbra\otimes\ket{\psi_3}\bra{\psi_3}\otimes\ldots\otimes\ket{\psi_n}\bra{\psi_n}$, for instance. It is clear that $\tr{\rho'_{L_\mathrm{err}}}=(1-\epsilon/L_\mathrm{err})^{(L_\mathrm{err}-1)n}<1$ whenever $L_\mathrm{err}>1$, since discarding noncoincidental data means that every action by a subciruit unitary operator $W_l$ loses information about the error component at every dilution layer. From Eq.~\eqref{eq:full_loss_product}, the trace-normalized error component has degenerate eigenvalues according to the following multiplicities:
\begin{table}[h!] 
	\caption{Eigenvalues of $\rho_\mathrm{err}$ for the i.i.d. loss channel acting on a product state and their multiplicities.\label{tab:eig_err}}
	\newcolumntype{C}{>{\centering\arraybackslash}X}
	\begin{tabularx}{\textwidth}{CC}
		\toprule
		eigenvalues of $\rho_{L_\mathrm{err}}$	& multiplicity\\
		\midrule
		$\displaystyle0$ & $\displaystyle3^n-2^n+1$\\
		$\displaystyle\frac{\epsilon}{L_\mathrm{err}}\left(1-\frac{\epsilon}{L_\mathrm{err}}\right)^{n-1}$ & $\displaystyle\binom{n}{1}$\\[1ex]
		$\displaystyle\left(\frac{\epsilon}{L_\mathrm{err}}\right)^2\left(1-\frac{\epsilon}{L_\mathrm{err}}\right)^{n-2}$ & $\displaystyle\binom{n}{2}$\\[1ex]
		$\displaystyle\vdots$ & $\displaystyle\vdots$\\[1ex]
		$\displaystyle\left(\frac{\epsilon}{L_\mathrm{err}}\right)^{n-1}\left(1-\frac{\epsilon}{L_\mathrm{err}}\right)$ & $\displaystyle\binom{n}{n-1}$\\[1ex]
		$\displaystyle\left(\frac{\epsilon}{L_\mathrm{err}}\right)^{n}$ & $\displaystyle\binom{n}{n}$\\
		\bottomrule
	\end{tabularx}
\end{table}

From the definition of the Hellinger distance in~Eq.~\ref{eq:hellinger}, these eigenvalues and multiplicities lead to \eqref{eq:hellinger_iid_loss}.


\begin{thebibliography}{116}%
\makeatletter
\providecommand \@ifxundefined [1]{%
 \@ifx{#1\undefined}
}%
\providecommand \@ifnum [1]{%
 \ifnum #1\expandafter \@firstoftwo
 \else \expandafter \@secondoftwo
 \fi
}%
\providecommand \@ifx [1]{%
 \ifx #1\expandafter \@firstoftwo
 \else \expandafter \@secondoftwo
 \fi
}%
\providecommand \natexlab [1]{#1}%
\providecommand \enquote  [1]{``#1''}%
\providecommand \bibnamefont  [1]{#1}%
\providecommand \bibfnamefont [1]{#1}%
\providecommand \citenamefont [1]{#1}%
\providecommand \href@noop [0]{\@secondoftwo}%
\providecommand \href [0]{\begingroup \@sanitize@url \@href}%
\providecommand \@href[1]{\@@startlink{#1}\@@href}%
\providecommand \@@href[1]{\endgroup#1\@@endlink}%
\providecommand \@sanitize@url [0]{\catcode `\\12\catcode `\$12\catcode
  `\&12\catcode `\#12\catcode `\^12\catcode `\_12\catcode `\%12\relax}%
\providecommand \@@startlink[1]{}%
\providecommand \@@endlink[0]{}%
\providecommand \url  [0]{\begingroup\@sanitize@url \@url }%
\providecommand \@url [1]{\endgroup\@href {#1}{\urlprefix }}%
\providecommand \urlprefix  [0]{URL }%
\providecommand \Eprint [0]{\href }%
\providecommand \doibase [0]{https://doi.org/}%
\providecommand \selectlanguage [0]{\@gobble}%
\providecommand \bibinfo  [0]{\@secondoftwo}%
\providecommand \bibfield  [0]{\@secondoftwo}%
\providecommand \translation [1]{[#1]}%
\providecommand \BibitemOpen [0]{}%
\providecommand \bibitemStop [0]{}%
\providecommand \bibitemNoStop [0]{.\EOS\space}%
\providecommand \EOS [0]{\spacefactor3000\relax}%
\providecommand \BibitemShut  [1]{\csname bibitem#1\endcsname}%
\let\auto@bib@innerbib\@empty
\bibitem [{\citenamefont {Deutsch}\ \emph {et~al.}(1995)\citenamefont
  {Deutsch}, \citenamefont {Barenco},\ and\ \citenamefont
  {Ekert}}]{Deutsch:1995universality}%
  \BibitemOpen
  \bibfield  {author} {\bibinfo {author} {\bibfnamefont {D.~E.}\ \bibnamefont
  {Deutsch}}, \bibinfo {author} {\bibfnamefont {A.}~\bibnamefont {Barenco}},\
  and\ \bibinfo {author} {\bibfnamefont {A.}~\bibnamefont {Ekert}},\ }\bibfield
   {title} {\bibinfo {title} {Universality in quantum computation},\ }\href
  {https://doi.org/10.1098/rspa.1995.0065} {\bibfield  {journal} {\bibinfo
  {journal} {Proceedings of the Royal Society of London. Series A: Mathematical
  and Physical Sciences}\ }\textbf {\bibinfo {volume} {449}},\ \bibinfo {pages}
  {669} (\bibinfo {year} {1995})}\BibitemShut {NoStop}%
\bibitem [{\citenamefont {Barenco}\ \emph {et~al.}(1995)\citenamefont
  {Barenco}, \citenamefont {Bennett}, \citenamefont {Cleve}, \citenamefont
  {DiVincenzo}, \citenamefont {Margolus}, \citenamefont {Shor}, \citenamefont
  {Sleator}, \citenamefont {Smolin},\ and\ \citenamefont
  {Weinfurter}}]{Barenco:1995elementary}%
  \BibitemOpen
  \bibfield  {author} {\bibinfo {author} {\bibfnamefont {A.}~\bibnamefont
  {Barenco}}, \bibinfo {author} {\bibfnamefont {C.~H.}\ \bibnamefont
  {Bennett}}, \bibinfo {author} {\bibfnamefont {R.}~\bibnamefont {Cleve}},
  \bibinfo {author} {\bibfnamefont {D.~P.}\ \bibnamefont {DiVincenzo}},
  \bibinfo {author} {\bibfnamefont {N.}~\bibnamefont {Margolus}}, \bibinfo
  {author} {\bibfnamefont {P.}~\bibnamefont {Shor}}, \bibinfo {author}
  {\bibfnamefont {T.}~\bibnamefont {Sleator}}, \bibinfo {author} {\bibfnamefont
  {J.~A.}\ \bibnamefont {Smolin}},\ and\ \bibinfo {author} {\bibfnamefont
  {H.}~\bibnamefont {Weinfurter}},\ }\bibfield  {title} {\bibinfo {title}
  {Elementary gates for quantum computation},\ }\href
  {https://doi.org/10.1103/PhysRevA.52.3457} {\bibfield  {journal} {\bibinfo
  {journal} {Phys. Rev. A}\ }\textbf {\bibinfo {volume} {52}},\ \bibinfo
  {pages} {3457} (\bibinfo {year} {1995})}\BibitemShut {NoStop}%
\bibitem [{\citenamefont {Englert}\ \emph {et~al.}(2001)\citenamefont
  {Englert}, \citenamefont {Kurtsiefer},\ and\ \citenamefont
  {Weinfurter}}]{Englert:2001universal}%
  \BibitemOpen
  \bibfield  {author} {\bibinfo {author} {\bibfnamefont {B.-G.}\ \bibnamefont
  {Englert}}, \bibinfo {author} {\bibfnamefont {C.}~\bibnamefont
  {Kurtsiefer}},\ and\ \bibinfo {author} {\bibfnamefont {H.}~\bibnamefont
  {Weinfurter}},\ }\bibfield  {title} {\bibinfo {title} {Universal unitary gate
  for single-photon two-qubit states},\ }\href
  {https://doi.org/10.1103/PhysRevA.63.032303} {\bibfield  {journal} {\bibinfo
  {journal} {Phys. Rev. A}\ }\textbf {\bibinfo {volume} {63}},\ \bibinfo
  {pages} {032303} (\bibinfo {year} {2001})}\BibitemShut {NoStop}%
\bibitem [{\citenamefont {Bartlett}\ \emph {et~al.}(2002)\citenamefont
  {Bartlett}, \citenamefont {Sanders}, \citenamefont {Braunstein},\ and\
  \citenamefont {Nemoto}}]{Bartlett:2002efficient}%
  \BibitemOpen
  \bibfield  {author} {\bibinfo {author} {\bibfnamefont {S.~D.}\ \bibnamefont
  {Bartlett}}, \bibinfo {author} {\bibfnamefont {B.~C.}\ \bibnamefont
  {Sanders}}, \bibinfo {author} {\bibfnamefont {S.~L.}\ \bibnamefont
  {Braunstein}},\ and\ \bibinfo {author} {\bibfnamefont {K.}~\bibnamefont
  {Nemoto}},\ }\bibfield  {title} {\bibinfo {title} {{Efficient Classical
  Simulation of Continuous Variable Quantum Information Processes}},\ }\href
  {https://doi.org/10.1103/PhysRevLett.88.097904} {\bibfield  {journal}
  {\bibinfo  {journal} {Phys. Rev. Lett.}\ }\textbf {\bibinfo {volume} {88}},\
  \bibinfo {pages} {097904} (\bibinfo {year} {2002})}\BibitemShut {NoStop}%
\bibitem [{\citenamefont {Sawicki}\ \emph {et~al.}(2022)\citenamefont
  {Sawicki}, \citenamefont {Mattioli},\ and\ \citenamefont
  {Zimbor\'as}}]{Sawicki:2022universality}%
  \BibitemOpen
  \bibfield  {author} {\bibinfo {author} {\bibfnamefont {A.}~\bibnamefont
  {Sawicki}}, \bibinfo {author} {\bibfnamefont {L.}~\bibnamefont {Mattioli}},\
  and\ \bibinfo {author} {\bibfnamefont {Z.}~\bibnamefont {Zimbor\'as}},\
  }\bibfield  {title} {\bibinfo {title} {Universality verification for a set of
  quantum gates},\ }\href {https://doi.org/10.1103/PhysRevA.105.052602}
  {\bibfield  {journal} {\bibinfo  {journal} {Phys. Rev. A}\ }\textbf {\bibinfo
  {volume} {105}},\ \bibinfo {pages} {052602} (\bibinfo {year}
  {2022})}\BibitemShut {NoStop}%
\bibitem [{\citenamefont {Chuang}\ and\ \citenamefont
  {Nielsen}(2000)}]{Chuang:2000fk}%
  \BibitemOpen
  \bibfield  {author} {\bibinfo {author} {\bibfnamefont {I.}~\bibnamefont
  {Chuang}}\ and\ \bibinfo {author} {\bibfnamefont {M.}~\bibnamefont
  {Nielsen}},\ }\href@noop {} {\emph {\bibinfo {title} {Quantum Computation and
  Quantum Information}}}\ (\bibinfo  {publisher} {Cambridge University Press},\
  \bibinfo {address} {Cambridge},\ \bibinfo {year} {2000})\BibitemShut
  {NoStop}%
\bibitem [{\citenamefont {Ladd}\ \emph {et~al.}(2010)\citenamefont {Ladd},
  \citenamefont {Jelezko}, \citenamefont {Laflamme}, \citenamefont {Nakamura},
  \citenamefont {Monroe},\ and\ \citenamefont {O'Brien}}]{Ladd:2010aa}%
  \BibitemOpen
  \bibfield  {author} {\bibinfo {author} {\bibfnamefont {T.~D.}\ \bibnamefont
  {Ladd}}, \bibinfo {author} {\bibfnamefont {F.}~\bibnamefont {Jelezko}},
  \bibinfo {author} {\bibfnamefont {R.}~\bibnamefont {Laflamme}}, \bibinfo
  {author} {\bibfnamefont {Y.}~\bibnamefont {Nakamura}}, \bibinfo {author}
  {\bibfnamefont {C.}~\bibnamefont {Monroe}},\ and\ \bibinfo {author}
  {\bibfnamefont {J.~L.}\ \bibnamefont {O'Brien}},\ }\bibfield  {title}
  {\bibinfo {title} {Quantum computers},\ }\href
  {https://dx.doi.org/10.1038/nature08812} {\bibfield  {journal} {\bibinfo
  {journal} {Nature}\ }\textbf {\bibinfo {volume} {464}},\ \bibinfo {pages}
  {45} (\bibinfo {year} {2010})}\BibitemShut {NoStop}%
\bibitem [{\citenamefont {Campbell}\ \emph {et~al.}(2017)\citenamefont
  {Campbell}, \citenamefont {Terhal},\ and\ \citenamefont
  {Vuillot}}]{Campbell:2017aa}%
  \BibitemOpen
  \bibfield  {author} {\bibinfo {author} {\bibfnamefont {E.~T.}\ \bibnamefont
  {Campbell}}, \bibinfo {author} {\bibfnamefont {B.~M.}\ \bibnamefont
  {Terhal}},\ and\ \bibinfo {author} {\bibfnamefont {C.}~\bibnamefont
  {Vuillot}},\ }\bibfield  {title} {\bibinfo {title} {Roads towards
  fault-tolerant universal quantum computation},\ }\href
  {https://dx.doi.org/10.1038/nature23460} {\bibfield  {journal} {\bibinfo
  {journal} {Nature}\ }\textbf {\bibinfo {volume} {549}},\ \bibinfo {pages}
  {172} (\bibinfo {year} {2017})}\BibitemShut {NoStop}%
\bibitem [{\citenamefont {Lekitsch}\ \emph {et~al.}(2017)\citenamefont
  {Lekitsch}, \citenamefont {Weidt}, \citenamefont {Fowler}, \citenamefont
  {M{\o}lmer}, \citenamefont {Devitt}, \citenamefont {Wunderlich},\ and\
  \citenamefont {Hensinger}}]{Lekitsch:2017aa}%
  \BibitemOpen
  \bibfield  {author} {\bibinfo {author} {\bibfnamefont {B.}~\bibnamefont
  {Lekitsch}}, \bibinfo {author} {\bibfnamefont {S.}~\bibnamefont {Weidt}},
  \bibinfo {author} {\bibfnamefont {A.~G.}\ \bibnamefont {Fowler}}, \bibinfo
  {author} {\bibfnamefont {K.}~\bibnamefont {M{\o}lmer}}, \bibinfo {author}
  {\bibfnamefont {S.~J.}\ \bibnamefont {Devitt}}, \bibinfo {author}
  {\bibfnamefont {C.}~\bibnamefont {Wunderlich}},\ and\ \bibinfo {author}
  {\bibfnamefont {W.~K.}\ \bibnamefont {Hensinger}},\ }\bibfield  {title}
  {\bibinfo {title} {Blueprint for a microwave trapped ion quantum computer},\
  }\href {https://dx.doi.org/10.1126/sciadv.1601540} {\bibfield  {journal}
  {\bibinfo  {journal} {Sci. Adv.}\ }\textbf {\bibinfo {volume} {3}},\ \bibinfo
  {pages} {e1601540} (\bibinfo {year} {2017})}\BibitemShut {NoStop}%
\bibitem [{\citenamefont {Grover}(1996)}]{Grover:1996fast}%
  \BibitemOpen
  \bibfield  {author} {\bibinfo {author} {\bibfnamefont {L.~K.}\ \bibnamefont
  {Grover}},\ }\bibfield  {title} {\bibinfo {title} {A fast quantum mechanical
  algorithm for database search},\ }in\ \href
  {https://doi.org/10.1145/237814.237866} {\emph {\bibinfo {booktitle}
  {Proceedings of the Twenty-Eighth Annual ACM Symposium on Theory of
  Computing}}},\ \bibinfo {series and number} {STOC '96}\ (\bibinfo
  {publisher} {Association for Computing Machinery},\ \bibinfo {address} {New
  York, NY, USA},\ \bibinfo {year} {1996})\ p.\ \bibinfo {pages}
  {212–219}\BibitemShut {NoStop}%
\bibitem [{\citenamefont {Shor}(1997)}]{Shor:1997polynomial}%
  \BibitemOpen
  \bibfield  {author} {\bibinfo {author} {\bibfnamefont {P.~W.}\ \bibnamefont
  {Shor}},\ }\bibfield  {title} {\bibinfo {title} {{Polynomial-Time Algorithms
  for Prime Factorization and Discrete Logarithms on a Quantum Computer}},\
  }\href {https://doi.org/10.1137/S0097539795293172} {\bibfield  {journal}
  {\bibinfo  {journal} {SIAM Journal on Computing}\ }\textbf {\bibinfo {volume}
  {26}},\ \bibinfo {pages} {1484} (\bibinfo {year} {1997})}\BibitemShut
  {NoStop}%
\bibitem [{\citenamefont {Raussendorf}\ and\ \citenamefont
  {Briegel}(2001)}]{Raussendorf:2001one-way}%
  \BibitemOpen
  \bibfield  {author} {\bibinfo {author} {\bibfnamefont {R.}~\bibnamefont
  {Raussendorf}}\ and\ \bibinfo {author} {\bibfnamefont {H.~J.}\ \bibnamefont
  {Briegel}},\ }\bibfield  {title} {\bibinfo {title} {A one-way quantum
  computer},\ }\href {https://doi.org/10.1103/PhysRevLett.86.5188} {\bibfield
  {journal} {\bibinfo  {journal} {Phys. Rev. Lett.}\ }\textbf {\bibinfo
  {volume} {86}},\ \bibinfo {pages} {5188} (\bibinfo {year}
  {2001})}\BibitemShut {NoStop}%
\bibitem [{\citenamefont {Kitaev}(2003)}]{Kitaev:2003fault-tolerant}%
  \BibitemOpen
  \bibfield  {author} {\bibinfo {author} {\bibfnamefont {A.}~\bibnamefont
  {Kitaev}},\ }\bibfield  {title} {\bibinfo {title} {Fault-tolerant quantum
  computation by anyons},\ }\href
  {https://doi.org/https://doi.org/10.1016/S0003-4916(02)00018-0} {\bibfield
  {journal} {\bibinfo  {journal} {Ann. Phys.}\ }\textbf {\bibinfo {volume}
  {303}},\ \bibinfo {pages} {2} (\bibinfo {year} {2003})}\BibitemShut {NoStop}%
\bibitem [{\citenamefont {Raussendorf}\ \emph {et~al.}(2007)\citenamefont
  {Raussendorf}, \citenamefont {Harrington},\ and\ \citenamefont
  {Goyal}}]{Raussendorf:2007topological}%
  \BibitemOpen
  \bibfield  {author} {\bibinfo {author} {\bibfnamefont {R.}~\bibnamefont
  {Raussendorf}}, \bibinfo {author} {\bibfnamefont {J.}~\bibnamefont
  {Harrington}},\ and\ \bibinfo {author} {\bibfnamefont {K.}~\bibnamefont
  {Goyal}},\ }\bibfield  {title} {\bibinfo {title} {Topological fault-tolerance
  in cluster state quantum computation},\ }\href
  {https://doi.org/10.1088/1367-2630/9/6/199} {\bibfield  {journal} {\bibinfo
  {journal} {New J. Phys.}\ }\textbf {\bibinfo {volume} {9}},\ \bibinfo {pages}
  {199} (\bibinfo {year} {2007})}\BibitemShut {NoStop}%
\bibitem [{\citenamefont {Sehrawat}\ \emph {et~al.}(2011)\citenamefont
  {Sehrawat}, \citenamefont {Nguyen},\ and\ \citenamefont
  {Englert}}]{Sehrawat:2011test-state}%
  \BibitemOpen
  \bibfield  {author} {\bibinfo {author} {\bibfnamefont {A.}~\bibnamefont
  {Sehrawat}}, \bibinfo {author} {\bibfnamefont {L.~H.}\ \bibnamefont
  {Nguyen}},\ and\ \bibinfo {author} {\bibfnamefont {B.-G.}\ \bibnamefont
  {Englert}},\ }\bibfield  {title} {\bibinfo {title} {Test-state approach to
  the quantum search problem},\ }\href
  {https://doi.org/10.1103/PhysRevA.83.052311} {\bibfield  {journal} {\bibinfo
  {journal} {Phys. Rev. A}\ }\textbf {\bibinfo {volume} {83}},\ \bibinfo
  {pages} {052311} (\bibinfo {year} {2011})}\BibitemShut {NoStop}%
\bibitem [{\citenamefont {Montanaro}(2016)}]{Montanaro:2016quantum}%
  \BibitemOpen
  \bibfield  {author} {\bibinfo {author} {\bibfnamefont {A.}~\bibnamefont
  {Montanaro}},\ }\bibfield  {title} {\bibinfo {title} {Quantum algorithms: an
  overview},\ }\href {https://doi.org/10.1038/npjqi.2015.23} {\bibfield
  {journal} {\bibinfo  {journal} {npj Quantum Information}\ }\textbf {\bibinfo
  {volume} {2}},\ \bibinfo {pages} {15023} (\bibinfo {year}
  {2016})}\BibitemShut {NoStop}%
\bibitem [{\citenamefont {Knill}\ \emph {et~al.}(1998)\citenamefont {Knill},
  \citenamefont {Laflamme},\ and\ \citenamefont {Zurek}}]{Knill:1998resilient}%
  \BibitemOpen
  \bibfield  {author} {\bibinfo {author} {\bibfnamefont {E.}~\bibnamefont
  {Knill}}, \bibinfo {author} {\bibfnamefont {R.}~\bibnamefont {Laflamme}},\
  and\ \bibinfo {author} {\bibfnamefont {W.~H.}\ \bibnamefont {Zurek}},\
  }\bibfield  {title} {\bibinfo {title} {{Resilient Quantum Computation}},\
  }\href {https://doi.org/10.1126/science.279.5349.342} {\bibfield  {journal}
  {\bibinfo  {journal} {Science}\ }\textbf {\bibinfo {volume} {279}},\ \bibinfo
  {pages} {342} (\bibinfo {year} {1998})}\BibitemShut {NoStop}%
\bibitem [{\citenamefont {Franklin}\ and\ \citenamefont
  {Chong}(2004)}]{Franklin:2004challenges}%
  \BibitemOpen
  \bibfield  {author} {\bibinfo {author} {\bibfnamefont {D.}~\bibnamefont
  {Franklin}}\ and\ \bibinfo {author} {\bibfnamefont {F.~T.}\ \bibnamefont
  {Chong}},\ }\bibinfo {title} {Challenges in reliable quantum computing},\ in\
  \href {https://doi.org/10.1007/1-4020-8068-9_8} {\emph {\bibinfo {booktitle}
  {Nano, Quantum and Molecular Computing: Implications to High Level Design and
  Validation}}},\ \bibinfo {editor} {edited by\ \bibinfo {editor}
  {\bibfnamefont {S.~K.}\ \bibnamefont {Shukla}}\ and\ \bibinfo {editor}
  {\bibfnamefont {R.~I.}\ \bibnamefont {Bahar}}}\ (\bibinfo  {publisher}
  {Springer US},\ \bibinfo {address} {Boston, MA},\ \bibinfo {year} {2004})\
  pp.\ \bibinfo {pages} {247--266}\BibitemShut {NoStop}%
\bibitem [{\citenamefont {Aharonov}\ and\ \citenamefont
  {Ben-Or}(2008)}]{Aharonov:2008fault-tolerant}%
  \BibitemOpen
  \bibfield  {author} {\bibinfo {author} {\bibfnamefont {D.}~\bibnamefont
  {Aharonov}}\ and\ \bibinfo {author} {\bibfnamefont {M.}~\bibnamefont
  {Ben-Or}},\ }\bibfield  {title} {\bibinfo {title} {{Fault-Tolerant Quantum
  Computation with Constant Error Rate}},\ }\href
  {https://doi.org/10.1137/S0097539799359385} {\bibfield  {journal} {\bibinfo
  {journal} {SIAM Journal on Computing}\ }\textbf {\bibinfo {volume} {38}},\
  \bibinfo {pages} {1207} (\bibinfo {year} {2008})}\BibitemShut {NoStop}%
\bibitem [{\citenamefont {Knill}(1995)}]{knill1995approximation}%
  \BibitemOpen
  \bibfield  {author} {\bibinfo {author} {\bibfnamefont {E.}~\bibnamefont
  {Knill}},\ }\href@noop {} {\bibinfo {title} {Approximation by quantum
  circuits}} (\bibinfo {year} {1995}),\ \Eprint
  {https://arxiv.org/abs/quant-ph/9508006} {arXiv:quant-ph/9508006 [quant-ph]}
  \BibitemShut {NoStop}%
\bibitem [{\citenamefont {Preskill}(2018)}]{Preskill2018quantumcomputingin}%
  \BibitemOpen
  \bibfield  {author} {\bibinfo {author} {\bibfnamefont {J.}~\bibnamefont
  {Preskill}},\ }\bibfield  {title} {\bibinfo {title} {Quantum {C}omputing in
  the {NISQ} era and beyond},\ }\href
  {https://doi.org/10.22331/q-2018-08-06-79} {\bibfield  {journal} {\bibinfo
  {journal} {{Quantum}}\ }\textbf {\bibinfo {volume} {2}},\ \bibinfo {pages}
  {79} (\bibinfo {year} {2018})}\BibitemShut {NoStop}%
\bibitem [{\citenamefont {Bromley}\ \emph {et~al.}(2020)\citenamefont
  {Bromley}, \citenamefont {Arrazola}, \citenamefont {Jahangiri}, \citenamefont
  {Izaac}, \citenamefont {Quesada}, \citenamefont {Gran}, \citenamefont
  {Schuld}, \citenamefont {Swinarton}, \citenamefont {Zabaneh},\ and\
  \citenamefont {Killoran}}]{Bromley:2020applications}%
  \BibitemOpen
  \bibfield  {author} {\bibinfo {author} {\bibfnamefont {T.~R.}\ \bibnamefont
  {Bromley}}, \bibinfo {author} {\bibfnamefont {J.~M.}\ \bibnamefont
  {Arrazola}}, \bibinfo {author} {\bibfnamefont {S.}~\bibnamefont {Jahangiri}},
  \bibinfo {author} {\bibfnamefont {J.}~\bibnamefont {Izaac}}, \bibinfo
  {author} {\bibfnamefont {N.}~\bibnamefont {Quesada}}, \bibinfo {author}
  {\bibfnamefont {A.~D.}\ \bibnamefont {Gran}}, \bibinfo {author}
  {\bibfnamefont {M.}~\bibnamefont {Schuld}}, \bibinfo {author} {\bibfnamefont
  {J.}~\bibnamefont {Swinarton}}, \bibinfo {author} {\bibfnamefont
  {Z.}~\bibnamefont {Zabaneh}},\ and\ \bibinfo {author} {\bibfnamefont
  {N.}~\bibnamefont {Killoran}},\ }\bibfield  {title} {\bibinfo {title}
  {Applications of near-term photonic quantum computers: software and
  algorithms},\ }\href {https://doi.org/10.1088/2058-9565/ab8504} {\bibfield
  {journal} {\bibinfo  {journal} {Quantum Sci. Technol.}\ }\textbf {\bibinfo
  {volume} {5}},\ \bibinfo {pages} {034010} (\bibinfo {year}
  {2020})}\BibitemShut {NoStop}%
\bibitem [{\citenamefont {Bharti}\ \emph {et~al.}(2022)\citenamefont {Bharti},
  \citenamefont {Cervera-Lierta}, \citenamefont {Kyaw}, \citenamefont {Haug},
  \citenamefont {Alperin-Lea}, \citenamefont {Anand}, \citenamefont {Degroote},
  \citenamefont {Heimonen}, \citenamefont {Kottmann}, \citenamefont {Menke},
  \citenamefont {Mok}, \citenamefont {Sim}, \citenamefont {Kwek},\ and\
  \citenamefont {Aspuru-Guzik}}]{Bharti:2022noisy}%
  \BibitemOpen
  \bibfield  {author} {\bibinfo {author} {\bibfnamefont {K.}~\bibnamefont
  {Bharti}}, \bibinfo {author} {\bibfnamefont {A.}~\bibnamefont
  {Cervera-Lierta}}, \bibinfo {author} {\bibfnamefont {T.~H.}\ \bibnamefont
  {Kyaw}}, \bibinfo {author} {\bibfnamefont {T.}~\bibnamefont {Haug}}, \bibinfo
  {author} {\bibfnamefont {S.}~\bibnamefont {Alperin-Lea}}, \bibinfo {author}
  {\bibfnamefont {A.}~\bibnamefont {Anand}}, \bibinfo {author} {\bibfnamefont
  {M.}~\bibnamefont {Degroote}}, \bibinfo {author} {\bibfnamefont
  {H.}~\bibnamefont {Heimonen}}, \bibinfo {author} {\bibfnamefont {J.~S.}\
  \bibnamefont {Kottmann}}, \bibinfo {author} {\bibfnamefont {T.}~\bibnamefont
  {Menke}}, \bibinfo {author} {\bibfnamefont {W.-K.}\ \bibnamefont {Mok}},
  \bibinfo {author} {\bibfnamefont {S.}~\bibnamefont {Sim}}, \bibinfo {author}
  {\bibfnamefont {L.-C.}\ \bibnamefont {Kwek}},\ and\ \bibinfo {author}
  {\bibfnamefont {A.}~\bibnamefont {Aspuru-Guzik}},\ }\bibfield  {title}
  {\bibinfo {title} {Noisy intermediate-scale quantum algorithms},\ }\href
  {https://doi.org/10.1103/RevModPhys.94.015004} {\bibfield  {journal}
  {\bibinfo  {journal} {Rev. Mod. Phys.}\ }\textbf {\bibinfo {volume} {94}},\
  \bibinfo {pages} {015004} (\bibinfo {year} {2022})}\BibitemShut {NoStop}%
\bibitem [{\citenamefont {Finnila}\ \emph {et~al.}(1994)\citenamefont
  {Finnila}, \citenamefont {Gomez}, \citenamefont {Sebenik}, \citenamefont
  {Stenson},\ and\ \citenamefont {Doll}}]{Finnila:1994quantum}%
  \BibitemOpen
  \bibfield  {author} {\bibinfo {author} {\bibfnamefont {A.}~\bibnamefont
  {Finnila}}, \bibinfo {author} {\bibfnamefont {M.}~\bibnamefont {Gomez}},
  \bibinfo {author} {\bibfnamefont {C.}~\bibnamefont {Sebenik}}, \bibinfo
  {author} {\bibfnamefont {C.}~\bibnamefont {Stenson}},\ and\ \bibinfo {author}
  {\bibfnamefont {J.}~\bibnamefont {Doll}},\ }\bibfield  {title} {\bibinfo
  {title} {Quantum annealing: A new method for minimizing multidimensional
  functions},\ }\href
  {https://doi.org/https://doi.org/10.1016/0009-2614(94)00117-0} {\bibfield
  {journal} {\bibinfo  {journal} {Chemical Physics Letters}\ }\textbf {\bibinfo
  {volume} {219}},\ \bibinfo {pages} {343} (\bibinfo {year}
  {1994})}\BibitemShut {NoStop}%
\bibitem [{\citenamefont {Kadowaki}\ and\ \citenamefont
  {Nishimori}(1998)}]{Kadowaki:1998quantum}%
  \BibitemOpen
  \bibfield  {author} {\bibinfo {author} {\bibfnamefont {T.}~\bibnamefont
  {Kadowaki}}\ and\ \bibinfo {author} {\bibfnamefont {H.}~\bibnamefont
  {Nishimori}},\ }\bibfield  {title} {\bibinfo {title} {Quantum annealing in
  the transverse ising model},\ }\href
  {https://doi.org/10.1103/PhysRevE.58.5355} {\bibfield  {journal} {\bibinfo
  {journal} {Phys. Rev. E}\ }\textbf {\bibinfo {volume} {58}},\ \bibinfo
  {pages} {5355} (\bibinfo {year} {1998})}\BibitemShut {NoStop}%
\bibitem [{\citenamefont {Aaronson}\ and\ \citenamefont
  {Arkhipov}(2011)}]{Aaronson:2011computational}%
  \BibitemOpen
  \bibfield  {author} {\bibinfo {author} {\bibfnamefont {S.}~\bibnamefont
  {Aaronson}}\ and\ \bibinfo {author} {\bibfnamefont {A.}~\bibnamefont
  {Arkhipov}},\ }\bibfield  {title} {\bibinfo {title} {The computational
  complexity of linear optics},\ }in\ \href
  {https://doi.org/10.1145/1993636.1993682} {\emph {\bibinfo {booktitle}
  {Proceedings of the Forty-Third Annual ACM Symposium on Theory of
  Computing}}},\ \bibinfo {series and number} {STOC '11}\ (\bibinfo
  {publisher} {Association for Computing Machinery},\ \bibinfo {address} {New
  York, NY, USA},\ \bibinfo {year} {2011})\ p.\ \bibinfo {pages}
  {333–342}\BibitemShut {NoStop}%
\bibitem [{\citenamefont {Aaronson}(2011)}]{Aaronson:2011linear-optical}%
  \BibitemOpen
  \bibfield  {author} {\bibinfo {author} {\bibfnamefont {S.}~\bibnamefont
  {Aaronson}},\ }\bibfield  {title} {\bibinfo {title} {A linear-optical proof
  that the permanent is~\#{P}-hard},\ }\href
  {https://doi.org/10.1098/rspa.2011.0232} {\bibfield  {journal} {\bibinfo
  {journal} {Proceedings of the Royal Society A: Mathematical, Physical and
  Engineering Sciences}\ }\textbf {\bibinfo {volume} {467}},\ \bibinfo {pages}
  {3393} (\bibinfo {year} {2011})}\BibitemShut {NoStop}%
\bibitem [{\citenamefont {Hamilton}\ \emph {et~al.}(2017)\citenamefont
  {Hamilton}, \citenamefont {Kruse}, \citenamefont {Sansoni}, \citenamefont
  {Barkhofen}, \citenamefont {Silberhorn},\ and\ \citenamefont
  {Jex}}]{Hamilton:2017gaussian}%
  \BibitemOpen
  \bibfield  {author} {\bibinfo {author} {\bibfnamefont {C.~S.}\ \bibnamefont
  {Hamilton}}, \bibinfo {author} {\bibfnamefont {R.}~\bibnamefont {Kruse}},
  \bibinfo {author} {\bibfnamefont {L.}~\bibnamefont {Sansoni}}, \bibinfo
  {author} {\bibfnamefont {S.}~\bibnamefont {Barkhofen}}, \bibinfo {author}
  {\bibfnamefont {C.}~\bibnamefont {Silberhorn}},\ and\ \bibinfo {author}
  {\bibfnamefont {I.}~\bibnamefont {Jex}},\ }\bibfield  {title} {\bibinfo
  {title} {Gaussian boson sampling},\ }\href
  {https://doi.org/10.1103/PhysRevLett.119.170501} {\bibfield  {journal}
  {\bibinfo  {journal} {Phys. Rev. Lett.}\ }\textbf {\bibinfo {volume} {119}},\
  \bibinfo {pages} {170501} (\bibinfo {year} {2017})}\BibitemShut {NoStop}%
\bibitem [{\citenamefont {Trabesinger}(2012)}]{Trabesinger:2012quantum}%
  \BibitemOpen
  \bibfield  {author} {\bibinfo {author} {\bibfnamefont {A.}~\bibnamefont
  {Trabesinger}},\ }\bibfield  {title} {\bibinfo {title} {Quantum simulation},\
  }\href {https://doi.org/10.1038/nphys2258} {\bibfield  {journal} {\bibinfo
  {journal} {Nature Physics}\ }\textbf {\bibinfo {volume} {8}},\ \bibinfo
  {pages} {263} (\bibinfo {year} {2012})}\BibitemShut {NoStop}%
\bibitem [{\citenamefont {Georgescu}\ \emph {et~al.}(2014)\citenamefont
  {Georgescu}, \citenamefont {Ashhab},\ and\ \citenamefont
  {Nori}}]{Georgescu:2014quantum}%
  \BibitemOpen
  \bibfield  {author} {\bibinfo {author} {\bibfnamefont {I.~M.}\ \bibnamefont
  {Georgescu}}, \bibinfo {author} {\bibfnamefont {S.}~\bibnamefont {Ashhab}},\
  and\ \bibinfo {author} {\bibfnamefont {F.}~\bibnamefont {Nori}},\ }\bibfield
  {title} {\bibinfo {title} {Quantum simulation},\ }\href
  {https://doi.org/10.1103/RevModPhys.86.153} {\bibfield  {journal} {\bibinfo
  {journal} {Rev. Mod. Phys.}\ }\textbf {\bibinfo {volume} {86}},\ \bibinfo
  {pages} {153} (\bibinfo {year} {2014})}\BibitemShut {NoStop}%
\bibitem [{\citenamefont {Biamonte}(2021)}]{Biamonte:2021universal}%
  \BibitemOpen
  \bibfield  {author} {\bibinfo {author} {\bibfnamefont {J.}~\bibnamefont
  {Biamonte}},\ }\bibfield  {title} {\bibinfo {title} {Universal variational
  quantum computation},\ }\href {https://doi.org/10.1103/PhysRevA.103.L030401}
  {\bibfield  {journal} {\bibinfo  {journal} {Phys. Rev. A}\ }\textbf {\bibinfo
  {volume} {103}},\ \bibinfo {pages} {L030401} (\bibinfo {year}
  {2021})}\BibitemShut {NoStop}%
\bibitem [{\citenamefont {Cerezo}\ \emph {et~al.}(2021)\citenamefont {Cerezo},
  \citenamefont {Arrasmith}, \citenamefont {Babbush}, \citenamefont {Benjamin},
  \citenamefont {Endo}, \citenamefont {Fujii}, \citenamefont {McClean},
  \citenamefont {Mitarai}, \citenamefont {Yuan}, \citenamefont {Cincio},\ and\
  \citenamefont {Coles}}]{Cerezo:2021variational}%
  \BibitemOpen
  \bibfield  {author} {\bibinfo {author} {\bibfnamefont {M.}~\bibnamefont
  {Cerezo}}, \bibinfo {author} {\bibfnamefont {A.}~\bibnamefont {Arrasmith}},
  \bibinfo {author} {\bibfnamefont {R.}~\bibnamefont {Babbush}}, \bibinfo
  {author} {\bibfnamefont {S.~C.}\ \bibnamefont {Benjamin}}, \bibinfo {author}
  {\bibfnamefont {S.}~\bibnamefont {Endo}}, \bibinfo {author} {\bibfnamefont
  {K.}~\bibnamefont {Fujii}}, \bibinfo {author} {\bibfnamefont {J.~R.}\
  \bibnamefont {McClean}}, \bibinfo {author} {\bibfnamefont {K.}~\bibnamefont
  {Mitarai}}, \bibinfo {author} {\bibfnamefont {X.}~\bibnamefont {Yuan}},
  \bibinfo {author} {\bibfnamefont {L.}~\bibnamefont {Cincio}},\ and\ \bibinfo
  {author} {\bibfnamefont {P.~J.}\ \bibnamefont {Coles}},\ }\bibfield  {title}
  {\bibinfo {title} {Variational quantum algorithms},\ }\href
  {https://doi.org/10.1038/s42254-021-00348-9} {\bibfield  {journal} {\bibinfo
  {journal} {Nature Reviews Physics}\ }\textbf {\bibinfo {volume} {3}},\
  \bibinfo {pages} {625} (\bibinfo {year} {2021})}\BibitemShut {NoStop}%
\bibitem [{\citenamefont {Cao}\ \emph {et~al.}(2019)\citenamefont {Cao},
  \citenamefont {Romero}, \citenamefont {Olson}, \citenamefont {Degroote},
  \citenamefont {Johnson}, \citenamefont {Kieferová}, \citenamefont
  {Kivlichan}, \citenamefont {Menke}, \citenamefont {Peropadre}, \citenamefont
  {Sawaya}, \citenamefont {Sim}, \citenamefont {Veis},\ and\ \citenamefont
  {Aspuru-Guzik}}]{Cao:2019quantum}%
  \BibitemOpen
  \bibfield  {author} {\bibinfo {author} {\bibfnamefont {Y.}~\bibnamefont
  {Cao}}, \bibinfo {author} {\bibfnamefont {J.}~\bibnamefont {Romero}},
  \bibinfo {author} {\bibfnamefont {J.~P.}\ \bibnamefont {Olson}}, \bibinfo
  {author} {\bibfnamefont {M.}~\bibnamefont {Degroote}}, \bibinfo {author}
  {\bibfnamefont {P.~D.}\ \bibnamefont {Johnson}}, \bibinfo {author}
  {\bibfnamefont {M.}~\bibnamefont {Kieferová}}, \bibinfo {author}
  {\bibfnamefont {I.~D.}\ \bibnamefont {Kivlichan}}, \bibinfo {author}
  {\bibfnamefont {T.}~\bibnamefont {Menke}}, \bibinfo {author} {\bibfnamefont
  {B.}~\bibnamefont {Peropadre}}, \bibinfo {author} {\bibfnamefont {N.~P.~D.}\
  \bibnamefont {Sawaya}}, \bibinfo {author} {\bibfnamefont {S.}~\bibnamefont
  {Sim}}, \bibinfo {author} {\bibfnamefont {L.}~\bibnamefont {Veis}},\ and\
  \bibinfo {author} {\bibfnamefont {A.}~\bibnamefont {Aspuru-Guzik}},\
  }\bibfield  {title} {\bibinfo {title} {{Quantum Chemistry in the Age of
  Quantum Computing}},\ }\href {https://doi.org/10.1021/acs.chemrev.8b00803}
  {\bibfield  {journal} {\bibinfo  {journal} {Chemical Reviews}\ }\textbf
  {\bibinfo {volume} {119}},\ \bibinfo {pages} {10856} (\bibinfo {year}
  {2019})},\ \bibinfo {note} {pMID: 31469277}\BibitemShut {NoStop}%
\bibitem [{\citenamefont {Endo}\ \emph {et~al.}(2021)\citenamefont {Endo},
  \citenamefont {Cai}, \citenamefont {Benjamin},\ and\ \citenamefont
  {Yuan}}]{Endo:2021hybrid}%
  \BibitemOpen
  \bibfield  {author} {\bibinfo {author} {\bibfnamefont {S.}~\bibnamefont
  {Endo}}, \bibinfo {author} {\bibfnamefont {Z.}~\bibnamefont {Cai}}, \bibinfo
  {author} {\bibfnamefont {S.~C.}\ \bibnamefont {Benjamin}},\ and\ \bibinfo
  {author} {\bibfnamefont {X.}~\bibnamefont {Yuan}},\ }\bibfield  {title}
  {\bibinfo {title} {Hybrid quantum-classical algorithms and quantum error
  mitigation},\ }\href {https://doi.org/10.7566/jpsj.90.032001} {\bibfield
  {journal} {\bibinfo  {journal} {Journal of the Physical Society of Japan}\
  }\textbf {\bibinfo {volume} {90}},\ \bibinfo {pages} {032001} (\bibinfo
  {year} {2021})}\BibitemShut {NoStop}%
\bibitem [{\citenamefont {McArdle}\ \emph {et~al.}(2020)\citenamefont
  {McArdle}, \citenamefont {Endo}, \citenamefont {Aspuru-Guzik}, \citenamefont
  {Benjamin},\ and\ \citenamefont {Yuan}}]{McArdle:2020quantum}%
  \BibitemOpen
  \bibfield  {author} {\bibinfo {author} {\bibfnamefont {S.}~\bibnamefont
  {McArdle}}, \bibinfo {author} {\bibfnamefont {S.}~\bibnamefont {Endo}},
  \bibinfo {author} {\bibfnamefont {A.}~\bibnamefont {Aspuru-Guzik}}, \bibinfo
  {author} {\bibfnamefont {S.~C.}\ \bibnamefont {Benjamin}},\ and\ \bibinfo
  {author} {\bibfnamefont {X.}~\bibnamefont {Yuan}},\ }\bibfield  {title}
  {\bibinfo {title} {Quantum computational chemistry},\ }\href
  {https://doi.org/10.1103/RevModPhys.92.015003} {\bibfield  {journal}
  {\bibinfo  {journal} {Rev. Mod. Phys.}\ }\textbf {\bibinfo {volume} {92}},\
  \bibinfo {pages} {015003} (\bibinfo {year} {2020})}\BibitemShut {NoStop}%
\bibitem [{\citenamefont {Teo}(2022)}]{Teo:2022optimized}%
  \BibitemOpen
  \bibfield  {author} {\bibinfo {author} {\bibfnamefont {Y.~S.}\ \bibnamefont
  {Teo}},\ }\href {https://doi.org/10.48550/ARXIV.2206.12643} {\bibinfo {title}
  {Optimized gradient and hessian estimators for scalable variational quantum
  algorithms}} (\bibinfo {year} {2022})\BibitemShut {NoStop}%
\bibitem [{\citenamefont {Peruzzo}\ \emph {et~al.}(2014)\citenamefont
  {Peruzzo}, \citenamefont {McClean}, \citenamefont {Shadbolt}, \citenamefont
  {Yung}, \citenamefont {Zhou}, \citenamefont {Love}, \citenamefont
  {Aspuru-Guzik},\ and\ \citenamefont {O'Brien}}]{Peruzzo:2014variational}%
  \BibitemOpen
  \bibfield  {author} {\bibinfo {author} {\bibfnamefont {A.}~\bibnamefont
  {Peruzzo}}, \bibinfo {author} {\bibfnamefont {J.}~\bibnamefont {McClean}},
  \bibinfo {author} {\bibfnamefont {P.}~\bibnamefont {Shadbolt}}, \bibinfo
  {author} {\bibfnamefont {M.-H.}\ \bibnamefont {Yung}}, \bibinfo {author}
  {\bibfnamefont {X.-Q.}\ \bibnamefont {Zhou}}, \bibinfo {author}
  {\bibfnamefont {P.~J.}\ \bibnamefont {Love}}, \bibinfo {author}
  {\bibfnamefont {A.}~\bibnamefont {Aspuru-Guzik}},\ and\ \bibinfo {author}
  {\bibfnamefont {J.~L.}\ \bibnamefont {O'Brien}},\ }\bibfield  {title}
  {\bibinfo {title} {A variational eigenvalue solver on a photonic quantum
  processor},\ }\href {https://doi.org/10.1038/ncomms5213} {\bibfield
  {journal} {\bibinfo  {journal} {Nature Communications}\ }\textbf {\bibinfo
  {volume} {5}},\ \bibinfo {pages} {4213} (\bibinfo {year} {2014})}\BibitemShut
  {NoStop}%
\bibitem [{\citenamefont {Wecker}\ \emph {et~al.}(2015)\citenamefont {Wecker},
  \citenamefont {Hastings},\ and\ \citenamefont
  {Troyer}}]{Wecker:2015progress}%
  \BibitemOpen
  \bibfield  {author} {\bibinfo {author} {\bibfnamefont {D.}~\bibnamefont
  {Wecker}}, \bibinfo {author} {\bibfnamefont {M.~B.}\ \bibnamefont
  {Hastings}},\ and\ \bibinfo {author} {\bibfnamefont {M.}~\bibnamefont
  {Troyer}},\ }\bibfield  {title} {\bibinfo {title} {Progress towards practical
  quantum variational algorithms},\ }\href
  {https://doi.org/10.1103/PhysRevA.92.042303} {\bibfield  {journal} {\bibinfo
  {journal} {Phys. Rev. A}\ }\textbf {\bibinfo {volume} {92}},\ \bibinfo
  {pages} {042303} (\bibinfo {year} {2015})}\BibitemShut {NoStop}%
\bibitem [{\citenamefont {McClean}\ \emph {et~al.}(2016)\citenamefont
  {McClean}, \citenamefont {Romero}, \citenamefont {Babbush},\ and\
  \citenamefont {Aspuru-Guzik}}]{McClean:2016theory}%
  \BibitemOpen
  \bibfield  {author} {\bibinfo {author} {\bibfnamefont {J.~R.}\ \bibnamefont
  {McClean}}, \bibinfo {author} {\bibfnamefont {J.}~\bibnamefont {Romero}},
  \bibinfo {author} {\bibfnamefont {R.}~\bibnamefont {Babbush}},\ and\ \bibinfo
  {author} {\bibfnamefont {A.}~\bibnamefont {Aspuru-Guzik}},\ }\bibfield
  {title} {\bibinfo {title} {The theory of variational hybrid quantum-classical
  algorithms},\ }\href {https://doi.org/10.1088/1367-2630/18/2/023023}
  {\bibfield  {journal} {\bibinfo  {journal} {New J. Phys.}\ }\textbf {\bibinfo
  {volume} {18}},\ \bibinfo {pages} {023023} (\bibinfo {year}
  {2016})}\BibitemShut {NoStop}%
\bibitem [{\citenamefont {Farhi}\ \emph {et~al.}(2014)\citenamefont {Farhi},
  \citenamefont {Goldstone},\ and\ \citenamefont
  {Gutmann}}]{Farhi:2014quantum}%
  \BibitemOpen
  \bibfield  {author} {\bibinfo {author} {\bibfnamefont {E.}~\bibnamefont
  {Farhi}}, \bibinfo {author} {\bibfnamefont {J.}~\bibnamefont {Goldstone}},\
  and\ \bibinfo {author} {\bibfnamefont {S.}~\bibnamefont {Gutmann}},\
  }\href@noop {} {\bibinfo {title} {A quantum approximate optimization
  algorithm}} (\bibinfo {year} {2014}),\ \Eprint
  {https://arxiv.org/abs/arXiv:1411.4028} {arXiv:arXiv:1411.4028 [quant-ph]}
  \BibitemShut {NoStop}%
\bibitem [{\citenamefont {Zhou}\ \emph {et~al.}(2020)\citenamefont {Zhou},
  \citenamefont {Wang}, \citenamefont {Choi}, \citenamefont {Pichler},\ and\
  \citenamefont {Lukin}}]{Zhou:2020quantum}%
  \BibitemOpen
  \bibfield  {author} {\bibinfo {author} {\bibfnamefont {L.}~\bibnamefont
  {Zhou}}, \bibinfo {author} {\bibfnamefont {S.-T.}\ \bibnamefont {Wang}},
  \bibinfo {author} {\bibfnamefont {S.}~\bibnamefont {Choi}}, \bibinfo {author}
  {\bibfnamefont {H.}~\bibnamefont {Pichler}},\ and\ \bibinfo {author}
  {\bibfnamefont {M.~D.}\ \bibnamefont {Lukin}},\ }\bibfield  {title} {\bibinfo
  {title} {Quantum approximate optimization algorithm: Performance, mechanism,
  and implementation on near-term devices},\ }\href
  {https://doi.org/10.1103/PhysRevX.10.021067} {\bibfield  {journal} {\bibinfo
  {journal} {Phys. Rev. X}\ }\textbf {\bibinfo {volume} {10}},\ \bibinfo
  {pages} {021067} (\bibinfo {year} {2020})}\BibitemShut {NoStop}%
\bibitem [{\citenamefont {Schuld}\ \emph {et~al.}(2015)\citenamefont {Schuld},
  \citenamefont {Sinayskiy},\ and\ \citenamefont
  {Petruccione}}]{Schuld:2015introduction}%
  \BibitemOpen
  \bibfield  {author} {\bibinfo {author} {\bibfnamefont {M.}~\bibnamefont
  {Schuld}}, \bibinfo {author} {\bibfnamefont {I.}~\bibnamefont {Sinayskiy}},\
  and\ \bibinfo {author} {\bibfnamefont {F.}~\bibnamefont {Petruccione}},\
  }\bibfield  {title} {\bibinfo {title} {An introduction to quantum machine
  learning},\ }\href {https://doi.org/10.1080/00107514.2014.964942} {\bibfield
  {journal} {\bibinfo  {journal} {Contemporary Physics}\ }\textbf {\bibinfo
  {volume} {56}},\ \bibinfo {pages} {172} (\bibinfo {year} {2015})}\BibitemShut
  {NoStop}%
\bibitem [{\citenamefont {Schuld}\ and\ \citenamefont
  {Killoran}(2019)}]{Schuld:2019quantum}%
  \BibitemOpen
  \bibfield  {author} {\bibinfo {author} {\bibfnamefont {M.}~\bibnamefont
  {Schuld}}\ and\ \bibinfo {author} {\bibfnamefont {N.}~\bibnamefont
  {Killoran}},\ }\bibfield  {title} {\bibinfo {title} {Quantum machine learning
  in feature hilbert spaces},\ }\href
  {https://doi.org/10.1103/PhysRevLett.122.040504} {\bibfield  {journal}
  {\bibinfo  {journal} {Phys. Rev. Lett.}\ }\textbf {\bibinfo {volume} {122}},\
  \bibinfo {pages} {040504} (\bibinfo {year} {2019})}\BibitemShut {NoStop}%
\bibitem [{\citenamefont {Carleo}\ \emph {et~al.}(2019)\citenamefont {Carleo},
  \citenamefont {Cirac}, \citenamefont {Cranmer}, \citenamefont {Daudet},
  \citenamefont {Schuld}, \citenamefont {Tishby}, \citenamefont
  {Vogt-Maranto},\ and\ \citenamefont {Zdeborov\'a}}]{Carleo:2019machine}%
  \BibitemOpen
  \bibfield  {author} {\bibinfo {author} {\bibfnamefont {G.}~\bibnamefont
  {Carleo}}, \bibinfo {author} {\bibfnamefont {I.}~\bibnamefont {Cirac}},
  \bibinfo {author} {\bibfnamefont {K.}~\bibnamefont {Cranmer}}, \bibinfo
  {author} {\bibfnamefont {L.}~\bibnamefont {Daudet}}, \bibinfo {author}
  {\bibfnamefont {M.}~\bibnamefont {Schuld}}, \bibinfo {author} {\bibfnamefont
  {N.}~\bibnamefont {Tishby}}, \bibinfo {author} {\bibfnamefont
  {L.}~\bibnamefont {Vogt-Maranto}},\ and\ \bibinfo {author} {\bibfnamefont
  {L.}~\bibnamefont {Zdeborov\'a}},\ }\bibfield  {title} {\bibinfo {title}
  {Machine learning and the physical sciences},\ }\href
  {https://doi.org/10.1103/RevModPhys.91.045002} {\bibfield  {journal}
  {\bibinfo  {journal} {Rev. Mod. Phys.}\ }\textbf {\bibinfo {volume} {91}},\
  \bibinfo {pages} {045002} (\bibinfo {year} {2019})}\BibitemShut {NoStop}%
\bibitem [{\citenamefont {Date}(2020)}]{date2020quantum}%
  \BibitemOpen
  \bibfield  {author} {\bibinfo {author} {\bibfnamefont {P.}~\bibnamefont
  {Date}},\ }\href@noop {} {\bibinfo {title} {Quantum discriminator for binary
  classification}} (\bibinfo {year} {2020}),\ \Eprint
  {https://arxiv.org/abs/2009.01235} {arXiv:2009.01235 [quant-ph]} \BibitemShut
  {NoStop}%
\bibitem [{\citenamefont {P{\'e}rez-Salinas}\ \emph {et~al.}(2020)\citenamefont
  {P{\'e}rez-Salinas}, \citenamefont {Cervera-Lierta}, \citenamefont
  {Gil-Fuster},\ and\ \citenamefont {Latorre}}]{Perez-Salinas:2020aa}%
  \BibitemOpen
  \bibfield  {author} {\bibinfo {author} {\bibfnamefont {A.}~\bibnamefont
  {P{\'e}rez-Salinas}}, \bibinfo {author} {\bibfnamefont {A.}~\bibnamefont
  {Cervera-Lierta}}, \bibinfo {author} {\bibfnamefont {E.}~\bibnamefont
  {Gil-Fuster}},\ and\ \bibinfo {author} {\bibfnamefont {J.~I.}\ \bibnamefont
  {Latorre}},\ }\bibfield  {title} {\bibinfo {title} {Data re-uploading for a
  universal quantum classifier},\ }\href
  {https://doi.org/10.22331/q-2020-02-06-226} {\bibfield  {journal} {\bibinfo
  {journal} {Quantum}\ }\textbf {\bibinfo {volume} {4}},\ \bibinfo {pages}
  {226} (\bibinfo {year} {2020})}\BibitemShut {NoStop}%
\bibitem [{\citenamefont {Dutta}\ \emph {et~al.}(2021)\citenamefont {Dutta},
  \citenamefont {P{\'e}rez-Salinas}, \citenamefont {Cheng}, \citenamefont
  {Latorre},\ and\ \citenamefont {Mukherjee}}]{dutta2021singlequbit}%
  \BibitemOpen
  \bibfield  {author} {\bibinfo {author} {\bibfnamefont {T.}~\bibnamefont
  {Dutta}}, \bibinfo {author} {\bibfnamefont {A.}~\bibnamefont
  {P{\'e}rez-Salinas}}, \bibinfo {author} {\bibfnamefont {J.~P.~S.}\
  \bibnamefont {Cheng}}, \bibinfo {author} {\bibfnamefont {J.~I.}\ \bibnamefont
  {Latorre}},\ and\ \bibinfo {author} {\bibfnamefont {M.}~\bibnamefont
  {Mukherjee}},\ }\href@noop {} {\bibinfo {title} {Single-qubit universal
  classifier implemented on an ion-trap quantum device}} (\bibinfo {year}
  {2021}),\ \Eprint {https://arxiv.org/abs/2106.14059} {arXiv:2106.14059
  [quant-ph]} \BibitemShut {NoStop}%
\bibitem [{\citenamefont {Goto}\ \emph {et~al.}(2021)\citenamefont {Goto},
  \citenamefont {Tran},\ and\ \citenamefont {Nakajima}}]{Goto:2021universal}%
  \BibitemOpen
  \bibfield  {author} {\bibinfo {author} {\bibfnamefont {T.}~\bibnamefont
  {Goto}}, \bibinfo {author} {\bibfnamefont {Q.~H.}\ \bibnamefont {Tran}},\
  and\ \bibinfo {author} {\bibfnamefont {K.}~\bibnamefont {Nakajima}},\
  }\bibfield  {title} {\bibinfo {title} {Universal approximation property of
  quantum machine learning models in quantum-enhanced feature spaces},\ }\href
  {https://doi.org/10.1103/PhysRevLett.127.090506} {\bibfield  {journal}
  {\bibinfo  {journal} {Phys. Rev. Lett.}\ }\textbf {\bibinfo {volume} {127}},\
  \bibinfo {pages} {090506} (\bibinfo {year} {2021})}\BibitemShut {NoStop}%
\bibitem [{\citenamefont {Shin}\ \emph {et~al.}(2022)\citenamefont {Shin},
  \citenamefont {Teo},\ and\ \citenamefont {Jeong}}]{Shin:2022exponential}%
  \BibitemOpen
  \bibfield  {author} {\bibinfo {author} {\bibfnamefont {S.}~\bibnamefont
  {Shin}}, \bibinfo {author} {\bibfnamefont {Y.~S.}\ \bibnamefont {Teo}},\ and\
  \bibinfo {author} {\bibfnamefont {H.}~\bibnamefont {Jeong}},\ }\href
  {https://doi.org/10.48550/ARXIV.2206.12105} {\bibinfo {title} {Exponential
  data encoding for quantum supervised learning}} (\bibinfo {year}
  {2022})\BibitemShut {NoStop}%
\bibitem [{\citenamefont {Paris}\ and\ \citenamefont
  {\v{R}eh\'a\v{c}ek}(2004)}]{lnp:2004uq}%
  \BibitemOpen
  \bibinfo {editor} {\bibfnamefont {M.~G.~A.}\ \bibnamefont {Paris}}\ and\
  \bibinfo {editor} {\bibfnamefont {J.}~\bibnamefont {\v{R}eh\'a\v{c}ek}},\
  eds.,\ \href@noop {} {\emph {\bibinfo {title} {Quantum State Estimation}}},\
  \bibinfo {series} {Lect. Not. Phys.}, Vol.\ \bibinfo {volume} {649}\
  (\bibinfo  {publisher} {Springer},\ \bibinfo {address} {Berlin},\ \bibinfo
  {year} {2004})\BibitemShut {NoStop}%
\bibitem [{\citenamefont {Teo}(2015)}]{Teo:2015introduction}%
  \BibitemOpen
  \bibfield  {author} {\bibinfo {author} {\bibfnamefont {Y.~S.}\ \bibnamefont
  {Teo}},\ }\href {https://doi.org/10.1142/9617} {\emph {\bibinfo {title}
  {Introduction to Quantum-State Estimation}}}\ (\bibinfo  {publisher} {WORLD
  SCIENTIFIC},\ \bibinfo {year} {2015})\ \Eprint
  {https://arxiv.org/abs/https://www.worldscientific.com/doi/pdf/10.1142/9617}
  {https://www.worldscientific.com/doi/pdf/10.1142/9617} \BibitemShut {NoStop}%
\bibitem [{\citenamefont {Gross}\ \emph {et~al.}(2010)\citenamefont {Gross},
  \citenamefont {Liu}, \citenamefont {Flammia}, \citenamefont {Becker},\ and\
  \citenamefont {Eisert}}]{Gross:2010quantum}%
  \BibitemOpen
  \bibfield  {author} {\bibinfo {author} {\bibfnamefont {D.}~\bibnamefont
  {Gross}}, \bibinfo {author} {\bibfnamefont {Y.-K.}\ \bibnamefont {Liu}},
  \bibinfo {author} {\bibfnamefont {S.~T.}\ \bibnamefont {Flammia}}, \bibinfo
  {author} {\bibfnamefont {S.}~\bibnamefont {Becker}},\ and\ \bibinfo {author}
  {\bibfnamefont {J.}~\bibnamefont {Eisert}},\ }\bibfield  {title} {\bibinfo
  {title} {Quantum state tomography via compressed sensing},\ }\href
  {https://doi.org/10.1103/PhysRevLett.105.150401} {\bibfield  {journal}
  {\bibinfo  {journal} {Phys. Rev. Lett.}\ }\textbf {\bibinfo {volume} {105}},\
  \bibinfo {pages} {150401} (\bibinfo {year} {2010})}\BibitemShut {NoStop}%
\bibitem [{\citenamefont {Ahn}\ \emph {et~al.}(2019)\citenamefont {Ahn},
  \citenamefont {Teo}, \citenamefont {Jeong}, \citenamefont {Bouchard},
  \citenamefont {Hufnagel}, \citenamefont {Karimi}, \citenamefont {Koutn\'y},
  \citenamefont {\ifmmode \check{R}\else \v{R}\fi{}eh\'a\ifmmode~\check{c}\else
  \v{c}\fi{}ek}, \citenamefont {Hradil}, \citenamefont {Leuchs},\ and\
  \citenamefont {S\'anchez-Soto}}]{Ahn:2019adaptive}%
  \BibitemOpen
  \bibfield  {author} {\bibinfo {author} {\bibfnamefont {D.}~\bibnamefont
  {Ahn}}, \bibinfo {author} {\bibfnamefont {Y.~S.}\ \bibnamefont {Teo}},
  \bibinfo {author} {\bibfnamefont {H.}~\bibnamefont {Jeong}}, \bibinfo
  {author} {\bibfnamefont {F.}~\bibnamefont {Bouchard}}, \bibinfo {author}
  {\bibfnamefont {F.}~\bibnamefont {Hufnagel}}, \bibinfo {author}
  {\bibfnamefont {E.}~\bibnamefont {Karimi}}, \bibinfo {author} {\bibfnamefont
  {D.}~\bibnamefont {Koutn\'y}}, \bibinfo {author} {\bibfnamefont
  {J.}~\bibnamefont {\ifmmode \check{R}\else
  \v{R}\fi{}eh\'a\ifmmode~\check{c}\else \v{c}\fi{}ek}}, \bibinfo {author}
  {\bibfnamefont {Z.}~\bibnamefont {Hradil}}, \bibinfo {author} {\bibfnamefont
  {G.}~\bibnamefont {Leuchs}},\ and\ \bibinfo {author} {\bibfnamefont {L.~L.}\
  \bibnamefont {S\'anchez-Soto}},\ }\bibfield  {title} {\bibinfo {title}
  {Adaptive compressive tomography with no a priori information},\ }\href
  {https://doi.org/10.1103/PhysRevLett.122.100404} {\bibfield  {journal}
  {\bibinfo  {journal} {Phys. Rev. Lett.}\ }\textbf {\bibinfo {volume} {122}},\
  \bibinfo {pages} {100404} (\bibinfo {year} {2019})}\BibitemShut {NoStop}%
\bibitem [{\citenamefont {Teo}\ and\ \citenamefont
  {S\'{a}nchez-Soto}(2021)}]{Teo:2021modern}%
  \BibitemOpen
  \bibfield  {author} {\bibinfo {author} {\bibfnamefont {Y.~S.}\ \bibnamefont
  {Teo}}\ and\ \bibinfo {author} {\bibfnamefont {L.~L.}\ \bibnamefont
  {S\'{a}nchez-Soto}},\ }\bibfield  {title} {\bibinfo {title} {Modern
  compressive tomography for quantum information science},\ }\href
  {https://doi.org/10.1142/S0219749921400037} {\bibfield  {journal} {\bibinfo
  {journal} {International Journal of Quantum Information}\ }\textbf {\bibinfo
  {volume} {19}},\ \bibinfo {pages} {2140003} (\bibinfo {year} {2021})},\
  \Eprint {https://arxiv.org/abs/https://doi.org/10.1142/S0219749921400037}
  {https://doi.org/10.1142/S0219749921400037} \BibitemShut {NoStop}%
\bibitem [{\citenamefont {Qin}\ \emph {et~al.}(2022)\citenamefont {Qin},
  \citenamefont {Xu},\ and\ \citenamefont {Li}}]{Qin:2022overview}%
  \BibitemOpen
  \bibfield  {author} {\bibinfo {author} {\bibfnamefont {D.}~\bibnamefont
  {Qin}}, \bibinfo {author} {\bibfnamefont {X.}~\bibnamefont {Xu}},\ and\
  \bibinfo {author} {\bibfnamefont {Y.}~\bibnamefont {Li}},\ }\bibfield
  {title} {\bibinfo {title} {An overview of quantum error mitigation
  formulas},\ }\href {https://doi.org/10.1088/1674-1056/ac7b1e} {\bibfield
  {journal} {\bibinfo  {journal} {Chinese Physics B}\ }\textbf {\bibinfo
  {volume} {31}},\ \bibinfo {pages} {090306} (\bibinfo {year}
  {2022})}\BibitemShut {NoStop}%
\bibitem [{\citenamefont {Li}\ and\ \citenamefont
  {Benjamin}(2017)}]{Li:2017efficient}%
  \BibitemOpen
  \bibfield  {author} {\bibinfo {author} {\bibfnamefont {Y.}~\bibnamefont
  {Li}}\ and\ \bibinfo {author} {\bibfnamefont {S.~C.}\ \bibnamefont
  {Benjamin}},\ }\bibfield  {title} {\bibinfo {title} {Efficient variational
  quantum simulator incorporating active error minimization},\ }\href
  {https://doi.org/10.1103/PhysRevX.7.021050} {\bibfield  {journal} {\bibinfo
  {journal} {Phys. Rev. X}\ }\textbf {\bibinfo {volume} {7}},\ \bibinfo {pages}
  {021050} (\bibinfo {year} {2017})}\BibitemShut {NoStop}%
\bibitem [{\citenamefont {Temme}\ \emph {et~al.}(2017)\citenamefont {Temme},
  \citenamefont {Bravyi},\ and\ \citenamefont {Gambetta}}]{Temme:2017error}%
  \BibitemOpen
  \bibfield  {author} {\bibinfo {author} {\bibfnamefont {K.}~\bibnamefont
  {Temme}}, \bibinfo {author} {\bibfnamefont {S.}~\bibnamefont {Bravyi}},\ and\
  \bibinfo {author} {\bibfnamefont {J.~M.}\ \bibnamefont {Gambetta}},\
  }\bibfield  {title} {\bibinfo {title} {Error mitigation for short-depth
  quantum circuits},\ }\href {https://doi.org/10.1103/PhysRevLett.119.180509}
  {\bibfield  {journal} {\bibinfo  {journal} {Phys. Rev. Lett.}\ }\textbf
  {\bibinfo {volume} {119}},\ \bibinfo {pages} {180509} (\bibinfo {year}
  {2017})}\BibitemShut {NoStop}%
\bibitem [{\citenamefont {Giurgica-Tiron}\ \emph {et~al.}(2020)\citenamefont
  {Giurgica-Tiron}, \citenamefont {Hindy}, \citenamefont {LaRose},
  \citenamefont {Mari},\ and\ \citenamefont
  {Zeng}}]{Giurgica-Tiron:2020digital}%
  \BibitemOpen
  \bibfield  {author} {\bibinfo {author} {\bibfnamefont {T.}~\bibnamefont
  {Giurgica-Tiron}}, \bibinfo {author} {\bibfnamefont {Y.}~\bibnamefont
  {Hindy}}, \bibinfo {author} {\bibfnamefont {R.}~\bibnamefont {LaRose}},
  \bibinfo {author} {\bibfnamefont {A.}~\bibnamefont {Mari}},\ and\ \bibinfo
  {author} {\bibfnamefont {W.~J.}\ \bibnamefont {Zeng}},\ }\bibfield  {title}
  {\bibinfo {title} {Digital zero noise extrapolation for quantum error
  mitigation},\ }in\ \href {https://doi.org/10.1109/QCE49297.2020.00045} {\emph
  {\bibinfo {booktitle} {2020 IEEE International Conference on Quantum
  Computing and Engineering (QCE)}}}\ (\bibinfo {year} {2020})\ pp.\ \bibinfo
  {pages} {306--316}\BibitemShut {NoStop}%
\bibitem [{\citenamefont {Greenbaum}(2015)}]{Greenbaum:2015introduction}%
  \BibitemOpen
  \bibfield  {author} {\bibinfo {author} {\bibfnamefont {D.}~\bibnamefont
  {Greenbaum}},\ }\href {https://doi.org/10.48550/ARXIV.1509.02921} {\bibinfo
  {title} {Introduction to quantum gate set tomography}} (\bibinfo {year}
  {2015})\BibitemShut {NoStop}%
\bibitem [{\citenamefont {Song}\ \emph {et~al.}(2019)\citenamefont {Song},
  \citenamefont {Cui}, \citenamefont {Wang}, \citenamefont {Hao}, \citenamefont
  {Feng},\ and\ \citenamefont {Li}}]{Chao:2019quantum}%
  \BibitemOpen
  \bibfield  {author} {\bibinfo {author} {\bibfnamefont {C.}~\bibnamefont
  {Song}}, \bibinfo {author} {\bibfnamefont {J.}~\bibnamefont {Cui}}, \bibinfo
  {author} {\bibfnamefont {H.}~\bibnamefont {Wang}}, \bibinfo {author}
  {\bibfnamefont {J.}~\bibnamefont {Hao}}, \bibinfo {author} {\bibfnamefont
  {H.}~\bibnamefont {Feng}},\ and\ \bibinfo {author} {\bibfnamefont
  {Y.}~\bibnamefont {Li}},\ }\bibfield  {title} {\bibinfo {title} {Quantum
  computation with universal error mitigation on a superconducting quantum
  processor},\ }\href {https://doi.org/10.1126/sciadv.aaw5686} {\bibfield
  {journal} {\bibinfo  {journal} {Science Advances}\ }\textbf {\bibinfo
  {volume} {5}},\ \bibinfo {pages} {eaaw5686} (\bibinfo {year} {2019})},\
  \Eprint
  {https://arxiv.org/abs/https://www.science.org/doi/pdf/10.1126/sciadv.aaw5686}
  {https://www.science.org/doi/pdf/10.1126/sciadv.aaw5686} \BibitemShut
  {NoStop}%
\bibitem [{\citenamefont {Zhang}\ \emph {et~al.}(2020)\citenamefont {Zhang},
  \citenamefont {Lu}, \citenamefont {Zhang}, \citenamefont {Chen},
  \citenamefont {Li}, \citenamefont {Zhang},\ and\ \citenamefont
  {Kim}}]{Zhang:2020error-mitigated}%
  \BibitemOpen
  \bibfield  {author} {\bibinfo {author} {\bibfnamefont {S.}~\bibnamefont
  {Zhang}}, \bibinfo {author} {\bibfnamefont {Y.}~\bibnamefont {Lu}}, \bibinfo
  {author} {\bibfnamefont {K.}~\bibnamefont {Zhang}}, \bibinfo {author}
  {\bibfnamefont {W.}~\bibnamefont {Chen}}, \bibinfo {author} {\bibfnamefont
  {Y.}~\bibnamefont {Li}}, \bibinfo {author} {\bibfnamefont {J.-N.}\
  \bibnamefont {Zhang}},\ and\ \bibinfo {author} {\bibfnamefont
  {K.}~\bibnamefont {Kim}},\ }\bibfield  {title} {\bibinfo {title}
  {Error-mitigated quantum gates exceeding physical fidelities in a trapped-ion
  system},\ }\href {https://doi.org/10.1038/s41467-020-14376-z} {\bibfield
  {journal} {\bibinfo  {journal} {Nature Communications}\ }\textbf {\bibinfo
  {volume} {11}},\ \bibinfo {pages} {587} (\bibinfo {year} {2020})}\BibitemShut
  {NoStop}%
\bibitem [{\citenamefont {Kwon}\ and\ \citenamefont
  {Bae}(2021)}]{Kwon:2021hybrid}%
  \BibitemOpen
  \bibfield  {author} {\bibinfo {author} {\bibfnamefont {H.}~\bibnamefont
  {Kwon}}\ and\ \bibinfo {author} {\bibfnamefont {J.}~\bibnamefont {Bae}},\
  }\bibfield  {title} {\bibinfo {title} {A hybrid quantum-classical approach to
  mitigating measurement errors in quantum algorithms},\ }\href
  {https://doi.org/10.1109/TC.2020.3009664} {\bibfield  {journal} {\bibinfo
  {journal} {IEEE Transactions on Computers}\ }\textbf {\bibinfo {volume}
  {70}},\ \bibinfo {pages} {1401} (\bibinfo {year} {2021})}\BibitemShut
  {NoStop}%
\bibitem [{\citenamefont {McClean}\ \emph {et~al.}(2017)\citenamefont
  {McClean}, \citenamefont {Kimchi-Schwartz}, \citenamefont {Carter},\ and\
  \citenamefont {de~Jong}}]{McClean:2017hybrid}%
  \BibitemOpen
  \bibfield  {author} {\bibinfo {author} {\bibfnamefont {J.~R.}\ \bibnamefont
  {McClean}}, \bibinfo {author} {\bibfnamefont {M.~E.}\ \bibnamefont
  {Kimchi-Schwartz}}, \bibinfo {author} {\bibfnamefont {J.}~\bibnamefont
  {Carter}},\ and\ \bibinfo {author} {\bibfnamefont {W.~A.}\ \bibnamefont
  {de~Jong}},\ }\bibfield  {title} {\bibinfo {title} {Hybrid quantum-classical
  hierarchy for mitigation of decoherence and determination of excited
  states},\ }\href {https://doi.org/10.1103/PhysRevA.95.042308} {\bibfield
  {journal} {\bibinfo  {journal} {Phys. Rev. A}\ }\textbf {\bibinfo {volume}
  {95}},\ \bibinfo {pages} {042308} (\bibinfo {year} {2017})}\BibitemShut
  {NoStop}%
\bibitem [{\citenamefont {McClean}\ \emph {et~al.}(2020)\citenamefont
  {McClean}, \citenamefont {Jiang}, \citenamefont {Rubin}, \citenamefont
  {Babbush},\ and\ \citenamefont {Neven}}]{McClean:2020decoding}%
  \BibitemOpen
  \bibfield  {author} {\bibinfo {author} {\bibfnamefont {J.~R.}\ \bibnamefont
  {McClean}}, \bibinfo {author} {\bibfnamefont {Z.}~\bibnamefont {Jiang}},
  \bibinfo {author} {\bibfnamefont {N.~C.}\ \bibnamefont {Rubin}}, \bibinfo
  {author} {\bibfnamefont {R.}~\bibnamefont {Babbush}},\ and\ \bibinfo {author}
  {\bibfnamefont {H.}~\bibnamefont {Neven}},\ }\bibfield  {title} {\bibinfo
  {title} {Decoding quantum errors with subspace expansions},\ }\href
  {https://doi.org/10.1038/s41467-020-14341-w} {\bibfield  {journal} {\bibinfo
  {journal} {Nature Communications}\ }\textbf {\bibinfo {volume} {11}},\
  \bibinfo {pages} {636} (\bibinfo {year} {2020})}\BibitemShut {NoStop}%
\bibitem [{\citenamefont {Suchsland}\ \emph {et~al.}(2021)\citenamefont
  {Suchsland}, \citenamefont {Tacchino}, \citenamefont {Fischer}, \citenamefont
  {Neupert}, \citenamefont {Barkoutsos},\ and\ \citenamefont
  {Tavernelli}}]{Suchsland:2021algorithmic}%
  \BibitemOpen
  \bibfield  {author} {\bibinfo {author} {\bibfnamefont {P.}~\bibnamefont
  {Suchsland}}, \bibinfo {author} {\bibfnamefont {F.}~\bibnamefont {Tacchino}},
  \bibinfo {author} {\bibfnamefont {M.~H.}\ \bibnamefont {Fischer}}, \bibinfo
  {author} {\bibfnamefont {T.}~\bibnamefont {Neupert}}, \bibinfo {author}
  {\bibfnamefont {P.~K.}\ \bibnamefont {Barkoutsos}},\ and\ \bibinfo {author}
  {\bibfnamefont {I.}~\bibnamefont {Tavernelli}},\ }\bibfield  {title}
  {\bibinfo {title} {Algorithmic error mitigation scheme for current quantum
  processors},\ }\href {https://doi.org/10.22331/q-2021-07-01-492} {\bibfield
  {journal} {\bibinfo  {journal} {Quantum}\ }\textbf {\bibinfo {volume} {5}},\
  \bibinfo {pages} {492} (\bibinfo {year} {2021})}\BibitemShut {NoStop}%
\bibitem [{\citenamefont {Wang}\ \emph {et~al.}(2021)\citenamefont {Wang},
  \citenamefont {Chen},\ and\ \citenamefont {Wang}}]{Wang:2021measurement}%
  \BibitemOpen
  \bibfield  {author} {\bibinfo {author} {\bibfnamefont {K.}~\bibnamefont
  {Wang}}, \bibinfo {author} {\bibfnamefont {Y.-A.}\ \bibnamefont {Chen}},\
  and\ \bibinfo {author} {\bibfnamefont {X.}~\bibnamefont {Wang}},\ }\href
  {https://doi.org/10.48550/ARXIV.2103.13856} {\bibinfo {title} {Measurement
  error mitigation via truncated neumann series}} (\bibinfo {year}
  {2021})\BibitemShut {NoStop}%
\bibitem [{\citenamefont {Koczor}(2021)}]{Koczor:2021exponential}%
  \BibitemOpen
  \bibfield  {author} {\bibinfo {author} {\bibfnamefont {B.}~\bibnamefont
  {Koczor}},\ }\bibfield  {title} {\bibinfo {title} {Exponential error
  suppression for near-term quantum devices},\ }\href
  {https://doi.org/10.1103/PhysRevX.11.031057} {\bibfield  {journal} {\bibinfo
  {journal} {Phys. Rev. X}\ }\textbf {\bibinfo {volume} {11}},\ \bibinfo
  {pages} {031057} (\bibinfo {year} {2021})}\BibitemShut {NoStop}%
\bibitem [{\citenamefont {Huggins}\ \emph {et~al.}(2021)\citenamefont
  {Huggins}, \citenamefont {McArdle}, \citenamefont {O'Brien}, \citenamefont
  {Lee}, \citenamefont {Rubin}, \citenamefont {Boixo}, \citenamefont {Whaley},
  \citenamefont {Babbush},\ and\ \citenamefont
  {McClean}}]{Huggins:2021virtual}%
  \BibitemOpen
  \bibfield  {author} {\bibinfo {author} {\bibfnamefont {W.~J.}\ \bibnamefont
  {Huggins}}, \bibinfo {author} {\bibfnamefont {S.}~\bibnamefont {McArdle}},
  \bibinfo {author} {\bibfnamefont {T.~E.}\ \bibnamefont {O'Brien}}, \bibinfo
  {author} {\bibfnamefont {J.}~\bibnamefont {Lee}}, \bibinfo {author}
  {\bibfnamefont {N.~C.}\ \bibnamefont {Rubin}}, \bibinfo {author}
  {\bibfnamefont {S.}~\bibnamefont {Boixo}}, \bibinfo {author} {\bibfnamefont
  {K.~B.}\ \bibnamefont {Whaley}}, \bibinfo {author} {\bibfnamefont
  {R.}~\bibnamefont {Babbush}},\ and\ \bibinfo {author} {\bibfnamefont {J.~R.}\
  \bibnamefont {McClean}},\ }\bibfield  {title} {\bibinfo {title} {Virtual
  distillation for quantum error mitigation},\ }\href
  {https://doi.org/10.1103/PhysRevX.11.041036} {\bibfield  {journal} {\bibinfo
  {journal} {Phys. Rev. X}\ }\textbf {\bibinfo {volume} {11}},\ \bibinfo
  {pages} {041036} (\bibinfo {year} {2021})}\BibitemShut {NoStop}%
\bibitem [{\citenamefont {Yamamoto}\ \emph {et~al.}(2021)\citenamefont
  {Yamamoto}, \citenamefont {Endo}, \citenamefont {Hakoshima}, \citenamefont
  {Matsuzaki},\ and\ \citenamefont {Tokunaga}}]{Yamamoto:2021error-mitigated}%
  \BibitemOpen
  \bibfield  {author} {\bibinfo {author} {\bibfnamefont {K.}~\bibnamefont
  {Yamamoto}}, \bibinfo {author} {\bibfnamefont {S.}~\bibnamefont {Endo}},
  \bibinfo {author} {\bibfnamefont {H.}~\bibnamefont {Hakoshima}}, \bibinfo
  {author} {\bibfnamefont {Y.}~\bibnamefont {Matsuzaki}},\ and\ \bibinfo
  {author} {\bibfnamefont {Y.}~\bibnamefont {Tokunaga}},\ }\href
  {https://doi.org/10.48550/ARXIV.2112.01850} {\bibinfo {title}
  {Error-mitigated quantum metrology via virtual purification}} (\bibinfo
  {year} {2021})\BibitemShut {NoStop}%
\bibitem [{\citenamefont {Seif}\ \emph {et~al.}(2022)\citenamefont {Seif},
  \citenamefont {Cian}, \citenamefont {Zhou}, \citenamefont {Chen},\ and\
  \citenamefont {Jiang}}]{Seif:2022shadow}%
  \BibitemOpen
  \bibfield  {author} {\bibinfo {author} {\bibfnamefont {A.}~\bibnamefont
  {Seif}}, \bibinfo {author} {\bibfnamefont {Z.-P.}\ \bibnamefont {Cian}},
  \bibinfo {author} {\bibfnamefont {S.}~\bibnamefont {Zhou}}, \bibinfo {author}
  {\bibfnamefont {S.}~\bibnamefont {Chen}},\ and\ \bibinfo {author}
  {\bibfnamefont {L.}~\bibnamefont {Jiang}},\ }\href
  {https://doi.org/10.48550/ARXIV.2203.07309} {\bibinfo {title} {Shadow
  distillation: Quantum error mitigation with classical shadows for near-term
  quantum processors}} (\bibinfo {year} {2022})\BibitemShut {NoStop}%
\bibitem [{\citenamefont {Mogilevtsev}\ and\ \citenamefont
  {Shchesnovich}(2010)}]{Mogilevtsev:2010single-photon}%
  \BibitemOpen
  \bibfield  {author} {\bibinfo {author} {\bibfnamefont {D.}~\bibnamefont
  {Mogilevtsev}}\ and\ \bibinfo {author} {\bibfnamefont {V.~S.}\ \bibnamefont
  {Shchesnovich}},\ }\bibfield  {title} {\bibinfo {title} {Single-photon
  generation by correlated loss in a three-core optical fiber},\ }\href
  {https://doi.org/10.1364/OL.35.003375} {\bibfield  {journal} {\bibinfo
  {journal} {Opt. Lett.}\ }\textbf {\bibinfo {volume} {35}},\ \bibinfo {pages}
  {3375} (\bibinfo {year} {2010})}\BibitemShut {NoStop}%
\bibitem [{\citenamefont {Carpenter}\ \emph {et~al.}(2013)\citenamefont
  {Carpenter}, \citenamefont {Xiong}, \citenamefont {Collins}, \citenamefont
  {Li}, \citenamefont {Krauss}, \citenamefont {Eggleton}, \citenamefont
  {Clark},\ and\ \citenamefont {Schr\"{o}der}}]{Carpenter:2013nonlinear}%
  \BibitemOpen
  \bibfield  {author} {\bibinfo {author} {\bibfnamefont {J.}~\bibnamefont
  {Carpenter}}, \bibinfo {author} {\bibfnamefont {C.}~\bibnamefont {Xiong}},
  \bibinfo {author} {\bibfnamefont {M.~J.}\ \bibnamefont {Collins}}, \bibinfo
  {author} {\bibfnamefont {J.}~\bibnamefont {Li}}, \bibinfo {author}
  {\bibfnamefont {T.~F.}\ \bibnamefont {Krauss}}, \bibinfo {author}
  {\bibfnamefont {B.~J.}\ \bibnamefont {Eggleton}}, \bibinfo {author}
  {\bibfnamefont {A.~S.}\ \bibnamefont {Clark}},\ and\ \bibinfo {author}
  {\bibfnamefont {J.}~\bibnamefont {Schr\"{o}der}},\ }\bibfield  {title}
  {\bibinfo {title} {Mode multiplexed single-photon and classical channels in a
  few-mode fiber},\ }\href {https://doi.org/10.1364/OE.21.028794} {\bibfield
  {journal} {\bibinfo  {journal} {Opt. Express}\ }\textbf {\bibinfo {volume}
  {21}},\ \bibinfo {pages} {28794} (\bibinfo {year} {2013})}\BibitemShut
  {NoStop}%
\bibitem [{\citenamefont {Shomroni}\ \emph {et~al.}(2014)\citenamefont
  {Shomroni}, \citenamefont {Rosenblum}, \citenamefont {Lovsky}, \citenamefont
  {Bechler}, \citenamefont {Guendelman},\ and\ \citenamefont
  {Dayan}}]{Shomroni:2014all-optical}%
  \BibitemOpen
  \bibfield  {author} {\bibinfo {author} {\bibfnamefont {I.}~\bibnamefont
  {Shomroni}}, \bibinfo {author} {\bibfnamefont {S.}~\bibnamefont {Rosenblum}},
  \bibinfo {author} {\bibfnamefont {Y.}~\bibnamefont {Lovsky}}, \bibinfo
  {author} {\bibfnamefont {O.}~\bibnamefont {Bechler}}, \bibinfo {author}
  {\bibfnamefont {G.}~\bibnamefont {Guendelman}},\ and\ \bibinfo {author}
  {\bibfnamefont {B.}~\bibnamefont {Dayan}},\ }\bibfield  {title} {\bibinfo
  {title} {All-optical routing of single photons by a one-atom switch
  controlled by a single photon},\ }\href
  {https://doi.org/10.1126/science.1254699} {\bibfield  {journal} {\bibinfo
  {journal} {Science}\ }\textbf {\bibinfo {volume} {345}},\ \bibinfo {pages}
  {903} (\bibinfo {year} {2014})},\ \Eprint
  {https://arxiv.org/abs/https://www.science.org/doi/pdf/10.1126/science.1254699}
  {https://www.science.org/doi/pdf/10.1126/science.1254699} \BibitemShut
  {NoStop}%
\bibitem [{\citenamefont {Bonneau}\ \emph {et~al.}(2015)\citenamefont
  {Bonneau}, \citenamefont {Mendoza}, \citenamefont {O'Brien},\ and\
  \citenamefont {Thompson}}]{Bonneau:2015on-demand}%
  \BibitemOpen
  \bibfield  {author} {\bibinfo {author} {\bibfnamefont {D.}~\bibnamefont
  {Bonneau}}, \bibinfo {author} {\bibfnamefont {G.~J.}\ \bibnamefont
  {Mendoza}}, \bibinfo {author} {\bibfnamefont {J.~L.}\ \bibnamefont
  {O'Brien}},\ and\ \bibinfo {author} {\bibfnamefont {M.~G.}\ \bibnamefont
  {Thompson}},\ }\bibfield  {title} {\bibinfo {title} {Effect of loss on
  multiplexed single-photon sources},\ }\href
  {https://doi.org/10.1088/1367-2630/17/4/043057} {\bibfield  {journal}
  {\bibinfo  {journal} {New Journal of Physics}\ }\textbf {\bibinfo {volume}
  {17}},\ \bibinfo {pages} {043057} (\bibinfo {year} {2015})}\BibitemShut
  {NoStop}%
\bibitem [{\citenamefont {Mendoza}\ \emph {et~al.}(2016)\citenamefont
  {Mendoza}, \citenamefont {Santagati}, \citenamefont {Munns}, \citenamefont
  {Hemsley}, \citenamefont {Piekarek}, \citenamefont {Mart\'{i}n-L\'{o}pez},
  \citenamefont {Marshall}, \citenamefont {Bonneau}, \citenamefont {Thompson},\
  and\ \citenamefont {O'Brien}}]{Mendoza:2016active}%
  \BibitemOpen
  \bibfield  {author} {\bibinfo {author} {\bibfnamefont {G.~J.}\ \bibnamefont
  {Mendoza}}, \bibinfo {author} {\bibfnamefont {R.}~\bibnamefont {Santagati}},
  \bibinfo {author} {\bibfnamefont {J.}~\bibnamefont {Munns}}, \bibinfo
  {author} {\bibfnamefont {E.}~\bibnamefont {Hemsley}}, \bibinfo {author}
  {\bibfnamefont {M.}~\bibnamefont {Piekarek}}, \bibinfo {author}
  {\bibfnamefont {E.}~\bibnamefont {Mart\'{i}n-L\'{o}pez}}, \bibinfo {author}
  {\bibfnamefont {G.~D.}\ \bibnamefont {Marshall}}, \bibinfo {author}
  {\bibfnamefont {D.}~\bibnamefont {Bonneau}}, \bibinfo {author} {\bibfnamefont
  {M.~G.}\ \bibnamefont {Thompson}},\ and\ \bibinfo {author} {\bibfnamefont
  {J.~L.}\ \bibnamefont {O'Brien}},\ }\bibfield  {title} {\bibinfo {title}
  {Active temporal and spatial multiplexing of photons},\ }\href
  {https://doi.org/10.1364/OPTICA.3.000127} {\bibfield  {journal} {\bibinfo
  {journal} {Optica}\ }\textbf {\bibinfo {volume} {3}},\ \bibinfo {pages} {127}
  (\bibinfo {year} {2016})}\BibitemShut {NoStop}%
\bibitem [{\citenamefont {Jones}\ \emph {et~al.}(2018)\citenamefont {Jones},
  \citenamefont {Kirby},\ and\ \citenamefont
  {Brodsky}}]{Jones:2018PolarizationDL}%
  \BibitemOpen
  \bibfield  {author} {\bibinfo {author} {\bibfnamefont {D.~E.}\ \bibnamefont
  {Jones}}, \bibinfo {author} {\bibfnamefont {B.~T.}\ \bibnamefont {Kirby}},\
  and\ \bibinfo {author} {\bibfnamefont {M.}~\bibnamefont {Brodsky}},\
  }\bibfield  {title} {\bibinfo {title} {Polarization dependent loss in optical
  fibers — does it help or ruin photon entanglement distribution?},\
  }\href@noop {} {\bibfield  {journal} {\bibinfo  {journal} {2018 Optical Fiber
  Communications Conference and Exposition (OFC)}\ ,\ \bibinfo {pages} {1}}
  (\bibinfo {year} {2018})}\BibitemShut {NoStop}%
\bibitem [{\citenamefont {Omkar}\ \emph {et~al.}(2020)\citenamefont {Omkar},
  \citenamefont {Teo},\ and\ \citenamefont
  {Jeong}}]{Omkar:2020resource-efficient}%
  \BibitemOpen
  \bibfield  {author} {\bibinfo {author} {\bibfnamefont {S.}~\bibnamefont
  {Omkar}}, \bibinfo {author} {\bibfnamefont {Y.~S.}\ \bibnamefont {Teo}},\
  and\ \bibinfo {author} {\bibfnamefont {H.}~\bibnamefont {Jeong}},\ }\bibfield
   {title} {\bibinfo {title} {Resource-efficient topological fault-tolerant
  quantum computation with hybrid entanglement of light},\ }\href
  {https://doi.org/10.1103/PhysRevLett.125.060501} {\bibfield  {journal}
  {\bibinfo  {journal} {Phys. Rev. Lett.}\ }\textbf {\bibinfo {volume} {125}},\
  \bibinfo {pages} {060501} (\bibinfo {year} {2020})}\BibitemShut {NoStop}%
\bibitem [{\citenamefont {Bartolucci}\ \emph {et~al.}(2021)\citenamefont
  {Bartolucci}, \citenamefont {Birchall}, \citenamefont {Bonneau},
  \citenamefont {Cable}, \citenamefont {Gimeno-Segovia}, \citenamefont
  {Kieling}, \citenamefont {Nickerson}, \citenamefont {Rudolph},\ and\
  \citenamefont {Sparrow}}]{Bartolucci:2021switch}%
  \BibitemOpen
  \bibfield  {author} {\bibinfo {author} {\bibfnamefont {S.}~\bibnamefont
  {Bartolucci}}, \bibinfo {author} {\bibfnamefont {P.}~\bibnamefont
  {Birchall}}, \bibinfo {author} {\bibfnamefont {D.}~\bibnamefont {Bonneau}},
  \bibinfo {author} {\bibfnamefont {H.}~\bibnamefont {Cable}}, \bibinfo
  {author} {\bibfnamefont {M.}~\bibnamefont {Gimeno-Segovia}}, \bibinfo
  {author} {\bibfnamefont {K.}~\bibnamefont {Kieling}}, \bibinfo {author}
  {\bibfnamefont {N.}~\bibnamefont {Nickerson}}, \bibinfo {author}
  {\bibfnamefont {T.}~\bibnamefont {Rudolph}},\ and\ \bibinfo {author}
  {\bibfnamefont {C.}~\bibnamefont {Sparrow}},\ }\href
  {https://doi.org/10.48550/ARXIV.2109.13760} {\bibinfo {title} {Switch
  networks for photonic fusion-based quantum computing}} (\bibinfo {year}
  {2021})\BibitemShut {NoStop}%
\bibitem [{\citenamefont {Omkar}\ \emph {et~al.}(2022)\citenamefont {Omkar},
  \citenamefont {Lee}, \citenamefont {Teo}, \citenamefont {Lee},\ and\
  \citenamefont {Jeong}}]{Omkar:2022all-photonic}%
  \BibitemOpen
  \bibfield  {author} {\bibinfo {author} {\bibfnamefont {S.}~\bibnamefont
  {Omkar}}, \bibinfo {author} {\bibfnamefont {S.-H.}\ \bibnamefont {Lee}},
  \bibinfo {author} {\bibfnamefont {Y.~S.}\ \bibnamefont {Teo}}, \bibinfo
  {author} {\bibfnamefont {S.-W.}\ \bibnamefont {Lee}},\ and\ \bibinfo {author}
  {\bibfnamefont {H.}~\bibnamefont {Jeong}},\ }\bibfield  {title} {\bibinfo
  {title} {All-photonic architecture for scalable quantum computing with
  greenberger-horne-zeilinger states},\ }\href
  {https://doi.org/10.1103/PRXQuantum.3.030309} {\bibfield  {journal} {\bibinfo
   {journal} {PRX Quantum}\ }\textbf {\bibinfo {volume} {3}},\ \bibinfo {pages}
  {030309} (\bibinfo {year} {2022})}\BibitemShut {NoStop}%
\bibitem [{\citenamefont {Kim}\ \emph {et~al.}(2022)\citenamefont {Kim},
  \citenamefont {Chae}, \citenamefont {Jeong},\ and\ \citenamefont
  {Kim}}]{Kim:2022quantum}%
  \BibitemOpen
  \bibfield  {author} {\bibinfo {author} {\bibfnamefont {J.-H.}\ \bibnamefont
  {Kim}}, \bibinfo {author} {\bibfnamefont {J.-W.}\ \bibnamefont {Chae}},
  \bibinfo {author} {\bibfnamefont {Y.-C.}\ \bibnamefont {Jeong}},\ and\
  \bibinfo {author} {\bibfnamefont {Y.-H.}\ \bibnamefont {Kim}},\ }\bibfield
  {title} {\bibinfo {title} {Quantum communication with time-bin entanglement
  over a wavelength-multiplexed fiber network},\ }\href
  {https://doi.org/10.1063/5.0073040} {\bibfield  {journal} {\bibinfo
  {journal} {APL Photonics}\ }\textbf {\bibinfo {volume} {7}},\ \bibinfo
  {pages} {016106} (\bibinfo {year} {2022})},\ \Eprint
  {https://arxiv.org/abs/https://doi.org/10.1063/5.0073040}
  {https://doi.org/10.1063/5.0073040} \BibitemShut {NoStop}%
\bibitem [{\citenamefont {Dragan}\ and\ \citenamefont
  {W\'odkiewicz}(2005)}]{Dragan:2005depolarization}%
  \BibitemOpen
  \bibfield  {author} {\bibinfo {author} {\bibfnamefont {A.}~\bibnamefont
  {Dragan}}\ and\ \bibinfo {author} {\bibfnamefont {K.}~\bibnamefont
  {W\'odkiewicz}},\ }\bibfield  {title} {\bibinfo {title} {Depolarization
  channels with zero-bandwidth noises},\ }\href
  {https://doi.org/10.1103/PhysRevA.71.012322} {\bibfield  {journal} {\bibinfo
  {journal} {Phys. Rev. A}\ }\textbf {\bibinfo {volume} {71}},\ \bibinfo
  {pages} {012322} (\bibinfo {year} {2005})}\BibitemShut {NoStop}%
\bibitem [{\citenamefont {Bayat}\ \emph {et~al.}(2006)\citenamefont {Bayat},
  \citenamefont {Karimipour},\ and\ \citenamefont
  {Marvian}}]{Bayat:2006threshold}%
  \BibitemOpen
  \bibfield  {author} {\bibinfo {author} {\bibfnamefont {A.}~\bibnamefont
  {Bayat}}, \bibinfo {author} {\bibfnamefont {V.}~\bibnamefont {Karimipour}},\
  and\ \bibinfo {author} {\bibfnamefont {I.}~\bibnamefont {Marvian}},\
  }\bibfield  {title} {\bibinfo {title} {Threshold distances for transmission
  of epr pairs through pauli channels},\ }\href
  {https://doi.org/https://doi.org/10.1016/j.physleta.2006.02.009} {\bibfield
  {journal} {\bibinfo  {journal} {Physics Letters A}\ }\textbf {\bibinfo
  {volume} {355}},\ \bibinfo {pages} {81} (\bibinfo {year} {2006})}\BibitemShut
  {NoStop}%
\bibitem [{\citenamefont {Karpi\'{n}ski}\ \emph {et~al.}(2008)\citenamefont
  {Karpi\'{n}ski}, \citenamefont {Radzewicz},\ and\ \citenamefont
  {Banaszek}}]{Karpinski:2008fiber}%
  \BibitemOpen
  \bibfield  {author} {\bibinfo {author} {\bibfnamefont {M.}~\bibnamefont
  {Karpi\'{n}ski}}, \bibinfo {author} {\bibfnamefont {C.}~\bibnamefont
  {Radzewicz}},\ and\ \bibinfo {author} {\bibfnamefont {K.}~\bibnamefont
  {Banaszek}},\ }\bibfield  {title} {\bibinfo {title} {Fiber-optic realization
  of anisotropic depolarizing quantum channels},\ }\href
  {https://doi.org/10.1364/JOSAB.25.000668} {\bibfield  {journal} {\bibinfo
  {journal} {J. Opt. Soc. Am. B}\ }\textbf {\bibinfo {volume} {25}},\ \bibinfo
  {pages} {668} (\bibinfo {year} {2008})}\BibitemShut {NoStop}%
\bibitem [{\citenamefont {Amaral}\ and\ \citenamefont
  {Tempor{\~a}o}(2019)}]{Amaral:2019characterization}%
  \BibitemOpen
  \bibfield  {author} {\bibinfo {author} {\bibfnamefont {G.~C.}\ \bibnamefont
  {Amaral}}\ and\ \bibinfo {author} {\bibfnamefont {G.~P.}\ \bibnamefont
  {Tempor{\~a}o}},\ }\bibfield  {title} {\bibinfo {title} {Characterization of
  depolarizing channels using two-photon interference},\ }\href
  {https://doi.org/10.1007/s11128-019-2445-9} {\bibfield  {journal} {\bibinfo
  {journal} {Quantum Information Processing}\ }\textbf {\bibinfo {volume}
  {18}},\ \bibinfo {pages} {342} (\bibinfo {year} {2019})}\BibitemShut
  {NoStop}%
\bibitem [{\citenamefont {Harrow}\ and\ \citenamefont
  {Low}(2009)}]{harrow_random_2009}%
  \BibitemOpen
  \bibfield  {author} {\bibinfo {author} {\bibfnamefont {A.~W.}\ \bibnamefont
  {Harrow}}\ and\ \bibinfo {author} {\bibfnamefont {R.~A.}\ \bibnamefont
  {Low}},\ }\bibfield  {title} {\bibinfo {title} {Random {Quantum} {Circuits}
  are {Approximate} 2-designs},\ }\href
  {https://doi.org/10.1007/s00220-009-0873-6} {\bibfield  {journal} {\bibinfo
  {journal} {Communications in Mathematical Physics}\ }\textbf {\bibinfo
  {volume} {291}},\ \bibinfo {pages} {257} (\bibinfo {year}
  {2009})}\BibitemShut {NoStop}%
\bibitem [{\citenamefont {von Mises}\ and\ \citenamefont
  {Pollaczek-Geiringer}(1929)}]{vonMises:1929praktische}%
  \BibitemOpen
  \bibfield  {author} {\bibinfo {author} {\bibfnamefont {R.}~\bibnamefont {von
  Mises}}\ and\ \bibinfo {author} {\bibfnamefont {H.}~\bibnamefont
  {Pollaczek-Geiringer}},\ }\bibfield  {title} {\bibinfo {title} {Praktische
  verfahren der gleichungsaufl{\"o}sung},\ }\href@noop {} {\bibfield  {journal}
  {\bibinfo  {journal} {Zeitschrift f{\"u}r Angewandte Mathematik und
  Mechanik}\ }\textbf {\bibinfo {volume} {9}},\ \bibinfo {pages} {152}
  (\bibinfo {year} {1929})}\BibitemShut {NoStop}%
\bibitem [{\citenamefont {Lowe}\ \emph {et~al.}(2021)\citenamefont {Lowe},
  \citenamefont {Gordon}, \citenamefont {Czarnik}, \citenamefont {Arrasmith},
  \citenamefont {Coles},\ and\ \citenamefont {Cincio}}]{Lowe:2021unified}%
  \BibitemOpen
  \bibfield  {author} {\bibinfo {author} {\bibfnamefont {A.}~\bibnamefont
  {Lowe}}, \bibinfo {author} {\bibfnamefont {M.~H.}\ \bibnamefont {Gordon}},
  \bibinfo {author} {\bibfnamefont {P.}~\bibnamefont {Czarnik}}, \bibinfo
  {author} {\bibfnamefont {A.}~\bibnamefont {Arrasmith}}, \bibinfo {author}
  {\bibfnamefont {P.~J.}\ \bibnamefont {Coles}},\ and\ \bibinfo {author}
  {\bibfnamefont {L.}~\bibnamefont {Cincio}},\ }\bibfield  {title} {\bibinfo
  {title} {Unified approach to data-driven quantum error mitigation},\ }\href
  {https://doi.org/10.1103/PhysRevResearch.3.033098} {\bibfield  {journal}
  {\bibinfo  {journal} {Phys. Rev. Research}\ }\textbf {\bibinfo {volume}
  {3}},\ \bibinfo {pages} {033098} (\bibinfo {year} {2021})}\BibitemShut
  {NoStop}%
\bibitem [{\citenamefont {Cotler}\ \emph {et~al.}(2019)\citenamefont {Cotler},
  \citenamefont {Choi}, \citenamefont {Lukin}, \citenamefont {Gharibyan},
  \citenamefont {Grover}, \citenamefont {Tai}, \citenamefont {Rispoli},
  \citenamefont {Schittko}, \citenamefont {Preiss}, \citenamefont {Kaufman},
  \citenamefont {Greiner}, \citenamefont {Pichler},\ and\ \citenamefont
  {Hayden}}]{Cotler:2019quantum}%
  \BibitemOpen
  \bibfield  {author} {\bibinfo {author} {\bibfnamefont {J.}~\bibnamefont
  {Cotler}}, \bibinfo {author} {\bibfnamefont {S.}~\bibnamefont {Choi}},
  \bibinfo {author} {\bibfnamefont {A.}~\bibnamefont {Lukin}}, \bibinfo
  {author} {\bibfnamefont {H.}~\bibnamefont {Gharibyan}}, \bibinfo {author}
  {\bibfnamefont {T.}~\bibnamefont {Grover}}, \bibinfo {author} {\bibfnamefont
  {M.~E.}\ \bibnamefont {Tai}}, \bibinfo {author} {\bibfnamefont
  {M.}~\bibnamefont {Rispoli}}, \bibinfo {author} {\bibfnamefont
  {R.}~\bibnamefont {Schittko}}, \bibinfo {author} {\bibfnamefont {P.~M.}\
  \bibnamefont {Preiss}}, \bibinfo {author} {\bibfnamefont {A.~M.}\
  \bibnamefont {Kaufman}}, \bibinfo {author} {\bibfnamefont {M.}~\bibnamefont
  {Greiner}}, \bibinfo {author} {\bibfnamefont {H.}~\bibnamefont {Pichler}},\
  and\ \bibinfo {author} {\bibfnamefont {P.}~\bibnamefont {Hayden}},\
  }\bibfield  {title} {\bibinfo {title} {Quantum virtual cooling},\ }\href
  {https://doi.org/10.1103/PhysRevX.9.031013} {\bibfield  {journal} {\bibinfo
  {journal} {Phys. Rev. X}\ }\textbf {\bibinfo {volume} {9}},\ \bibinfo {pages}
  {031013} (\bibinfo {year} {2019})}\BibitemShut {NoStop}%
\bibitem [{\citenamefont {Cai}(2021)}]{Cai:2021resource-efficient}%
  \BibitemOpen
  \bibfield  {author} {\bibinfo {author} {\bibfnamefont {Z.}~\bibnamefont
  {Cai}},\ }\href {https://doi.org/10.48550/ARXIV.2107.07279} {\bibinfo {title}
  {Resource-efficient purification-based quantum error mitigation}} (\bibinfo
  {year} {2021})\BibitemShut {NoStop}%
\bibitem [{\citenamefont {Huo}\ and\ \citenamefont
  {Li}(2022)}]{Huo:2022dual-state}%
  \BibitemOpen
  \bibfield  {author} {\bibinfo {author} {\bibfnamefont {M.}~\bibnamefont
  {Huo}}\ and\ \bibinfo {author} {\bibfnamefont {Y.}~\bibnamefont {Li}},\
  }\bibfield  {title} {\bibinfo {title} {Dual-state purification for practical
  quantum error mitigation},\ }\href
  {https://doi.org/10.1103/PhysRevA.105.022427} {\bibfield  {journal} {\bibinfo
   {journal} {Phys. Rev. A}\ }\textbf {\bibinfo {volume} {105}},\ \bibinfo
  {pages} {022427} (\bibinfo {year} {2022})}\BibitemShut {NoStop}%
\bibitem [{\citenamefont {Czarnik}\ \emph {et~al.}(2021)\citenamefont
  {Czarnik}, \citenamefont {Arrasmith}, \citenamefont {Cincio},\ and\
  \citenamefont {Coles}}]{Czarnik:2021qubit-efficient}%
  \BibitemOpen
  \bibfield  {author} {\bibinfo {author} {\bibfnamefont {P.}~\bibnamefont
  {Czarnik}}, \bibinfo {author} {\bibfnamefont {A.}~\bibnamefont {Arrasmith}},
  \bibinfo {author} {\bibfnamefont {L.}~\bibnamefont {Cincio}},\ and\ \bibinfo
  {author} {\bibfnamefont {P.~J.}\ \bibnamefont {Coles}},\ }\href
  {https://doi.org/10.48550/ARXIV.2102.06056} {\bibinfo {title}
  {Qubit-efficient exponential suppression of errors}} (\bibinfo {year}
  {2021})\BibitemShut {NoStop}%
\bibitem [{\citenamefont {Xiong}\ \emph {et~al.}(2022)\citenamefont {Xiong},
  \citenamefont {Ng},\ and\ \citenamefont {Hanzo}}]{Xiong:2022quantum}%
  \BibitemOpen
  \bibfield  {author} {\bibinfo {author} {\bibfnamefont {Y.}~\bibnamefont
  {Xiong}}, \bibinfo {author} {\bibfnamefont {S.~X.}\ \bibnamefont {Ng}},\ and\
  \bibinfo {author} {\bibfnamefont {L.}~\bibnamefont {Hanzo}},\ }\bibfield
  {title} {\bibinfo {title} {Quantum error mitigation relying on permutation
  filtering},\ }\href {https://doi.org/10.1109/TCOMM.2021.3132914} {\bibfield
  {journal} {\bibinfo  {journal} {IEEE Transactions on Communications}\
  }\textbf {\bibinfo {volume} {70}},\ \bibinfo {pages} {1927} (\bibinfo {year}
  {2022})}\BibitemShut {NoStop}%
\bibitem [{\citenamefont {Aaronson}(2017)}]{Aaronson:2017shadow}%
  \BibitemOpen
  \bibfield  {author} {\bibinfo {author} {\bibfnamefont {S.}~\bibnamefont
  {Aaronson}},\ }\href {https://doi.org/10.48550/ARXIV.1711.01053} {\bibinfo
  {title} {Shadow tomography of quantum states}} (\bibinfo {year}
  {2017})\BibitemShut {NoStop}%
\bibitem [{\citenamefont {Huang}\ \emph {et~al.}(2020)\citenamefont {Huang},
  \citenamefont {Kueng},\ and\ \citenamefont
  {Preskill}}]{Huang:2020predicting}%
  \BibitemOpen
  \bibfield  {author} {\bibinfo {author} {\bibfnamefont {H.-Y.}\ \bibnamefont
  {Huang}}, \bibinfo {author} {\bibfnamefont {R.}~\bibnamefont {Kueng}},\ and\
  \bibinfo {author} {\bibfnamefont {J.}~\bibnamefont {Preskill}},\ }\bibfield
  {title} {\bibinfo {title} {Predicting many properties of a quantum system
  from very few measurements},\ }\href
  {https://doi.org/10.1038/s41567-020-0932-7} {\bibfield  {journal} {\bibinfo
  {journal} {Nature Physics}\ }\textbf {\bibinfo {volume} {16}},\ \bibinfo
  {pages} {1050} (\bibinfo {year} {2020})}\BibitemShut {NoStop}%
\bibitem [{\citenamefont {Paini}\ \emph {et~al.}(2021)\citenamefont {Paini},
  \citenamefont {Kalev}, \citenamefont {Padilha},\ and\ \citenamefont
  {Ruck}}]{Paini:2021estimating}%
  \BibitemOpen
  \bibfield  {author} {\bibinfo {author} {\bibfnamefont {M.}~\bibnamefont
  {Paini}}, \bibinfo {author} {\bibfnamefont {A.}~\bibnamefont {Kalev}},
  \bibinfo {author} {\bibfnamefont {D.}~\bibnamefont {Padilha}},\ and\ \bibinfo
  {author} {\bibfnamefont {B.}~\bibnamefont {Ruck}},\ }\bibfield  {title}
  {\bibinfo {title} {Estimating expectation values using approximate quantum
  states},\ }\href {https://doi.org/10.22331/q-2021-03-16-413} {\bibfield
  {journal} {\bibinfo  {journal} {Quantum}\ }\textbf {\bibinfo {volume} {5}},\
  \bibinfo {pages} {413} (\bibinfo {year} {2021})}\BibitemShut {NoStop}%
\bibitem [{\citenamefont {Chen}\ \emph {et~al.}(2021)\citenamefont {Chen},
  \citenamefont {Yu}, \citenamefont {Zeng},\ and\ \citenamefont
  {Flammia}}]{Chen:2021robust}%
  \BibitemOpen
  \bibfield  {author} {\bibinfo {author} {\bibfnamefont {S.}~\bibnamefont
  {Chen}}, \bibinfo {author} {\bibfnamefont {W.}~\bibnamefont {Yu}}, \bibinfo
  {author} {\bibfnamefont {P.}~\bibnamefont {Zeng}},\ and\ \bibinfo {author}
  {\bibfnamefont {S.~T.}\ \bibnamefont {Flammia}},\ }\bibfield  {title}
  {\bibinfo {title} {Robust shadow estimation},\ }\href
  {https://doi.org/10.1103/PRXQuantum.2.030348} {\bibfield  {journal} {\bibinfo
   {journal} {PRX Quantum}\ }\textbf {\bibinfo {volume} {2}},\ \bibinfo {pages}
  {030348} (\bibinfo {year} {2021})}\BibitemShut {NoStop}%
\bibitem [{\citenamefont {Phoenix}(1990)}]{Phoenix:1990wave-packet}%
  \BibitemOpen
  \bibfield  {author} {\bibinfo {author} {\bibfnamefont {S.~J.~D.}\
  \bibnamefont {Phoenix}},\ }\bibfield  {title} {\bibinfo {title} {Wave-packet
  evolution in the damped oscillator},\ }\href
  {https://doi.org/10.1103/PhysRevA.41.5132} {\bibfield  {journal} {\bibinfo
  {journal} {Phys. Rev. A}\ }\textbf {\bibinfo {volume} {41}},\ \bibinfo
  {pages} {5132} (\bibinfo {year} {1990})}\BibitemShut {NoStop}%
\bibitem [{\citenamefont {Fujiwara}\ and\ \citenamefont
  {Imai}(2003)}]{Fujiwara:2003quantum}%
  \BibitemOpen
  \bibfield  {author} {\bibinfo {author} {\bibfnamefont {A.}~\bibnamefont
  {Fujiwara}}\ and\ \bibinfo {author} {\bibfnamefont {H.}~\bibnamefont
  {Imai}},\ }\bibfield  {title} {\bibinfo {title} {Quantum parameter estimation
  of a generalized pauli channel},\ }\href
  {https://doi.org/10.1088/0305-4470/36/29/314} {\bibfield  {journal} {\bibinfo
   {journal} {Journal of Physics A: Mathematical and General}\ }\textbf
  {\bibinfo {volume} {36}},\ \bibinfo {pages} {8093} (\bibinfo {year}
  {2003})}\BibitemShut {NoStop}%
\bibitem [{\citenamefont {Chiuri}\ \emph {et~al.}(2011)\citenamefont {Chiuri},
  \citenamefont {Rosati}, \citenamefont {Vallone}, \citenamefont {P\'adua},
  \citenamefont {Imai}, \citenamefont {Giacomini}, \citenamefont
  {Macchiavello},\ and\ \citenamefont {Mataloni}}]{Chiuri:2011experimental}%
  \BibitemOpen
  \bibfield  {author} {\bibinfo {author} {\bibfnamefont {A.}~\bibnamefont
  {Chiuri}}, \bibinfo {author} {\bibfnamefont {V.}~\bibnamefont {Rosati}},
  \bibinfo {author} {\bibfnamefont {G.}~\bibnamefont {Vallone}}, \bibinfo
  {author} {\bibfnamefont {S.}~\bibnamefont {P\'adua}}, \bibinfo {author}
  {\bibfnamefont {H.}~\bibnamefont {Imai}}, \bibinfo {author} {\bibfnamefont
  {S.}~\bibnamefont {Giacomini}}, \bibinfo {author} {\bibfnamefont
  {C.}~\bibnamefont {Macchiavello}},\ and\ \bibinfo {author} {\bibfnamefont
  {P.}~\bibnamefont {Mataloni}},\ }\bibfield  {title} {\bibinfo {title}
  {Experimental realization of optimal noise estimation for a general pauli
  channel},\ }\href {https://doi.org/10.1103/PhysRevLett.107.253602} {\bibfield
   {journal} {\bibinfo  {journal} {Phys. Rev. Lett.}\ }\textbf {\bibinfo
  {volume} {107}},\ \bibinfo {pages} {253602} (\bibinfo {year}
  {2011})}\BibitemShut {NoStop}%
\bibitem [{\citenamefont {Omkar}\ \emph {et~al.}(2013)\citenamefont {Omkar},
  \citenamefont {Srikanth},\ and\ \citenamefont
  {Banerjee}}]{Omkar:2013dissipative}%
  \BibitemOpen
  \bibfield  {author} {\bibinfo {author} {\bibfnamefont {S.}~\bibnamefont
  {Omkar}}, \bibinfo {author} {\bibfnamefont {R.}~\bibnamefont {Srikanth}},\
  and\ \bibinfo {author} {\bibfnamefont {S.}~\bibnamefont {Banerjee}},\
  }\bibfield  {title} {\bibinfo {title} {Dissipative and non-dissipative
  single-qubit channels: dynamics and geometry},\ }\href
  {https://doi.org/10.1007/s11128-013-0628-3} {\bibfield  {journal} {\bibinfo
  {journal} {Quantum Information Processing}\ }\textbf {\bibinfo {volume}
  {12}},\ \bibinfo {pages} {3725} (\bibinfo {year} {2013})}\BibitemShut
  {NoStop}%
\bibitem [{\citenamefont {Flammia}\ and\ \citenamefont
  {Wallman}(2020)}]{Flammia:2020efficient}%
  \BibitemOpen
  \bibfield  {author} {\bibinfo {author} {\bibfnamefont {S.~T.}\ \bibnamefont
  {Flammia}}\ and\ \bibinfo {author} {\bibfnamefont {J.~J.}\ \bibnamefont
  {Wallman}},\ }\bibfield  {title} {\bibinfo {title} {Efficient estimation of
  pauli channels},\ }\bibfield  {journal} {\bibinfo  {journal} {ACM
  Transactions on Quantum Computing}\ }\textbf {\bibinfo {volume} {1}},\ \href
  {https://doi.org/10.1145/3408039} {10.1145/3408039} (\bibinfo {year}
  {2020})\BibitemShut {NoStop}%
\bibitem [{\citenamefont {Terhal}(2015)}]{Terhal:2015quantum}%
  \BibitemOpen
  \bibfield  {author} {\bibinfo {author} {\bibfnamefont {B.~M.}\ \bibnamefont
  {Terhal}},\ }\bibfield  {title} {\bibinfo {title} {Quantum error correction
  for quantum memories},\ }\href {https://doi.org/10.1103/RevModPhys.87.307}
  {\bibfield  {journal} {\bibinfo  {journal} {Rev. Mod. Phys.}\ }\textbf
  {\bibinfo {volume} {87}},\ \bibinfo {pages} {307} (\bibinfo {year}
  {2015})}\BibitemShut {NoStop}%
\bibitem [{\citenamefont {Wallman}\ and\ \citenamefont
  {Emerson}(2016)}]{Wallman:2016noise}%
  \BibitemOpen
  \bibfield  {author} {\bibinfo {author} {\bibfnamefont {J.~J.}\ \bibnamefont
  {Wallman}}\ and\ \bibinfo {author} {\bibfnamefont {J.}~\bibnamefont
  {Emerson}},\ }\bibfield  {title} {\bibinfo {title} {Noise tailoring for
  scalable quantum computation via randomized compiling},\ }\href
  {https://doi.org/10.1103/PhysRevA.94.052325} {\bibfield  {journal} {\bibinfo
  {journal} {Phys. Rev. A}\ }\textbf {\bibinfo {volume} {94}},\ \bibinfo
  {pages} {052325} (\bibinfo {year} {2016})}\BibitemShut {NoStop}%
\bibitem [{\citenamefont {Sanders}\ \emph {et~al.}(2015)\citenamefont
  {Sanders}, \citenamefont {Wallman},\ and\ \citenamefont
  {Sanders}}]{Sanders:2015bounding}%
  \BibitemOpen
  \bibfield  {author} {\bibinfo {author} {\bibfnamefont {Y.~R.}\ \bibnamefont
  {Sanders}}, \bibinfo {author} {\bibfnamefont {J.~J.}\ \bibnamefont
  {Wallman}},\ and\ \bibinfo {author} {\bibfnamefont {B.~C.}\ \bibnamefont
  {Sanders}},\ }\bibfield  {title} {\bibinfo {title} {Bounding quantum gate
  error rate based on reported average fidelity},\ }\href
  {https://doi.org/10.1088/1367-2630/18/1/012002} {\bibfield  {journal}
  {\bibinfo  {journal} {New Journal of Physics}\ }\textbf {\bibinfo {volume}
  {18}},\ \bibinfo {pages} {012002} (\bibinfo {year} {2015})}\BibitemShut
  {NoStop}%
\bibitem [{\citenamefont {Kueng}\ \emph {et~al.}(2016)\citenamefont {Kueng},
  \citenamefont {Long}, \citenamefont {Doherty},\ and\ \citenamefont
  {Flammia}}]{Kueng:2016comparing}%
  \BibitemOpen
  \bibfield  {author} {\bibinfo {author} {\bibfnamefont {R.}~\bibnamefont
  {Kueng}}, \bibinfo {author} {\bibfnamefont {D.~M.}\ \bibnamefont {Long}},
  \bibinfo {author} {\bibfnamefont {A.~C.}\ \bibnamefont {Doherty}},\ and\
  \bibinfo {author} {\bibfnamefont {S.~T.}\ \bibnamefont {Flammia}},\
  }\bibfield  {title} {\bibinfo {title} {Comparing experiments to the
  fault-tolerance threshold},\ }\href
  {https://doi.org/10.1103/PhysRevLett.117.170502} {\bibfield  {journal}
  {\bibinfo  {journal} {Phys. Rev. Lett.}\ }\textbf {\bibinfo {volume} {117}},\
  \bibinfo {pages} {170502} (\bibinfo {year} {2016})}\BibitemShut {NoStop}%
\bibitem [{\citenamefont {Huang}\ \emph {et~al.}(2019)\citenamefont {Huang},
  \citenamefont {Doherty},\ and\ \citenamefont
  {Flammia}}]{Huang:2019performance}%
  \BibitemOpen
  \bibfield  {author} {\bibinfo {author} {\bibfnamefont {E.}~\bibnamefont
  {Huang}}, \bibinfo {author} {\bibfnamefont {A.~C.}\ \bibnamefont {Doherty}},\
  and\ \bibinfo {author} {\bibfnamefont {S.}~\bibnamefont {Flammia}},\
  }\bibfield  {title} {\bibinfo {title} {Performance of quantum error
  correction with coherent errors},\ }\href
  {https://doi.org/10.1103/PhysRevA.99.022313} {\bibfield  {journal} {\bibinfo
  {journal} {Phys. Rev. A}\ }\textbf {\bibinfo {volume} {99}},\ \bibinfo
  {pages} {022313} (\bibinfo {year} {2019})}\BibitemShut {NoStop}%
\bibitem [{\citenamefont {Siudzi{\'{n}}ska}(2020)}]{Siudziska:2020classical}%
  \BibitemOpen
  \bibfield  {author} {\bibinfo {author} {\bibfnamefont {K.}~\bibnamefont
  {Siudzi{\'{n}}ska}},\ }\bibfield  {title} {\bibinfo {title} {Classical
  capacity of generalized pauli channels},\ }\href
  {https://doi.org/10.1088/1751-8121/abb276} {\bibfield  {journal} {\bibinfo
  {journal} {Journal of Physics A: Mathematical and Theoretical}\ }\textbf
  {\bibinfo {volume} {53}},\ \bibinfo {pages} {445301} (\bibinfo {year}
  {2020})}\BibitemShut {NoStop}%
\bibitem [{\citenamefont {Chen}\ \emph {et~al.}(2022)\citenamefont {Chen},
  \citenamefont {Zhou}, \citenamefont {Seif},\ and\ \citenamefont
  {Jiang}}]{Chen:2022quantum}%
  \BibitemOpen
  \bibfield  {author} {\bibinfo {author} {\bibfnamefont {S.}~\bibnamefont
  {Chen}}, \bibinfo {author} {\bibfnamefont {S.}~\bibnamefont {Zhou}}, \bibinfo
  {author} {\bibfnamefont {A.}~\bibnamefont {Seif}},\ and\ \bibinfo {author}
  {\bibfnamefont {L.}~\bibnamefont {Jiang}},\ }\bibfield  {title} {\bibinfo
  {title} {Quantum advantages for pauli channel estimation},\ }\href
  {https://doi.org/10.1103/PhysRevA.105.032435} {\bibfield  {journal} {\bibinfo
   {journal} {Phys. Rev. A}\ }\textbf {\bibinfo {volume} {105}},\ \bibinfo
  {pages} {032435} (\bibinfo {year} {2022})}\BibitemShut {NoStop}%
\bibitem [{\citenamefont {Dankert}\ \emph {et~al.}(2009)\citenamefont
  {Dankert}, \citenamefont {Cleve}, \citenamefont {Emerson},\ and\
  \citenamefont {Livine}}]{Dankert:2009exact}%
  \BibitemOpen
  \bibfield  {author} {\bibinfo {author} {\bibfnamefont {C.}~\bibnamefont
  {Dankert}}, \bibinfo {author} {\bibfnamefont {R.}~\bibnamefont {Cleve}},
  \bibinfo {author} {\bibfnamefont {J.}~\bibnamefont {Emerson}},\ and\ \bibinfo
  {author} {\bibfnamefont {E.}~\bibnamefont {Livine}},\ }\bibfield  {title}
  {\bibinfo {title} {Exact and approximate unitary 2-designs and their
  application to fidelity estimation},\ }\href
  {https://doi.org/10.1103/PhysRevA.80.012304} {\bibfield  {journal} {\bibinfo
  {journal} {Phys. Rev. A}\ }\textbf {\bibinfo {volume} {80}},\ \bibinfo
  {pages} {012304} (\bibinfo {year} {2009})}\BibitemShut {NoStop}%
\bibitem [{\citenamefont {Collins}\ and\ \citenamefont
  {{\'{S}}niady}(2006)}]{Collins:2006integration}%
  \BibitemOpen
  \bibfield  {author} {\bibinfo {author} {\bibfnamefont {B.}~\bibnamefont
  {Collins}}\ and\ \bibinfo {author} {\bibfnamefont {P.}~\bibnamefont
  {{\'{S}}niady}},\ }\bibfield  {title} {\bibinfo {title} {Integration with
  respect to the haar measure on unitary, orthogonal and symplectic group},\
  }\href {https://doi.org/10.1007/s00220-006-1554-3} {\bibfield  {journal}
  {\bibinfo  {journal} {Communications in Mathematical Physics}\ }\textbf
  {\bibinfo {volume} {264}},\ \bibinfo {pages} {773} (\bibinfo {year}
  {2006})}\BibitemShut {NoStop}%
\bibitem [{\citenamefont {Pucha{\l}a}\ and\ \citenamefont
  {Miszczak}(2017)}]{Puchala_Z._Symbolic_2017}%
  \BibitemOpen
  \bibfield  {author} {\bibinfo {author} {\bibfnamefont {Z.}~\bibnamefont
  {Pucha{\l}a}}\ and\ \bibinfo {author} {\bibfnamefont {J.}~\bibnamefont
  {Miszczak}},\ }\bibfield  {title} {\bibinfo {title} {Symbolic integration
  with respect to the haar measure on the unitary groups},\ }\href
  {https://doi.org/10.1515/bpasts-2017-0003} {\bibfield  {journal} {\bibinfo
  {journal} {Bulletin of the Polish Academy of Sciences: Technical Sciences}\
  }\textbf {\bibinfo {volume} {65}},\ \bibinfo {pages} {21} (\bibinfo {year}
  {2017})}\BibitemShut {NoStop}%
\bibitem [{\citenamefont {Holmes}\ \emph {et~al.}(2022)\citenamefont {Holmes},
  \citenamefont {Sharma}, \citenamefont {Cerezo},\ and\ \citenamefont
  {Coles}}]{Holmes:2022connecting}%
  \BibitemOpen
  \bibfield  {author} {\bibinfo {author} {\bibfnamefont {Z.}~\bibnamefont
  {Holmes}}, \bibinfo {author} {\bibfnamefont {K.}~\bibnamefont {Sharma}},
  \bibinfo {author} {\bibfnamefont {M.}~\bibnamefont {Cerezo}},\ and\ \bibinfo
  {author} {\bibfnamefont {P.~J.}\ \bibnamefont {Coles}},\ }\bibfield  {title}
  {\bibinfo {title} {Connecting ansatz expressibility to gradient magnitudes
  and barren plateaus},\ }\href {https://doi.org/10.1103/PRXQuantum.3.010313}
  {\bibfield  {journal} {\bibinfo  {journal} {PRX Quantum}\ }\textbf {\bibinfo
  {volume} {3}},\ \bibinfo {pages} {010313} (\bibinfo {year}
  {2022})}\BibitemShut {NoStop}%
\bibitem [{\citenamefont {Mezzadri}(2007)}]{Mezzadri:2007qr}%
  \BibitemOpen
  \bibfield  {author} {\bibinfo {author} {\bibfnamefont {F.}~\bibnamefont
  {Mezzadri}},\ }\bibfield  {title} {\bibinfo {title} {How to generate random
  matrices from the classical compact groups},\ }\href
  {http://www.ams.org/notices/200705/fea-mezzadri-web.pdf} {\bibfield
  {journal} {\bibinfo  {journal} {Notices of the AMS}\ }\textbf {\bibinfo
  {volume} {54}},\ \bibinfo {pages} {592} (\bibinfo {year} {2007})}\BibitemShut
  {NoStop}%
\bibitem [{Note1()}]{Note1}%
  \BibitemOpen
  \bibinfo {note} {The interested reader may even try taking the gradient in
  \protect \eqref {eq:ext_eqn} to set up a steepest-descent method and obtain
  the uniform probability distribution after many iterations.}\BibitemShut
  {Stop}%
\bibitem [{\citenamefont {Arrazola}\ \emph {et~al.}(2021)\citenamefont
  {Arrazola}, \citenamefont {Bergholm}, \citenamefont {Br{\'a}dler},
  \citenamefont {Bromley}, \citenamefont {Collins}, \citenamefont {Dhand},
  \citenamefont {Fumagalli}, \citenamefont {Gerrits}, \citenamefont {Goussev},
  \citenamefont {Helt}, \citenamefont {Hundal}, \citenamefont {Isacsson},
  \citenamefont {Israel}, \citenamefont {Izaac}, \citenamefont {Jahangiri},
  \citenamefont {Janik}, \citenamefont {Killoran}, \citenamefont {Kumar},
  \citenamefont {Lavoie}, \citenamefont {Lita}, \citenamefont {Mahler},
  \citenamefont {Menotti}, \citenamefont {Morrison}, \citenamefont {Nam},
  \citenamefont {Neuhaus}, \citenamefont {Qi}, \citenamefont {Quesada},
  \citenamefont {Repingon}, \citenamefont {Sabapathy}, \citenamefont {Schuld},
  \citenamefont {Su}, \citenamefont {Swinarton}, \citenamefont {Sz{\'a}va},
  \citenamefont {Tan}, \citenamefont {Tan}, \citenamefont {Vaidya},
  \citenamefont {Vernon}, \citenamefont {Zabaneh},\ and\ \citenamefont
  {Zhang}}]{Arrazola:2021quantum}%
  \BibitemOpen
  \bibfield  {author} {\bibinfo {author} {\bibfnamefont {J.~M.}\ \bibnamefont
  {Arrazola}}, \bibinfo {author} {\bibfnamefont {V.}~\bibnamefont {Bergholm}},
  \bibinfo {author} {\bibfnamefont {K.}~\bibnamefont {Br{\'a}dler}}, \bibinfo
  {author} {\bibfnamefont {T.~R.}\ \bibnamefont {Bromley}}, \bibinfo {author}
  {\bibfnamefont {M.~J.}\ \bibnamefont {Collins}}, \bibinfo {author}
  {\bibfnamefont {I.}~\bibnamefont {Dhand}}, \bibinfo {author} {\bibfnamefont
  {A.}~\bibnamefont {Fumagalli}}, \bibinfo {author} {\bibfnamefont
  {T.}~\bibnamefont {Gerrits}}, \bibinfo {author} {\bibfnamefont
  {A.}~\bibnamefont {Goussev}}, \bibinfo {author} {\bibfnamefont {L.~G.}\
  \bibnamefont {Helt}}, \bibinfo {author} {\bibfnamefont {J.}~\bibnamefont
  {Hundal}}, \bibinfo {author} {\bibfnamefont {T.}~\bibnamefont {Isacsson}},
  \bibinfo {author} {\bibfnamefont {R.~B.}\ \bibnamefont {Israel}}, \bibinfo
  {author} {\bibfnamefont {J.}~\bibnamefont {Izaac}}, \bibinfo {author}
  {\bibfnamefont {S.}~\bibnamefont {Jahangiri}}, \bibinfo {author}
  {\bibfnamefont {R.}~\bibnamefont {Janik}}, \bibinfo {author} {\bibfnamefont
  {N.}~\bibnamefont {Killoran}}, \bibinfo {author} {\bibfnamefont {S.~P.}\
  \bibnamefont {Kumar}}, \bibinfo {author} {\bibfnamefont {J.}~\bibnamefont
  {Lavoie}}, \bibinfo {author} {\bibfnamefont {A.~E.}\ \bibnamefont {Lita}},
  \bibinfo {author} {\bibfnamefont {D.~H.}\ \bibnamefont {Mahler}}, \bibinfo
  {author} {\bibfnamefont {M.}~\bibnamefont {Menotti}}, \bibinfo {author}
  {\bibfnamefont {B.}~\bibnamefont {Morrison}}, \bibinfo {author}
  {\bibfnamefont {S.~W.}\ \bibnamefont {Nam}}, \bibinfo {author} {\bibfnamefont
  {L.}~\bibnamefont {Neuhaus}}, \bibinfo {author} {\bibfnamefont {H.~Y.}\
  \bibnamefont {Qi}}, \bibinfo {author} {\bibfnamefont {N.}~\bibnamefont
  {Quesada}}, \bibinfo {author} {\bibfnamefont {A.}~\bibnamefont {Repingon}},
  \bibinfo {author} {\bibfnamefont {K.~K.}\ \bibnamefont {Sabapathy}}, \bibinfo
  {author} {\bibfnamefont {M.}~\bibnamefont {Schuld}}, \bibinfo {author}
  {\bibfnamefont {D.}~\bibnamefont {Su}}, \bibinfo {author} {\bibfnamefont
  {J.}~\bibnamefont {Swinarton}}, \bibinfo {author} {\bibfnamefont
  {A.}~\bibnamefont {Sz{\'a}va}}, \bibinfo {author} {\bibfnamefont
  {K.}~\bibnamefont {Tan}}, \bibinfo {author} {\bibfnamefont {P.}~\bibnamefont
  {Tan}}, \bibinfo {author} {\bibfnamefont {V.~D.}\ \bibnamefont {Vaidya}},
  \bibinfo {author} {\bibfnamefont {Z.}~\bibnamefont {Vernon}}, \bibinfo
  {author} {\bibfnamefont {Z.}~\bibnamefont {Zabaneh}},\ and\ \bibinfo {author}
  {\bibfnamefont {Y.}~\bibnamefont {Zhang}},\ }\bibfield  {title} {\bibinfo
  {title} {Quantum circuits with many photons on a programmable nanophotonic
  chip},\ }\href {https://doi.org/10.1038/s41586-021-03202-1} {\bibfield
  {journal} {\bibinfo  {journal} {Nature}\ }\textbf {\bibinfo {volume} {591}},\
  \bibinfo {pages} {54} (\bibinfo {year} {2021})}\BibitemShut {NoStop}%
\bibitem [{\citenamefont {Weihs}\ \emph {et~al.}(1998)\citenamefont {Weihs},
  \citenamefont {Jennewein}, \citenamefont {Simon}, \citenamefont
  {Weinfurter},\ and\ \citenamefont {Zeilinger}}]{Weihs:1998violation}%
  \BibitemOpen
  \bibfield  {author} {\bibinfo {author} {\bibfnamefont {G.}~\bibnamefont
  {Weihs}}, \bibinfo {author} {\bibfnamefont {T.}~\bibnamefont {Jennewein}},
  \bibinfo {author} {\bibfnamefont {C.}~\bibnamefont {Simon}}, \bibinfo
  {author} {\bibfnamefont {H.}~\bibnamefont {Weinfurter}},\ and\ \bibinfo
  {author} {\bibfnamefont {A.}~\bibnamefont {Zeilinger}},\ }\bibfield  {title}
  {\bibinfo {title} {Violation of bell's inequality under strict einstein
  locality conditions},\ }\href {https://doi.org/10.1103/PhysRevLett.81.5039}
  {\bibfield  {journal} {\bibinfo  {journal} {Phys. Rev. Lett.}\ }\textbf
  {\bibinfo {volume} {81}},\ \bibinfo {pages} {5039} (\bibinfo {year}
  {1998})}\BibitemShut {NoStop}%
\end{thebibliography}
\end{document}